\documentclass[apj,twocolumn, twocolappendix]{openjournal}
\usepackage[breaklinks,colorlinks,citecolor=blue,urlcolor=blue]{hyperref}

\usepackage{amsmath}
\usepackage{booktabs}
\usepackage{multirow,rotating}
\usepackage{color}
\usepackage{soul}
\usepackage{booktabs} 
\usepackage[dvipsnames]{xcolor}
\usepackage{listings}
\usepackage{color}
\definecolor{dkgreen}{rgb}{0,0.6,0}
\definecolor{gray}{rgb}{0.5,0.5,0.5}
\definecolor{mauve}{rgb}{0.58,0,0.82}
\definecolor{golden}{rgb}{0.86,0.65,0.01}
\usepackage{orcidlink}

\def\parcm{\hbox{$.\!^{\prime}$}}
\def\parcs{\hbox{$.\!\!^{\prime\prime}$}}
\def\surfb{{$\,$\text{mag}/$\Box^{\prime\prime}$}}

\begin{document}

\title{The LBT Satellites of Nearby Galaxies Survey (LBT-SONG): The Diffuse Satellite Population of Local Volume Hosts}

\author{A. Bianca Davis\,$^{1,2}$}
\author{Christopher T. Garling\,\orcidlink{0000-0001-9061-1697}$^{2,3,4}$}
\author{Anna M. Nierenberg\,\orcidlink{0000-0001-6809-2536}$^{5}$}
\author{Annika H. G. Peter\,\orcidlink{0000-0002-8040-6785
}$^{1,2,3,6}$}
\author{Amy Sardone\,\orcidlink{0000-0002-5783-145X}$^{2,3,8}$} 
\author{Christopher S. Kochanek\,\orcidlink{0000-0001-6017-2961}$^{2,3}$} 
\author{Adam K. Leroy\,\orcidlink{0000-0002-2545-1700}$^{2,3}$}
\author{Kirsten J. Casey\,\orcidlink{0000-0002-2991-9251}$^{1,2}$}
\author{Richard W. Pogge\,\orcidlink{0000-0003-1435-3053}$^{2,3}$}
\author{Daniella M. Roberts\,\orcidlink{0000-0002-8891-3466}$^{1,2}$}
\author{David J. Sand\,\orcidlink{0000-0003-4102-380X}$^{7}$}
\author{Johnny P. Greco\,\orcidlink{0000-0003-4970-2874}$^{1,2,8}$}

\affiliation{$^{1}$Department of Physics, The Ohio State University, 191 W. Woodruff Ave., Columbus OH 43210, USA}
\affiliation{$^{2}$Center for Cosmology and Astroparticle Physics, The Ohio State University, 191 W. Woodruff Ave., Columbus OH 43210, USA}
\affiliation{$^{3}$Department of Astronomy, The Ohio State University, 140 W. 18th Ave., Columbus OH 43210, USA}
\affiliation{$^{4}$Department of Astronomy, The University of Virginia, 530 McCormick Road, Charlottesville, VA 22904, USA}
\affiliation{$^{5}$Department of Physics, University of California Merced, 5200 North Lake Rd. Merced, CA 95343, USA}
\affiliation{$^{6}$School of Natural Sciences, Institute for Advanced Study, 1 Einstein Drive, Princeton, NJ 08540}
\affiliation{$^{7}$Steward Observatory, University of Arizona, 933 North Cherry Avenue, Rm. N204, Tucson, AZ 85721-0065, USA}
\affiliation{$^{8}$NSF Astronomy and Astrophysics Postdoctoral Fellow}
\email{davis.4811@osu.edu, peter.33@osu.edu}




\begin{abstract}
We present the results of the Large Binocular Telescope Satellites Of Nearby Galaxies Survey (LBT-SONG) ``Far Sample,''  including survey completeness estimates.  We find 10 satellite candidates in the inner virial regions of 13 star-forming galaxies outside the Local Group. The hosts are at distances between $\sim 5-11$ Mpc and have stellar masses in the little explored range of $\sim 5 \times 10^8 - 5\times 10^{10}~\text{M}_{\odot}$. Among the 10 satellite candidates, 3 are new discoveries in this survey.  In this paper, we characterize the properties of 8 low-mass satellite candidates, including the 3 new discoveries but excluding 2 well-studied massive satellites.  Of the 8 low-mass dwarfs, optical colors from the LBT imaging and measurements in the ultraviolet with GALEX suggest that 2 show signs of active star formation, and 6 are likely quenched (although some may still have H\textsc{i} gas reservoirs).  Notably, we report the discovery of an ultrafaint dwarf candidate, NGC 672 dwD, with $\text{M}_{\text{V}} = -6.6$ and an estimated stellar mass of $5.6 \times 10^4 ~\text{M}_{\odot}$ if its association with the host is confirmed.  It is spatially coincident with a weak detection of H\textsc{i}, with $\text{M}_{\text{HI}}/\text{M}_{\text{*}} \sim 1$.  
If confirmed, it would be the least luminous known ultrafaint satellite to be so gas-rich.  
The prevalence of quenched satellites in our sample suggests there are environmental effects at work in lower mass hosts that 
are similar to those at play in Milky Way-size hosts, although the preponderance of H\textsc{i} detections is at odds with the paucity of H\textsc{i} detections in Milky Way satellites. 
By robustly measuring our survey completeness function, we are able to compare our observational results to predictions from theory, 
finding good agreement with the Cold Dark Matter galaxy evolution paradigm.
\end{abstract}

\maketitle



\section{Introduction}

The study of dwarf galaxies ($\text{M}_{*}\lesssim10^9 \text{M}{_\odot}$) lies at the intersection of dark matter physics and the processes of galaxy formation and evolution. Dwarf galaxies reside in low-mass dark matter halos and therefore trace the dark matter distribution on small scales. Observations of dwarf galaxies place constraints on dark matter models, as these models make strong predictions for the cosmic abundance of small
halos \citep[e.g.,][]{Strigari_2007,Nierenberg2012,Horiuchi2014,Kennedy2014,Menci2016,Chau2017,Menci2017,nierenberg2016,Bose2017,Dooley17b,Jethwa_2017,Wang2017,Kim2018,Kim2021,Nadler2019microphys,Nadler2021,Newton2021,Ni2019,Read2019,Sameie2019,Dekker2022,Dekker2024,Hayashi2021,Roberts2021,Rudakovskyi2021,Mau2022,Esteban2023}. 

A key way to test dark matter models is to count halos by counting galaxies, but this depends on knowing how galaxies inhabit halos.  The statistics and global star-formation histories (SFH) of large galaxies are well-described by a simple halo abundance matching prescription that assigns galaxies to halos hierarchically in the context of $\Lambda\text{CDM}$, based on their stellar masses \citep{Tasitsiomi_2004,Conroy_2006,Behroozi13,behroozi2019,Reddick_2013,Moster13,moster2021,rodriguez-puebla2017,WechslerTinker2018,Wang2024}. The stellar mass function of galaxies down to the most massive dwarf galaxies (M$_* \gtrsim 10^7\,$M$_\odot$) is in line with extrapolations of these predictions \citep{Sales13,Carlsten2021,Carlsten2022Elves,Mao2021,Mao2024,driver2022}. Whether this relation continues to hold for lower-mass dwarfs is the subject of active research. While dwarf galaxies are expected to be the most abundant galaxies in the Universe \citep[e.g.,][]{Lan16}, the stellar mass function below $\text{M}_{*}\lesssim10^7\,\text{M}_\odot$ is almost completely uncharacterized outside the Local Group. This is because these galaxies are intrinsically faint and often have low surface brightness \citep{Baldry08,Baldry2012,Munoz15,danieli2018,Karachentsev2019,Carlsten2021size,driver2022}, making their discovery in the field difficult \citep[although not impossible;][]{mutlu-pakdil2021,sand2022,casey2023,li2024}.
The best studied low-mass galaxies are thus those around the Milky Way (MW).

On the smallest end, many ``ultrafaint" dwarf (UFD) galaxies \citep[$\text{M}_{*}\lesssim10^5\,\text{M}_{\odot}$ or $\text{M}_{\text{V}}\gtrsim-8$;][]{Willman05a} have been discovered around the MW in the past two decades. These galaxies are thought to be quenched by the global process of reionization regardless of environment \citep{Benson02,Bovill11,Brown14b,weisz2014b,rodriguezwimberly2019}, unlike their more massive ``classical dwarf'' ($10^5 \,\text{M}_\odot \lesssim \text{M}_{*}\lesssim10^7\,\text{M}_{\odot}$) counterparts, which experience star formation after reionization. The discovery of ultrafaint MW satellites in surveys such as SDSS, DES, Pan-STARRS, and HSC-SSP \citep[e.g.,][]{belokurov2008,belokurov2010,Bechtol15,DrlicaWagner2015,Kim15a,Kim15b,Koposov15,Laevens15,Homma2018,Homma2019,Koposov2018}, combined with completeness corrections \citep{Tollerud2008,Koposov2009,Walsh2009,Jethwa_2017,Newton2018,Kim2018,drlica-wagner2020,Nadler2020} suggest that the total number of MW satellites is likely in agreement with the $\Lambda\text{CDM}$ prediction. 

However, the more massive classical satellites -- the original motivators of the ``missing satellites problem" \citep{Moore_1999,Klypin99} -- are clearly affected by baryonic physics, even as they are intriguing testbeds for dark matter microphysics \citep[][]{Brooks2013,wetzel2016,Buck2019,Read2019,akins2021,applebaum2021,Hayashi2021,Santos-Santos2021,olsen2022}. Because environmental processes such as starvation \citep{larson1980} and tidal and ram-pressure stripping \citep{gunn1972,Garling2024,zhu2024} can quench star formation in dense environments but are not present in field environments, the mapping between the stellar mass of dwarf galaxies and their halos may depend on environment \citep{Christensen2024}.  

There is evidence of varied environmental impacts on dwarf galaxies when we look at the star-forming versus quenched fraction of satellites of MW-mass systems compared to field environments.  In our own MW, with the exception of the Magellanic Clouds, the classical dwarf satellites are devoid of neutral hydrogen, and are quenched. Based on HST-derived SFH and orbit modeling, a combination of ram-pressure stripping, tidal stripping, and outflows likely governs quenching of classical Local Group satellites \citep{mayer2006,nichols2011,slater2014,wetzel2015,fillingham2016,digby2019,garrison-kimmel2019}. The hot gas halo of the MW is thought to cause ram pressure stripping of satellites which may lead to their quenching \citep{gatto2013,slater2014,fillingham2016,akins2021}. Most MW-sized systems (in particular those studied in the SAGA \citep{geha2017saga,Geha2024} and ELVES \citep{Carlsten2022Elves,Greene2023} surveys) find a lower fraction of satellites to be quenched than the MW, although the MW is not an extreme outlier in quenched fraction \citep{karunakaran2019,karunakaran2022b,samuel2022,Samuel2023,zhu2023}.  \citet{jones2024} find evidence of ongoing ram-pressure stripping and an overall suppression in HI mass in SAGA satellites relative to the field.  
By contrast, massive dwarf galaxies ($\text{M}_{*}\gtrsim10^7\,\text{M}_{\odot}$) found in the field are overwhelmingly star-forming \citep{haines2007,geha2012,papastergis2012,kawinwanichakij2017}.  These findings indicate that environment plays a large role in quenching star formation and may therefore significantly affect the satellite luminosity function at these mass scales.

To explain this ensemble of results, a number of simulations find, for MW-mass hosts, that satellite quenching is a strong function of satellite stellar mass and the mass of the circumgalactic medium of the host \citep{akins2021,karunakaran2021,karunakaran2022,font2022,samuel2022}.  Massive satellites (M$_* \gtrsim 10^8\,$M$_\odot$) are almost universally star-forming, and small ones (M$_* \lesssim 10^6\,$M$_\odot$) are almost universally quenched with a short quenching time \citep{wheeler2014,wetzel2015}.  The key debate is exactly where the sharp transition in quenched fraction occurs.  \citet{karunakaran2019} suggested that the V-band absolute magnitude limit (M$_V = -12.1$) from the SAGA-I survey defines a dividing line for satellite galaxies: fainter galaxies are largely quenched, and brighter ones are largely star-forming \citep[see also][]{karunakaran2022b}.  

However, it is unclear if this trend and explanation are true more broadly for $\text{L}_{*}$ hosts \citep{karunakaran2021,samuel2022}, much less if this is true for less massive hosts. Hosts less massive than the MW have gentler tidal fields and are unlikely to have massive hot gas halos \citep{dekel2006}, making starvation (the cessation of gas accretion) and internal quenching mechanisms more important than ram-pressure stripping \citep{Stark2016}, although outflowing material from low-mass hosts may play a role in quenching satellites \citep{Garling2024}. Recent work suggests that all these mechanisms are at play for low-mass hosts, although their relative strengths may be different than for more massive hosts \citep{Carlin_2019,carlin2021,garling2019ddo113,garling2021,garling2022,jahn2021,Bhattacharyya2023,Samuel2023,Garling2024}.

To disentangle environmental effects on satellites and their connection to dark-matter halos, we need to look outside the Local Group and systematically explore a range of environments.  Specifically, surveys designed to probe the satellite luminosity functions of MW- (and lower-) mass hosts into the ultrafaint regime are needed to differentiate among these scenarios.  
In recent years, efforts have been undertaken to characterize satellite systems beyond the Local Group, but around nearby systems in the Local Volume ($< 11$ Mpc). In this distance range, one can use resolved stars or diffuse light to search for satellites, depending on the distance. The MW-like satellite systems that have been studied in the Local Volume include NGC 253 \citep{martinez-delgado2021,mutlu-pakdil2022n253,mutlu-pakdil2024}, NGC 4258 \citep{Spencer2014}, M94 \citep{smercina2018lonely}, M101 \citep{DanieliM101,bennet2019m101,Carlsten2020,bennet2020}, NGC 3175 \citep{Kondapally2018}, NGC 2950 and NGC 3245 \citep{Tanaka2018}. A statistical study of 30 MW-mass Local Volume systems has confirmed satellites based on surface brightness fluctuations (SBF) in the ELVES survey \citep{Carlsten2022Elves}. A handful of satellite systems around more massive galaxies have also been studied, such as the systems around M81 \citep{chiboucas2013confirmation} and Centaurus A \citep{crnojevic2019faint,Muller2018}. At much lower mass scales, the MADCASH  \citep{sand2015,Carlin16,Carlin_2019,hargis2019} and DELVE-DEEP \citep{delve2021,mcnanna2024} surveys study the satellites of LMC-mass hosts. 

We still need a study of intermediate mass hosts, those between LMC-mass and MW-sized hosts ($\sim 10^9\,$M$_\odot \lesssim\,$M$_* \lesssim 10^{11}\,$M$_\odot$). The LBT-SONG survey is an effort to statistically study a total of 19 hosts between 4-11 Mpc. Reprocessing data from a search for failed supernovae \citep{Kochanek08,Gerke2015,Adams2017,Neustadt2021}, we search deep images of the central regions of the target galaxies 
to identify and characterize dwarf galaxies. Hosts within 5 Mpc can be studied using resolved stars and were studied in the LBT-SONG Near Sample \citep{garling2021}. While no new dwarf galaxies were identified in the Near Sample, the two well-known massive dwarf spheroidals in the sample, DDO 113 \citep{garling2019ddo113} and LV J1228+4358 \citep{garling2021}, are gas-poor and were demonstrated to be environmentally quenched by their Magellanic Cloud-mass hosts. The number of satellites seen in the near part of survey was consistent with expectations, but the fact that the two identified satellites were nearly identical in luminosity (and bright) was not expected. 

In this paper, we present the results from the LBT-SONG Far Sample. The organization of this paper is as follows.  In Section \ref{sec:data} we describe our data and present the LBT-SONG Far Sample host galaxies. In Section \ref{sec:Pipeline} we summarize our data processing procedures and mock galaxy creation, insertion and recovery methods along with the completeness calculations, based on our earlier study of the NGC 628 system \citep{Davis2021}. We present the completeness results for all host galaxies in Section \ref{sec:completeness}. Dwarf galaxy candidates and their properties are presented in Section \ref{sec:dwarfCandidates}. We compare our satellite number counts to the expectations for the $\Lambda\text{CDM}$ model in Section \ref{sec:satelliteCounts}, and summarize our findings in Section \ref{sec:summary}.

\section{Data and Host Sample}\label{sec:data}

\subsection{Observations}

Our data are a subset of those originally obtained in a search for failed supernovae \citep{Kochanek08,Gerke2015,Adams2017,Neustadt2021}, which monitors 27 star-forming galaxies in the U, B, V, and R bands using the the Large Binocular Cameras \citep[LBC;][]{Ragazzoni2006,Giallongo2008,Neustadt2021} on the Large Binocular Telescope \citep[LBT;][]{Hill2010LBT}. Each galaxy is observed in a single LBC pointing, usually with the target galaxy lying on the central CCD of the camera. Each of the four $2048 \times 4096$ LBC chips covers 17\parcm3$\times$7\parcm7 with a pixel scale of 0\parcs225, resulting in a total area of approximately 23$^\prime \times \text{23}^\prime$. 

Data reduction, photometry, and calibration were done according to the methods described in \citet{garling2019ddo113} and summarized in \citet{Davis2021}. The final exposure times for each stacked image in each band can be found in Table \ref{table:hostprop}. The average final 5-$\sigma$ point-source depths of the stacked images are 25.0 mag in U and 27.0 mag in the BVR bands. 

\subsection{Overview of Host Galaxies}

The original data set contains 27 star-forming galaxies within $\sim 11$ Mpc. We remove those hosts which are interacting pairs or where the hosts span a prohibitively large fraction of the footprint (e.g. NGC 3628). We also remove hosts with high Galactic extinction (e.g. NGC 6946), complicated cirrus (e.g. M81), or that have already been extensively studied in the literature (M101). Of the remaining host galaxies, 6 were within 5 Mpc and suitable for resolved stellar population studies, and were studied in the LBT-SONG Near Sample \citep{garling2021}. The remaining 13 host galaxies at distances between $\sim 5-11$ Mpc make up the LBT-SONG Far Sample that we consider here. Their properties are summarized in Table \ref{table:hostprop}. The host galaxies range in stellar mass from $4.7 \times 10^{8} \,\text{M}_\odot$, roughly that of the SMC, to $5.4 \times 10^{10}\, \text{M}_\odot$, roughly that of the MW, thus capturing a range of intermediate masses. 

The LBT-SONG host galaxies offer the possibility of probing different environments, as their tidal indices $\theta_1$ (last column of Table~\ref{table:hostprop}), which take into account the distance and mass of the nearest significant neighbor, range from $-1.1\, \text{to}\, 3.4$ \citep{Karachentsev2018}.  Negative values correspond to isolated galaxies and positive values indicate group membership. Local Volume galaxies span $-3<\theta_1<7$. Roughly half our Far Sample hosts are isolated galaxies ($\theta_1<0$). Among those in groups, the most extreme is NGC 5474, which is gravitationally distorted by its larger companion, M101.

\begin{table*}
\setlength{\tabcolsep}{5pt}
\centering

 \label{table:hostprop}
 \begin{tabular}{lcccrcccccc}
  \hline
  Host & R.A. & Decl. & $\text{Distance}$ & $\text{M}_{\star}^{\text{a}}\hspace{0.5cm}$
 &  $\text{M}_{200}^{\text{b}}$ &$\text{R}_{200}^{\text{c}}$ & $\text{Area}^{\text{d}}$ &$\text{Exposure}^{\text{e}}$& $\text{E(B-V)}^{\text{f}}$&$\theta_{1}^{g}$\
   \cr & [J2000] &[J2000]  &  $\text{[Mpc]}$ & $[\text{log}_{10}\text{(M}_{\odot})]$ & $[\text{log}_{10}\text{(M}_{\odot})]$ & [kpc] & [$\text{kpc}^2$] & [s] &[mag] & \\\hline
  NGC 4826 & 12:56:43.64 &	+21:40:58.69 & 4.66$^{\text{h}}$ & $10.26\pm0.10$ & 11.6 & 167 & 979,849,398 & 5007,8612, & 0.037 & $-0.9$\\
  &&&&&&&& 6409,23033 &\\
  NGC 5474 & 14:05:01.61 &	+53:39:44.00 & 5.55$^{\text{i}}$ & $8.66\pm0.10$ & 10.9 & 86 & 1934,1392,1364 & 4804,5405, & 0.009 & 3.4\\
  &&&&&&&& 1802,17113 &\\
  NGC 6503 & 17:49:26.43 & +70:08:39.72 & 6.28$^{\text{i}}$ & $9.71\pm0.10$   & 11.3 & 124 & 1178,1279,1279 & 3364,3364,&0.028& 0.0\\
  &&&&&&&& 4085,10810 &\\
  NGC 4605 & 12:39:59.38 &	+61:36:33.10 & 6.55$^{\text{i}}$ & $9.52\pm0.10$   & 11.2 & 115 & 1389,1187,1027 & 2643,4325, & 0.013 & $-1.1$\\
  &&&&&&&& 4085,10090&\\
  NGC 4258 & 12:18:57.51 &	+47:18:14.30 & 6.82$^{\text{i}}$ & $10.57\pm0.10$ & 12.1 & 217 & 2097,1591,792 & 10614,17725, &0.014& $-0.1$\\ 
  &&&&&&&& 14021,31456 &\\
  NGC 3489 & 11:00:18.57 &	+13:54:04.40 & 7.18$^{\text{j}}$ & $9.80\pm0.10$  & 11.4 & 128 & 2324,2041,1717 & 4807,4807, & 0.014& $-0.1$\\
  &&&&&&&& 1602,9211 &\\
  NGC 672  & 01:47:54.52 &	+27:25:58.00 & 7.18$^{\text{k}}$ & $9.25\pm0.12$   & 11.1 & 104 & 2324,1964,1430 & 4804,7298, & 0.069& 0.2\\
  &&&&&&&& 7809,15912 &\\
  NGC 5194 & 13:29:52.71 &	+47:11:42.62 & 7.55$^{\text{i}}$ & $10.62\pm0.10$  & 12.1 & 229 & 2570,1482,1042 & 8011,10715, & 0.031& 0.1\\
  &&&&&&&& 9413,25832 &\\
  NGC 2903 & 09:32:10.11 &	+21:30:03.00 & 8.00$^{\text{l}}$ & $10.37\pm0.10$  & 11.8 & 175 & 2885,2317,1783 & 6607,9310, & 0.028 &$-0.3$\\
  &&&&&&&& 7569,24680 &\\
  NGC 3344 & 10:43:31.15 &	+24:55:20.00 & 8.28$^{\text{i}}$ & $10.13\pm0.10$  & 11.6 & 150 & 3091,2545,2065 & 2753,4255, & 0.029& $-0.3$\\
  &&&&&&&& 3254,7007 &\\
  NGC 925  & 02:27:16.88 &	+33:34:45.00 & 9.16$^{\text{m}}$ & $9.75\pm0.11$   & 11.4 & 126 & 3783,3185,1516 & 7207,7507, & 0.067& 1.0\\
  &&&&&&&& 7808,17714 &\\
  NGC 628 & 01:36:41.77 & +15:47:00.46 & 9.77$^{\text{n}}$ & $10.24\pm0.10$ & 11.7 & 160 & 4304,3051,1350& 6607,9310,  & 0.062 &0.2 \\
  &&&&&&&& 7569,24680 &\\
  NGC 3627 & 11:20:14.96 &	+12:59:29.54 & 10.70$^{\text{i}}$ & $10.68\pm0.10$ & 12.3 & 248 & 5162,3949,3213 & 4809,6612, & 0.029 & 0.6\\
  &&&&&&&& 6161,23736 &\\

  \hline
 \end{tabular}

 \caption{\label{table:hostprop}\textbf{Properties of the Host Galaxies.}  Host galaxies are ordered by increasing distance from the MW.  $^{\text{a}}$The stellar masses $\text{M}_*$ are from \citet{Leroy2019} and scaled to the distances in the 4th column. $^\text{b}$The $\text{M}_{200}$ halo masses are derived using the $\text{M}_*-\text{M}_\mathrm{halo}$ relation of \citet{Moster13}.  $^\text{c}$Virial radii $\text{R}_{200}$ are based on the \citet{Moster13} relation.  $^{\text{d}}$The three values in this column are the total unmasked area, and the available area after applying the first and the second, more conservative, masks respectively. $^{\text{e}}$The exposure times are for U, B, V, and R bands, respectively. $^{\text{f}}$\citet{Schlegel1998,Schlafly11}. $^{\text{g}}$Tidal index, as described and tabulated in \citet{Karachentsev2018}. $^{\text{h}}$\citet{Jacobs2009};$^{\text{i}}$\citet{Sabbi2018}; $^{\text{j}}$\citet{Theureau2007};$^{\text{k}}$\citet{Tully2013}; $^{\text{l}}$\citet{Carlsten2020};$^{\text{m}}$\citet{Karachentsev2003};$^{\text{n}}$\citet{Mcquinn2017Distance}.}

\end{table*}

\begin{figure}[t]
 \includegraphics[width=\columnwidth]{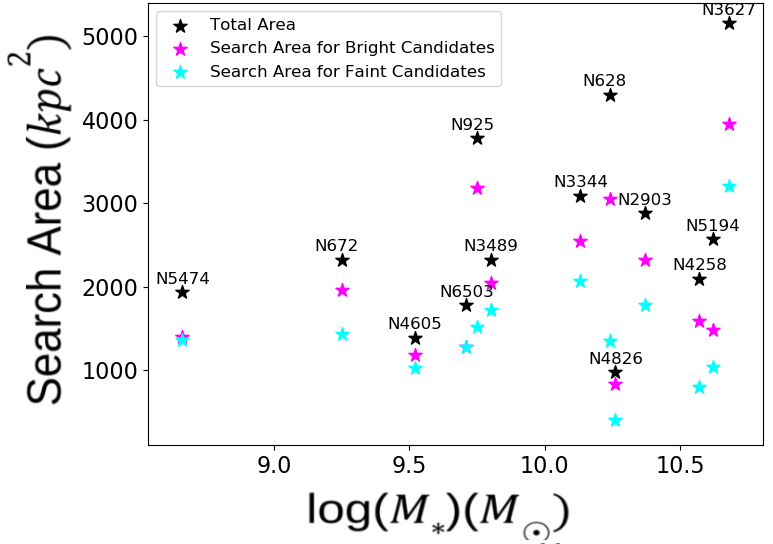}
 \caption{The physical search area for the host galaxies versus stellar mass, assuming the host distances shown in Table~\ref{table:hostprop}. The black stars represent the total area in the LBT pointing. The cyan and magenta stars represent the available search area to search for the bright and faint candidates, after masking, respectively. Masking and completeness are discussed in Section \ref{sec:completeness}.}
 \label{fig:searchArea}
\end{figure}

We can estimate what fraction of a host's satellites may lie in our footprint, a topic we return to in Section~\ref{sec:satelliteCounts}.  Because of the host distance range of 4.7 Mpc to 10.7 Mpc, the LBC footprint captures areas ranging from 30 kpc $\times$ 30 kpc to 70 kpc $\times$ 70 kpc. To determine what fraction of the galaxies' virial volumes we probe, we use the $\text{M}_*-\text{M}_\mathrm{halo}$ relation of \citet{Moster13} to obtain halo virial masses M$_{200}$ (defined to have an average density of 200 times the critical density of the Universe) and virial radii R$_{200}$.  We estimate the total unmasked search areas correspond to $\sim 0.1\%$ to $\sim 20\%$ of the virial volume of the hosts.
Depending on the radial distribution of satellites, we expect anywhere from 1$-$35\% (if satellites are distributed isothermally) to 5$-$60\% (if satellites trace the smooth halo of the host) of the bound satellites to lie within the LBC footprint for the whole LBT-SONG sample. These two choices of radial distribution bracket the extremes typically considered in the literature, and reflect the theoretical and observational uncertainties in satellite distributions \citep{Hargis14,Dooley2017a,Kim2018,Carlsten2020Rad,Mao2024}. If the satellites have the same radial distribution as the MW-mass SAGA galaxies \citep{Mao2024}, we expect $\sim 10-30\%$ of our hosts' satellites to lie in the LBT-SONG Far Sample. 

In Fig.~\ref{fig:searchArea} we show search areas for each galaxy, after masking, as a function of host stellar mass. We scale the stellar masses from \citet{Leroy2019} to the host distances used in this work (Table~\ref{table:hostprop}). The black, cyan and magenta stars represent the physical areas available in the LBT pointing, and then the area remaining after applying the masks for the bright and faint candidate searches, respectively.  As we discuss in  Section \ref{sec:Pipeline}, a more conservative masking of the host galaxies is necessary to identify faint, low surface brightness dwarfs than brighter or higher surface brightness galaxies.

\section{Methods} \label{sec:Pipeline}

To perform our search for satellite dwarf galaxies as diffuse objects, while simultaneously quantifying completeness, we employ the diffuse dwarf galaxy detection pipeline described in detail in \citet{Davis2021}. We provide a summary of this pipeline here and note a few changes made relative to that prior work.

We inject and then use {\tt{SExtractor}} \citep{Bertin1996} to recover mock galaxies to quantify our completeness while simultaneously searching for potential candidates in the images. We begin by simulating 96 simple-stellar-population (SSP) galaxy models with combinations of luminosities, surface brightnesses, ages and metallicities spanning the range of properties found for Local Group dwarfs \citep{McConnachie12} and summarized in Table \ref{table:mocks}. These models are defined as in \citet{Davis2021}. In short, we use \citet{Marigo17} SSPs with a Chabrier initial mass function \citep{Chabrier2001} to model stellar populations, and model the spatial distribution of stars with \citet{plummer1911} profiles.  A change with regard to \citet{Davis2021} is that we add mock galaxies with $\text{M}_{\text{V}}  = -10$ and drop those with $\text{M}_{\text{V}}  = -6$, as their small size often led to confusion with background compact galaxies.

\begin{table}
\centering 
 \begin{tabular}{ccc}
  \hline
  Property & Values \\
  \hline
  $\mu_{\rm eff}$ in V band ($\rm mag/\square''$)& 25, 26, 27, 28 \\
  $\text{M}_{\text{V}}$ & $-7$,$-8$,$-9$,$-10$ \\
  $\left[ \text{Fe} / \text{H} \right]$ & $-2$,$-1$ \\
  Age (Myr) & $10^2, 10^3, 10^4$ \\
  \hline
 \end{tabular}
\caption{\label{table:mocks}\textbf{The range of simulated dwarf galaxy properties.} Each mock galaxy is defined by its effective surface brightness in V band, V-band absolute magnitude, metallicity and age. We simulate galaxies with all possible combinations of these properties, resulting in 96 mock galaxy models.}
\end{table}

We create {\tt{SExtractor}} weight images and apply one of two masks, depending on the properties of the mock galaxies. The first mask is used for $-10 \ge \text{M}_{\text{V}} \ge -8$ and  $\mu_{\rm eff}$ brighter than 28 \surfb{}.  It mainly masks foreground stars and diffuse halo light. A more conservative set of masks is used for models with $\text{M}_{\text{V}}= -7$ or $\mu_{\rm eff} = 28$\surfb{}. These mask fainter foreground stars and background objects as well as extended cirrus or diffuse light near the host. 

Each mock galaxy model is injected into the images at the assumed host distance, on grids spanning the field. A range of {\tt{SExtractor}} DETECT{\_}MINAREA and ANALYSIS{\_}THRESH parameters, as well as all four detection filters are used to search for the mock galaxies. All other {\tt{SExtractor}} parameters are the same as in \citet{Davis2021}, with the exception of the convolution filter, where we instead used the tophat{\_}5.0{\_}5x5 filter because we found that it produced fewer spurious detections than the tophat{\_}1.5{\_}3x3 filter we used in our prior work. The range of DETECT{\_}MINAREA and ANALYSIS{\_}THRESH and the fixed {\tt{SExtractor}} parameters were chosen to minimize false positives and the fragmentation of the larger, more diffuse mock galaxies. To maximize the completeness while minimizing the number of false positives, we selected the combination of search parameters and detection filters which simultaneously found at least $90\%$ of the injected models, and had the smallest total number of additional {\tt{SExtractor}} detections. These optimal parameters are then used on the real data for the candidate search and to characterize the completeness.

After running {\tt{SExtractor}} with the optimized parameters, we remove candidates with more than 25\% of their pixels masked within their {\tt{SExtractor}} half-light radii. 
We next apply selection cuts to better discriminate the real mock galaxies from false positives. As in \citet{Davis2021}, we apply selection cuts on the apparent R band magnitude, the radii containing $50\%$ and $10\%$ of the light, $\text{r}_{0.5}$ and $\text{r}_{0.1}$, and the ratio of $\text{r}_{0.5}$ and $\text{r}_{0.1}$ to reject stars, and on the {\tt{SExtractor}} flags. Generally we rejected objects with FLAGS $\ge 3$, which corresponds to objects that are both deblended and have more than $10\%$ bad pixels in the aperture, or have saturated pixels or are too close to the image edge. A cutoff of $\text{r}_{0.5} \geq 6$ pixels was used for all hosts, as candidates below this threshold suffered from point source confusion. This selection cut removed a large number of false positives from our candidate catalogs while having very little impact on our completeness estimated from our injected mock galaxies.  
The values for all other selection cuts varied by host, as the optimal values depend on distance. 

These selection cuts are applied to obtain candidates for visual inspection and for calculating the completeness. We expect only a small number of dwarf satellites in the footprint, so the vast majority of candidates are spurious detections. The false positives are dominated by blends of point-like sources with background and foreground diffuse light (e.g., a nearby bright star, spiral arms of the host, massive low-z galaxies, or Galactic cirrus). 
 On average, application of the above cuts removed $91\%$ of the candidates while only removing $26\%$ of the mock galaxies. After applying the cuts, we are left with 300-1500 candidates for each host. The variation is mainly due to differences in foreground and background contamination and crowding. After visual inspection---both of the candidates from our pipeline as well as an inspection of all the images, to capture any large objects that may not have been flagged by the pipeline---we identify 10 high-confidence satellite candidates. Two of these are massive well-known satellites of their hosts (NGC 4248 and IC 1727), and 8 are faint dwarf satellites candidates.

\section{Completeness Results}\label{sec:completeness}

\begin{figure*}[tp]
\begin{centering}
 \includegraphics[width=\textwidth]{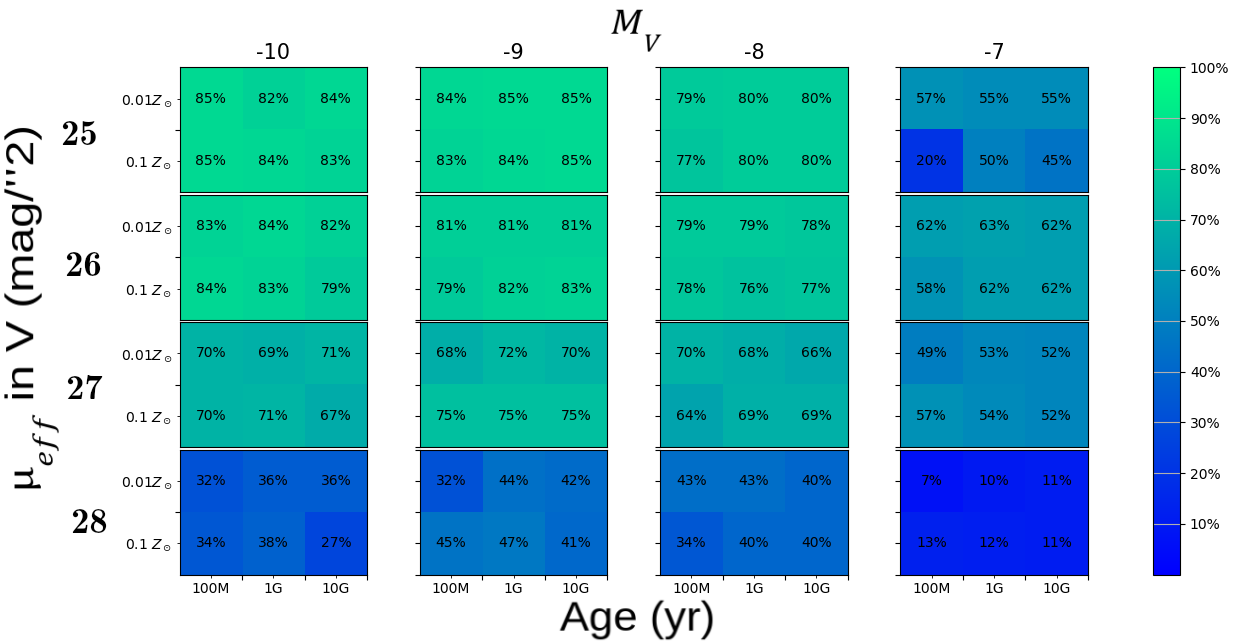}
 \caption{The completeness averaged over all hosts. This figure includes the incompleteness due to masking.  Each box is for galaxies of a single $\rm \text{M}_{\text{V}}$ and $\mu_{\text{V}}$, increasing in V-band absolute magnitude towards the right and fainter in surface brightness going downward. Within each box are six numbers for the completeness as a function of the age (columns) and metallicity (rows) of the SSP, as labeled along the bottom and left of the figure. 
 \label{fig:complRaw}}
\end{centering}
\end{figure*}

\begin{figure*}
\begin{centering}
 \includegraphics[width=\textwidth]{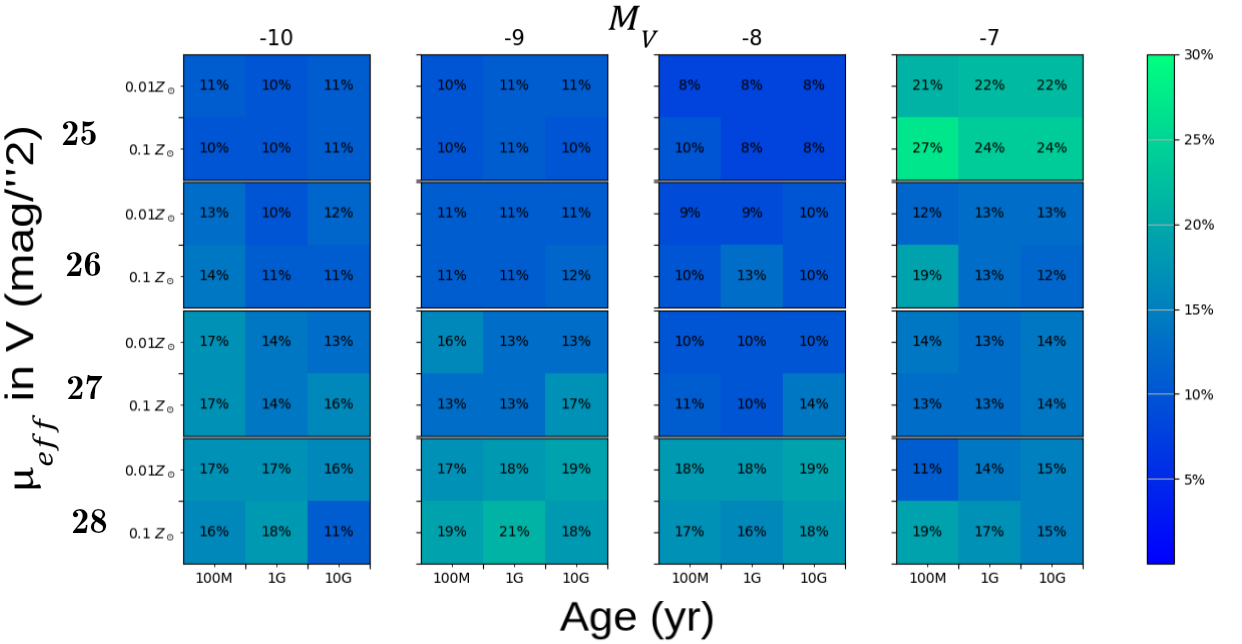}
 \caption{The standard deviation among the completeness results for all hosts as a function of mock galaxy model, including the effects of masking. This figure illustrates the variation in completeness from host to host.  The structure of the figure is the same as for Figure \ref{fig:complRaw}.}
 \label{fig:complRawErr}
\end{centering}
\end{figure*}

We calculate the completeness for every model galaxy for each host in order to compare theoretical predictions for the satellite luminosity functions to our final satellite sample (Section~\ref{sec:satelliteCounts}). For each figure in this section, beginning with Fig.~\ref{fig:complRaw}, each box represents a set of models with fixed $\rm \text{M}_{\text{V}}$ and $\mu_{\text{eff}}$, increasing in V band magnitude (fainter galaxies) towards the right and fainter in surface brightness downward. In each box, there are six numbers for the completeness given for the six combinations of age and metallicity used for the SSP models (see Table~\ref{table:mocks}).  
Fig.~\ref{fig:complRaw} shows the completeness averaged over the hosts, given by the ratio of the recovered galaxies that passed all selection cuts to the total number of injected galaxies, including the effect of masking. 
We show the host-to-host dispersion in the completeness percentage as a function of model in Fig.~\ref{fig:complRawErr}. 

As expected, we are more complete for the brighter, higher surface brightness galaxy models. In Fig.~\ref{fig:complRaw}, we recover nearly all galaxies in the unmasked regions with $\text{M}_{\text{V}} = -10$ and $-9$ and $\mu_{\text{eff}}$ = 25 and 26 \surfb{}. Because the recovery rate is essentially 100\% in the unmasked region for these models, the completeness describes the average masking of the image for bright satellite searches.  When we increase the V-band magnitude or the central surface brightness by one magnitude (making galaxies fainter and more diffuse), we recover roughly $90\%$ of the injected galaxies in the unmasked regions (and 70\%-80\% in the footprint overall) with little variation due to differences in age or metallicity. We are less complete for models with $\text{M}_{\text{V}} = -8$ with $\mu_{\text{eff}}$ = 27 and $\text{M}_{\text{V}} = -7$ with $\mu_{\text{eff}}$ = 26, recovering approximately $80\%$ of mock galaxies in the unmasked regions. Models with $\mu_{\text{eff}}$ = 28 as well as the fainter $\text{M}_{\text{V}} = -7$ with $\mu_{\text{eff}}$ = 27 have still lower completeness.  Because the hosts are more aggressively masked in searches for mock galaxies with very low surface brightness, we are relatively less complete to faint and low surface brightness galaxies just on the basis of masking.  Models with $\text{M}_{\text{V}} = -7$ with $\mu_{\text{eff}}$ = 28 were not visually distinguishable from background variations and are the detection limit in the LBT-SONG distant host sample. While fainter, more compact models such as those with $\text{M}_{\text{V}} = -7$ and $\mu_{\text{eff}}$ = 25 are easily detected, their half-light radii are often below our selection cut of 6 pixels, and would be easily confused with background galaxies.  However, as we discuss in Sec.~\ref{sec:dwarfCandidates}, such faint galaxies are not expected to have such bright surface brightnesses.  The large host-to-host variation in the completeness for these models in Fig.~\ref{fig:complRawErr} highlights the distance-dependence as to whether these faint, compact satellite candidates pass our angular size cut.  

These completeness measurements are a key input for our prediction for the $\Lambda$CDM prediction of the satellite population that we describe in Sec.~\ref{sec:satelliteCounts}.

\section{Dwarf galaxy candidates} \label{sec:dwarfCandidates}

\begin{figure*}[t]
\begin{centering}
 \includegraphics[width=\textwidth]{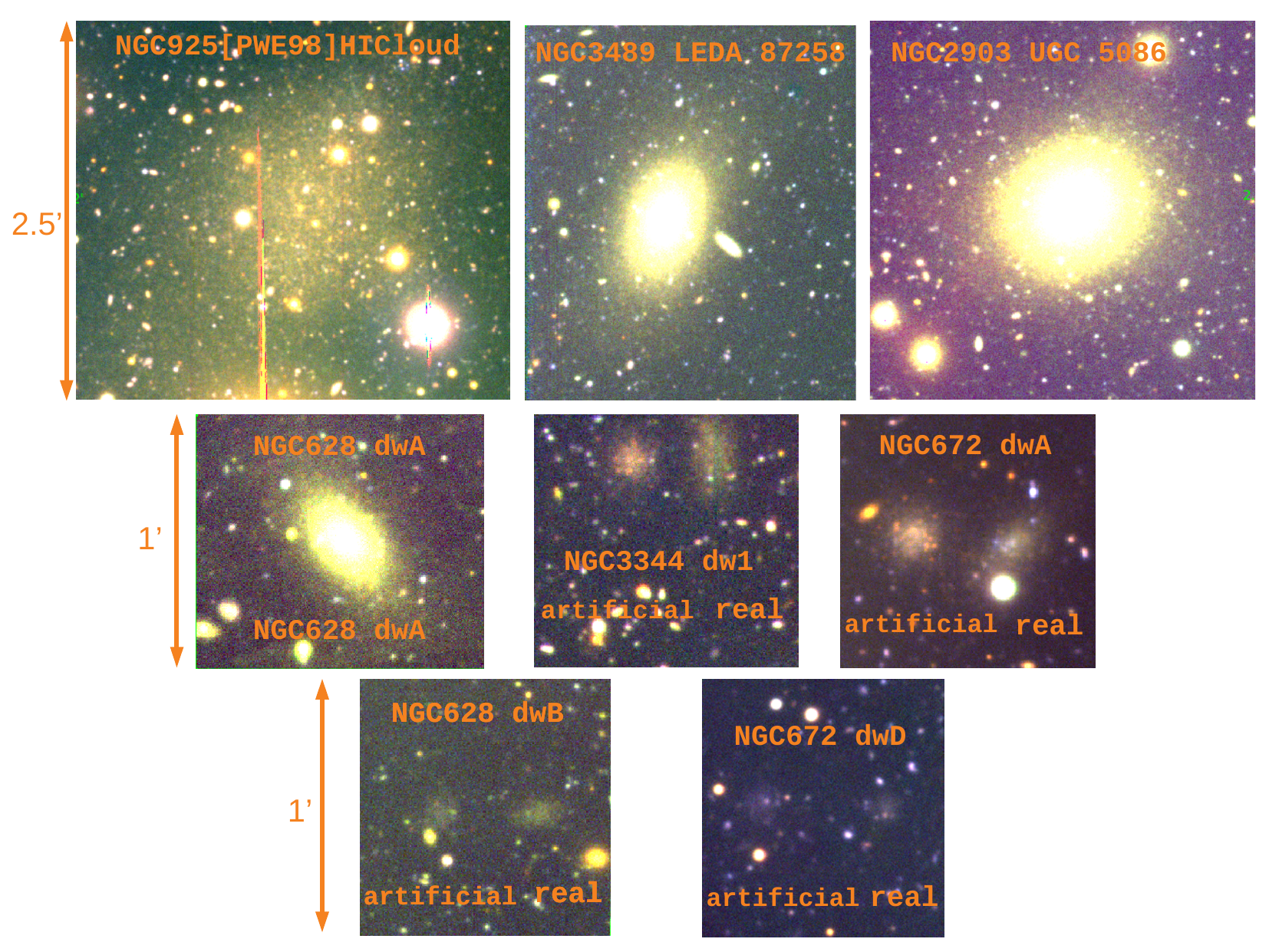}
 \caption{Composite LBC BVR images of the eight candidate low-mass dwarf satellites found in the LBT-SONG Far Sample, excluding IC 1727 and NGC 4248.
Candidates are grouped by angular half-light radius, with the most extended galaxies in the top row, and the smallest in the bottom row. Each image on the top row is 2\parcm{}5 on a side, while those on the middle and bottom rows are 1\parcm{}0 on a side. For the
four faintest galaxies $(\text{M}_\text{V}\gtrsim -10)$, we show the artificial dwarfs with the most similar $\text{M}_\text{V}$ and surface brightness at the host distance. The mock galaxy analogs for NGC 3344 dw1, NGC 672 dwA, NGC 628 dwB and NGC 672 dwD have ($\text{M}_\text{V},\mu_\text{V}$ in \surfb{}) $= (-9,\, 26),\, (-10,25),\ (-8,\, 27)$, and $(-7, 27)$, respectively. All four artificial galaxies have ages of 1 Gyr and metallicities $\text{Z} = 0.01 \text{M}_{\odot}$. There is an image artifact running
through the V band image of NGC 3344 dw1.}
 \label{fig:candidateGallery}
 \end{centering}
\end{figure*}

Around the 13 host galaxies in the LBT-SONG Far Sample, we find 10 satellite galaxy candidates. Of these 10, IC 1727 and NGC 4248 are well-characterized, massive (Magellanic Cloud-mass) satellites of their hosts.  Because other work has shown that quenching operates differently for such massive dwarf satellite galaxies in a wide range of environments than for less massive dwarf galaxies \citep{wheeler2014,stierwalt2015,pearson2016,karunakaran2021}, they are not the focus of this paper.   
We focus on the remaining 8 dwarf satellites and dwarf satellite candidates shown in Fig.~\ref{fig:candidateGallery}.  We will introduce these satellite candidates briefly here, present their measured and derived properties in Sec.~\ref{sec:compiled}, and compare their properties to those of well-studied Local Group dwarf galaxies in Sec.~\ref{sec:compare}.

Of the 8 (candidate) low-mass dwarf satellite candidates, 3 were discovered in the LBT-SONG survey.  Two of these were identified close in projection to NGC 628 and were presented in \citet{Davis2021}.  Since their discovery, the distance of the brighter of the pair (NGC 628 dwA) was measured, confirming its association with NGC 628 \citep{Carlsten2022Elves}.  We have obtained a preliminary surface brightness fluctuation distance measurement to the fainter candidate (NGC 628 dwB) that also suggests an association with NGC 628.  We will present the full description of the distance measurement in a separate paper.  In this work, we present the faintest satellite dwarf candidate discovered in the LBT-SONG survey, NGC 672 dwD.

The remaining 5 satellite candidates were previously discovered  \citep{schombert1992,Pisano1998,karachentsev2015spiral}.  UGC 5086 and NGC 3344dw1 have distances from surface brightness fluctuations that confirm their associations with NGC 2903 and NGC 3344 respectively \citep{Carlsten2021,Carlsten2022Elves}.  While the others do not yet have distance measurements, their visual appearance is more consistent with being Local Volume dwarf galaxies than luminous background galaxies, as they are not concentrated or smooth in their central regions.  For the rest of this paper, when we compare properties of these candidates with other dwarf galaxies or satellite systems, we assume that candidates are associated with the LBT-SONG hosts and located at similar distances.

The positions, photometric properties, and derived properties are summarized in Table \ref{table:candresults}. In this section we discuss the data and derived quantities for the 8 (candidate) dwarf satellite galaxies, and then compare the structural properties, derived stellar and HI masses and star formation rates to those of the satellites in the Local Group.  Measurements for the two dwarf galaxies near NGC 628 were previously presented in \citet{Davis2021}.

\subsection{Compiled candidate data}\label{sec:compiled}

We use a heterogeneous approach to the optical photometry of the low-mass dwarf galaxy candidates.  We use {\tt{SExtractor}} to measure the UBVR apparent magnitudes of the faintest candidates in the sample, NGC 672 dwB, NGC 672 dwA, and NGC 3344 dw1 (Table \ref{table:candresults}). The magnitudes are corrected for Galactic extinction. 
We estimated uncertainties on these results by injecting and recovering 
500 mock galaxies with the most similar magnitudes and surface brightness to each candidate and computing the standard deviation of the resultant magnitudes. We do this to account for contamination from background variations, which disproportionately affect the photometry of these candidates due to their small size and low surface brightness. For PWE98, due to its more diffuse structure and position near a bright foreground star, we used Imfit \citep{Erwin2015} to model the candidate. 
For UGC 5086 and LEDA 87258 we used the SDSS photometry from \citet{sdss2009}. We convert these magnitudes to the Johnson-Cousins UBVR band values in Table \ref{table:candresults} using the empirical color transformations from \citet{Jordi2006}. We estimate errors with a lower bound of 0.05 mag for the UBVR magitudes, given the low reported SDSS measurement uncertainties. Finally, we include measurements for NGC 628 dwA and NGC 628 dwB from \citet{Davis2021}. 

To search for on-going star formation we use the z=0 Multiwavelength Galaxy Synthesis (z0MGS) catalog and images \citep{Leroy2019}, which are based on WISE \citep{WISE2010A} and GALEX \citep{GALEX2005} photometry. We search for emission in the WISE W1, W2, W3, W4 bands as well as for the GALEX NUV and FUV bands. We measure the integrated flux in each band at the location of the candidates within one effective radius, as derived from the LBC photometry. If detected, we convert the fluxes to magnitudes and use the RMS of the flux throughout the image to calculate uncertainties, all listed in Table \ref{table:candresults}. Otherwise, we use an upper limit of $5\times$ the RMS of the flux in the image, and within the optical effective radius of the galaxy.

We report two stellar mass estimates in Table \ref{table:candresults}. We use the more accurate WISE W1 measurement and $\text{M}_*/\text{L}_{\text{W}1} = 0.5\, $M$_\odot/$L$_\odot$ for passive, low-redshift galaxies \citep{Leroy2019} where applicable (listed as method 2 in the table).  Because this often gives only an upper limit, we also use the measured $\text{M}_{\text{V}}$ for all candidates (method 1 in the table) to estimate stellar masses.  Here, we use $\text{M}_*/L_{\rm V} = 0.3-2\,$M$_\odot/$L$_\odot$  for the intermediate age and metallicity models found to be most similar to our candidates, based on the \citet{Marigo17} isochrones.  We obtain star formation rates (SFR) using the NUV flux and the scaling from \citet{Leroy2019}. For candidates with upper limits on the NUV flux we take $5 \times$ the RMS flux of the NUV image within the effective radius of the object as measured in the LBT data to obtain upper limits on the SFR. 

We compile HI measurements from a variety of data sets.  The upper limits on the HI masses for the two candidates near NGC 628 were derived from THINGS \citep{Leroy2008THINGS}, as described in \citet{Davis2021}.  The upper limit on HI mass for UGC 5086 comes from \citet{Irwin2009}, who used D-configuration VLA data to separate a possible signal from this galaxy from the HI disk of its host, NGC 2903.  The disk of the host is unusually large, and the low resolution of prior single-dish observations made it impossible to distinguish  emission from the satellite and the host.   

We perform measurements using HI spectra from the ALFALFA survey \citep{ALFALFA2011} to derive total HI masses for NGC 3344 dw1, PWE98, and LEDA 87258. HI masses and 5$\sigma$ HI mass upper limits were derived based on $M_{\mathrm{HI}} = 2.36 \times 10^5 \, D^2 \int_{v_1}^{v_2} S(v) dv \, \rm M_{\sun}$ under the assumption of optically thin HI emission. Here, $D$ is the distance in Mpc, $S(v)$ is the HI flux in Jy km/s, and $v$ is the velocity in km/s.  HI mass uncertainties were estimated under the conservative assumption of a 15\% flux uncertainty and a 10\% distance uncertainty.

For the two dwarf candidates in NGC 672, we inspected their locations in the HALOGAS \citep{Heald2011} 21-cm cubes covering NGC 672 and IC1727. There is a clear HI peak exactly at the location of NGC 672 dwA. At the location of the peak we fit a Gaussian line profile with $1\sigma$ width $\approx 6$ km s$^{-1}$  centered at velocity $\approx 372$ km s$^{-1}$ (LSRK frame). Integrating from 350 to 400 km s$^{-1}$ we find a total HI mass of $(3.0 \pm 0.6) \times 10^5$ M$_\odot$, calculated assuming optically thin 21-cm emission and assuming independent channels for the noise calculation. This mass estimate assumes the emission is concentrated in a single $30''$ (FWHM) beam, and therefore represents a lower limit. Inspecting the cube around the central peak, there are hints of extended HI emission around the peak. If we integrate over a broad 250 to 400 km s$^{-1}$ velocity range across a $60''\times60''$ cube centered on the dwarf location, we find a larger total HI mass of $\approx 2.1 \times 10^6$~M$_\odot$. Whether this fainter extended emission is real, and whether it is associated with the dwarf itself will also require more investigation. A fair amount of potential tidal material is also visible in this region of the HALOGAS cube, though mostly at locations and velocities distinct from the dwarf candidate.

NGC 672 dwD does not show a clear detection, but there is a hint of narrow ($1\sigma$ line width $\approx 5$ km s$^{-1}$) 21-cm emission near VLSR velocity 507 km/s at the location of the dwarf. The signal is not highly significant, yielding an integrated HI mass of $(1.5 \pm 0.4) \times 10^5$ M$_\odot$ when integrated from 495 to 520 km s$^{-1}$. However, it is suggestive and worth further investigation or deeper follow up observations for the reasons that we outline in the next section.

\begin{sidewaystable*}

\centering
 
 \begin{tabular}{ cccccccccc }
  \hline
  Candidate & NGC 672 dwD & NGC 672 dwA & NGC 3344 dw1 & [PWE98] & LEDA 87258 & UGC 5086 & NGC 628 dwB & NGC 628 dwA\\
  \hline
  RA [J2000] & 01:47:51.6 & 01:47:19.3 & 10:42:43.6 & 02:26:52.7 & 11:00:51.9 & 09:32:48.7 & 01:36:23.2 & 01:37:17.8\\
  Dec [J2000] & +27:37:53.0 & +27:15:17.4 & +25:01:29.2 & +33:25:36.2 & +13:52:56.2 & +21:27:59.0  & +15:57:53.0 & +15:37:58.2\\
  Host & NGC 672 & NGC 672 & NGC 3344 & NGC 925 & NGC 3489 & NGC 2903 & NGC 628 & NGC 628 \\
  Separation [$^\prime$] & 11.93 & 13.22 & 12.34 & 10.42 & 8.19 & 9.20 & 11.8 & 12.6\\
  Separation [kpc] & 24.92 & 27.61 & 29.72 & 27.76 & 17.11 & 21.41 & 34 & 36\\
  Distance [Mpc] & N/A & N/A & 10.57$^{+2.11,\, \text{a}}_{-1.79}$ & N/A & N/A & 8.7$\pm 0.9^{\text{c}}$ & N/A & 10.12$^{+1.87, \, \text{a}}_{-1.67}$\\ 
  $\text{m}_{\text{U}}$ (mag) & 23.55 $\pm$ 0.62 & 19.87 $\pm$ 0.16 & 21.53$\pm$0.28 & $17.81 \pm 0.1$ & 17.72 $\pm 0.05$ &17.14$\pm 0.05$ & 22.52 $\pm$ 0.25 & 17.49 $\pm$ 0.15\\
  $\text{m}_{\text{B}}$ (mag)& 23.81 $\pm$ 0.32 & 20.16 $\pm$ 0.13 &22.40$\pm$0.21 & $18.03 \pm 0.1$ & 17.38 $\pm 0.05$ &16.97 $\pm 0.05$ & 22.67 $\pm$ 0.14 & 18.34 $\pm$ 0.09\\
  $\text{m}_{\text{V}}$ (mag)& 23.13 $\pm$ 0.51 & 19.86 $\pm$ 0.14 &20.70$\pm$0.25 & $17.88 \pm 0.1$ & 16.78 $\pm 0.05$ &16.04 $\pm 0.05$ & 22.23 $\pm$ 0.13 & 17.80 $\pm$ 0.07\\
  $\text{m}_{\text{R}}$ (mag)& 23.18 $\pm$ 0.38 & 19.87 $\pm$ 0.15 &20.94$\pm$0.23 & $16.73 \pm 0.1$ & 16.49 $\pm 0.05$&15.58 $\pm 0.05$ & 22.12 $\pm$ 0.10 & 17.64 $\pm$ 0.06\\
  half light radius [arcsec] & 2.32 $\pm$0.5 &  5.80 $\pm$0.3 & 10.35 $\pm$ 0.3 & 32.3 $\pm$1 & 21.46 $\pm$ 0.5$^{\text{b}}$ & $16.7 \pm 0.7$ & 3.50 $\pm$ 0.02 & 7.79 $\pm$ 0.07\\
  W1 (mag)&  $>20.01$ & $>17.89$ & 17.28 $\pm$ 0.55 & $>15.96$ & 15.82 $\pm$ 0.05 &13.02 $\pm0.05$ & $>18.6$ & 16.33 $\pm 0.065$\\
  W2 (mag)& $>18.77$& $>16.64$ & 16.79 $\pm$ 0.80& $>14.77$& 15.66 $\pm$0.13 &13.38 $\pm0.2$ & $>17.3$ & 16.16 $\pm 0.198$\\
  W3 (mag)& $>14.33$ & $>12.20$ & 14.94 $\pm$3.00&$>10.24$ & $>12.26$ &$>11.37$ & $>12.9$ & $>12.63$\\
  W4 (mag)& $>10.86$ & $>9.04$ & $>10.46$& $>6.82$& $>9.08$ &$>7.81$ & $>9.8$ & $>9.05$ \\
  GALEX NUV (mag)& $>25.3$  & 20.67 $\pm 0.07$  & $>23.91$ & 20.26 $\pm 1.05$ & - & 19.52 $\pm 0.16$ & $>22.7$ & 21.66 $\pm 0.27$\\
  GALEX FUV (mag)& $>25.3$ & 22.63 $\pm 0.40$ &$>24.15$&$>19.74$& 16.70 $\pm 0.05$& $>22.07$ & $>22.7$ & $>22.18$\\
  $\text{M}_{\star} (1)(\text{M}_{\odot})$ & $(9.1-61)\times 10^3$ & $(2.5-17)\times 
 10^5$ &$(9.1-61)\times 10^4$ &$(5.3-35)\times 10^6$ & $(5.3-35)\times 10^6$ & $(5.3-35)\times 10^6$ & $(2.8-19)\times 10^4$ & $(1.8-12)\times 10^6$\\
 $\text{M}_{\star} (2)(\text{M}_{\odot})$ & $<5.6 \times 10^4$& $<3.9 \times 10^5$&($9.1\pm 3.6)\times 10^5$ & $<3.8\times 10^6$ & $(2.6\pm 0.1)\times 10^6$& $(4.32\pm 0.2)\times 10^7$ & $<2.6\times 10^6$& $(3.4\pm 1.4)\times 10^6$ \\
 Star Formation Rate [$\text{M}_{\odot}\text{yr}^{-1}$]& $<1.7 \times 10^{-6}$& $(1.2 \pm 1.1)\times 10^{-4}$& $<8.3\times10^{-6}$ & $(2.9 \pm 1.1)\times 10^{-4}$& $(3.6\pm0.2)\times10^{-3}$& $(4.4 \pm 3.8)\times 10^{-4}$ & $<5.4\times10^{-5}$& $(1.2 \pm 0.7)\times 10^{-4}$ \\
 $\text{M}_{\text{HI}}([\text{log}_{10}\text{(M}_{\odot})])$ & $5.2\pm 0.1$ & $5.5\pm 0.1$ & $7.58 \pm 0.065$ & $8.31 \pm 0.019$ & $<$6.75 & $<$5.75$^{\text{d}}$ &$<5.70$ & $<5.70$ \\
  \hline
 \end{tabular}
 
 \caption{\label{table:candresults}\textbf{Properties of the satellite candidates.} Candidates are ordered by host distance, and for hosts with more than one satellite candidate, candidates are further ordered by magnitude.  The magnitudes are corrected for Galactic extinction, and include errors derived from analog mock galaxy insertion tests. Infrared and UV data are from AllWISE \citep{ALLWISE} and GALEX \citep{GALEX2005} where available. The remaining values and upper limits are measured from the z=0 Multiwavelength Galaxy Synthesis (z0MGS) Data Access portal and combining the ALLWISE W1 and W4 and GALEX NUV and based on the results from \citet{Leroy2019}. NGC 628 dwA and NGC 628 dwB were presented in \citet{Davis2021} and their values from that paper are reproduced here. $^{\text{a}}$ \citet{Carlsten2022Elves}.$^{\text{b}}$ \citet{sdssdr13_2017} $^{\text{c}}$ \citet{Carlsten2021}. $^{\text{d}}$ \citet{Irwin2009}.
 }
\end{sidewaystable*}

\subsection{Comparison to other satellites}\label{sec:compare}

In this section, we compare the properties of the dwarf satellite candidates to galaxies of similar luminosity in the Local Group and Local Volume. First, we consider the morphologies of the galaxies, and where they lie in the space of absolute magnitude and surface brightness.  To compare the absolute magnitudes of the candidates with those of Local Group satellites, we use the measured distances to the candidates when available (Table \ref{table:candresults}), and adopt the host distance in Table \ref{table:hostprop} for the candidates without direct distance measurements. In cases where candidate distances are measured, they are consistent with the host distances. We obtain absolute V-band magnitudes that place two candidates in the ultrafaint regime: $\text{M}_{\text{V}} = -7.7$ for NGC 628 dwB, and $\text{M}_{\text{V}} = -6.6$ for NGC 672 dwD, both of which were discovered in the LBT-SONG survey. As we discuss in more detail below, six candidates (NGC 3344 dw1 with $\text{M}_{\text{V}} = -8.9$; NGC 672 dwA, $\text{M}_{\text{V}} = -10.2$, PWE98, $\text{M}_{\text{V}} = -12.0$; NGC 628 dwA, $\text{M}_{\text{V}} = -12.2$; LEDA 87258, $\text{M}_{\text{V}} = -13.3$; and UGC 5086, $\text{M}_{\text{V}} = -13.8$) are dwarf spheroidals (dSphs), dwarf transitionals (dTrans) or are galaxies that do not neatly fit in these categories.  Although we do not show the two massive dwarfs on the Local Group comparison plots, NGC 4248 ($\text{M}_{\text{V}} = -16.0$) and IC 1727 ($\text{M}_{\text{V}} = -17.6$) are morphologically distinct and bright.  The former is an irregular galaxy, while the latter is a dwarf barred spiral.  

\begin{figure*}
 \includegraphics[width=0.95\textwidth]{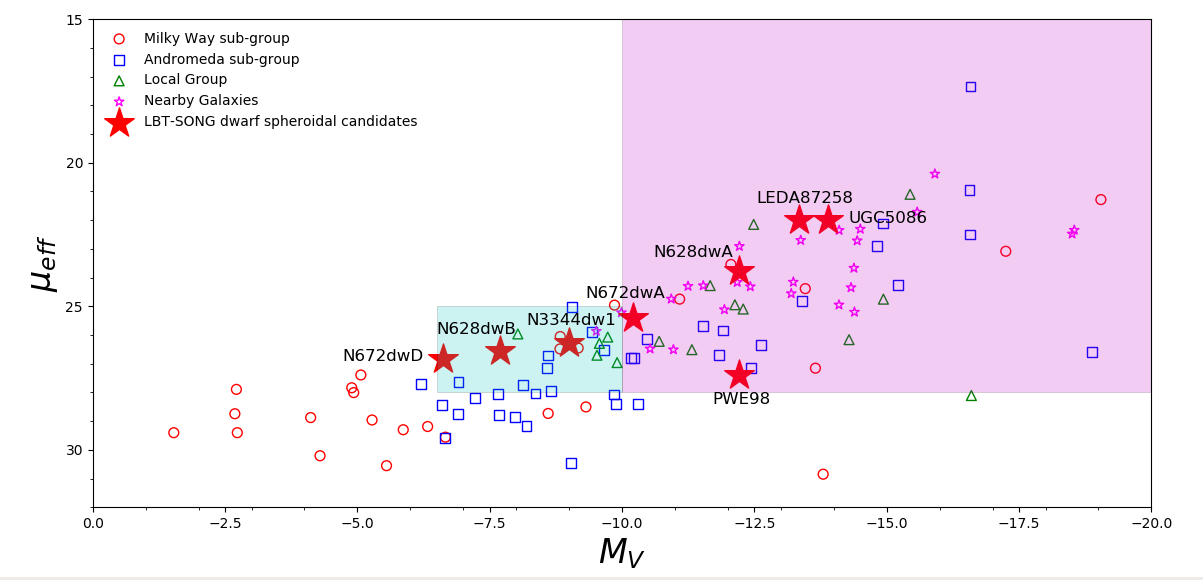}
 \caption{The V-band effective surface brightness and absolute magnitude $\text{M}_{\text{V}}$ for the 8 low-mass dwarf candidates are shown as red stars. We compare them to an updated version of the Local Group galaxy compilation from \citet{McConnachie12}. The symbols identify satellites of the MW, M31, Local Group and nearby galaxies. The smaller blue region represents the parameter range where the search method described in Sec \ref{sec:Pipeline} is applied, while the larger magenta region is the region covered by our visual inspection.}
 \label{fig:MWDwarfs}
\end{figure*}

Figure~\ref{fig:MWDwarfs} shows the effective surface brightness and absolute V-band magnitudes for the 8 low-mass dwarf candidates that are the focus of this section, along with an updated version of the compilation of Local Group and Local Volume dwarfs of \citet{McConnachie12}. All 8 candidates appear to have structural properties in line with those found in the Local Volume, albeit generally on the high surface brightness side for fixed absolute magnitude, presumably due to our selection function.

\begin{figure*}
 \includegraphics[width=0.95\textwidth]{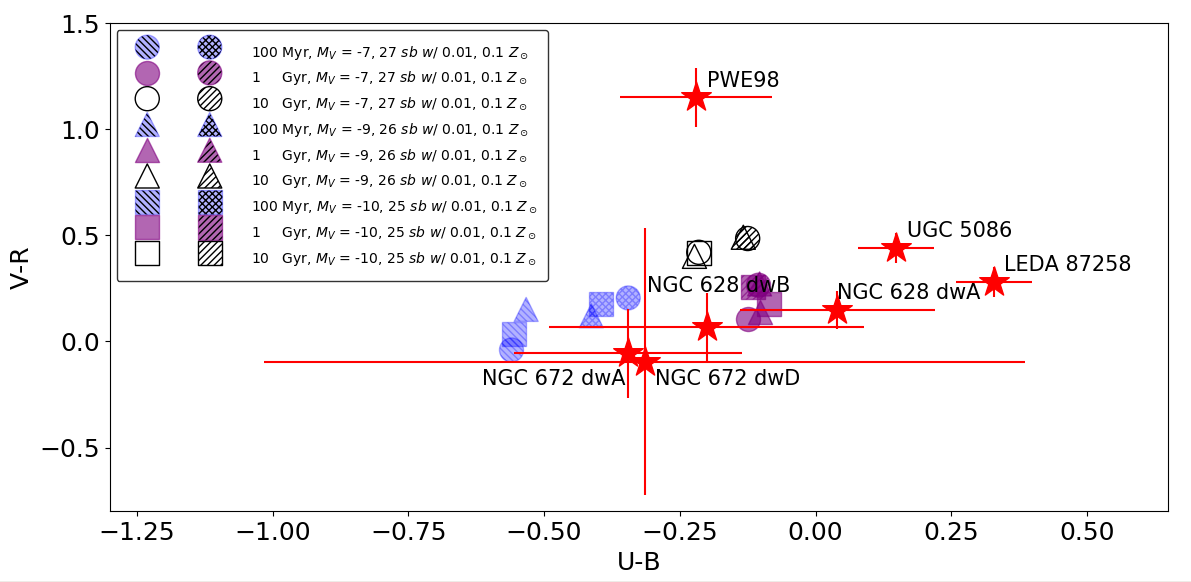}
 \caption{The Galactic extinction-corrected V$-$R and U$-$B colors of the 8 low-mass dwarf candidates (red stars) and 3 closest analog mock galaxy models in terms of $\text{M}_{\text{V}}$ and surface brightness for NGC 672 dwD (circles), NGC 3344 dw1 (triangles), and NGC 672 dwA (squares). We show these points to indicate where realistic SSPs for the faintest satellite candidates lie in this space.  We simulate each analog galaxy with each of the following ages: 100 Myr, 1 Gyr, 10 Gyr, and use metallicities of 0.01 $\text{Z}_\odot$ and 0.1 $\text{Z}_\odot$. We inject 100 mock galaxies for each model into the image of the corresponding satellite candidate. Error bars for the candidates are the standard deviation of the color measurements from these trials. We omit NGC 3344 dw1 as it is affected by an image artifact (Fig.~\ref{fig:candidateGallery}).}
 \label{fig:color-color}
\end{figure*}

We use the colors of the candidates to estimate ages. Figure \ref{fig:color-color} shows the $\text{V}-\text{R}$ and $\text{U}-\text{B}$ colors of the candidates along with those of the mock galaxies with a range of morphological properties, and the best analogs for NGC 672 dwD, NGC 3344 dw1, and NGC 672 dwA corresponding to the models with $\text{M}_{\text{V}} = -7$, 27\surfb{}, $\text{M}_{\text{V}} = -9$, 26 \surfb{}, and $\text{M}_{\text{V}} = -10$, 25 \surfb{} respectively. We show all the trial ages of (100 Myr, 1 Gyr, 10 Gyr) and metallicities (0.1 $\text{Z}_{\odot}$ and 0.01 $\text{Z}_{\odot}$) which are typical for Local Volume galaxies with similar absolute magnitudes \citep{Kirby13}. For each galaxy model, we measure the colors of 100 mock galaxies injected into the image in which each candidate is found. Although the uncertainties are larger for the intrinsically faintest candidates (NGC 672 dwD, NGC 628 dwB, and NGC 672 dwA), they have colors consistent with either young or intermediate-age stellar populations. We do not show NGC 3344 dw1 because of the image artifacts running through the galaxy that render the colors less certain (see Fig.~\ref{fig:candidateGallery}).  As we discuss below, the lack of UV flux from these objects sets limits on recent star formation activity, indicating that the stellar populations cannot be young.  NGC 628 dwA most closely resembles models of intermediate age, while the colors of the intrinsically brighter LEDA 87258 and UGC 5086 prefer intermediate or older stellar populations. While PWE98 is not well matched to any of the SSP models, its V$-$R color suggests that it has an older stellar populations. Note that none of our models include gas or dust, but that PWE98 in particular is embedded in a large known gas reservoir.  PWE98's colors are consistent with those of more massive but quenched galaxies \citep{assef2010}.

\begin{figure}
 \includegraphics[width=0.95\columnwidth]{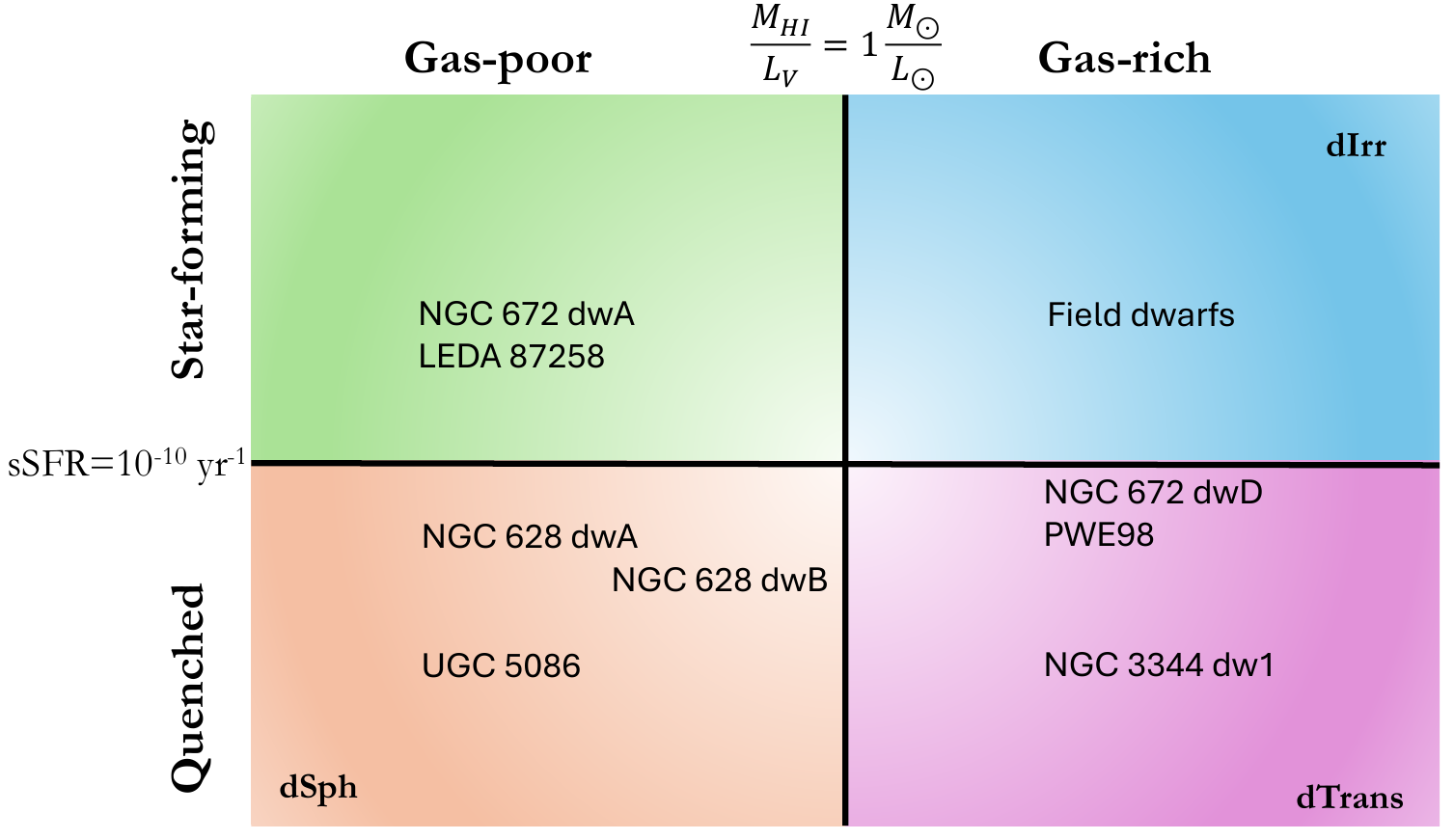}
 \caption{The categorization of the LBT-SONG Far Sample candidates as a function of star formation and atomic gas content. These determinations are made according to the specific star formation rates (sSFR), the presence of FUV flux and the $\text{M}_{\text{HI}}/\text{L}_{\text{V}}$ derived from the measurements in Table~\ref{table:candresults}. Candidates with sSFR $>10^{-10} \text{yr}^{-1}$ and FUV flux are likely to have recently ($\sim 100$ Myr) formed stars or to be actively forming stars and are categorized as ``star-forming". Otherwise we label the candidate as ``quenched". Candidates with $\text{M}_{\text{HI}}/\text{L}_{\text{V}} \gtrsim 1 \,\text{M}_\odot/\text{L}_\odot$ are categorized as gas-rich, and gas-poor if $\text{M}_{\text{HI}}/\text{L}_{\text{V}} \lesssim 1\, \text{M}_\odot/\text{L}_\odot$ \citep{Spekkens2014,Brandford2015,Putman2021}.}
 \label{fig:SFverGas}
\end{figure}

Dwarf galaxies are separated into three main morphological categories based on their gas content and SFR. These classes are thought to be associated with evolutionary processes \citep{weisz2011}. The first are dwarf irregulars (dIrr) which are irregularly shaped, gas-rich, and star-forming.  dIrr are the most common type of dwarfs found in the field by a wide margin.  Dwarf spheroidals (dSph) generally have a smooth morphology, with little to no gas, old stellar populations and little to no ongoing star formation. With the exception of the Magellanic Clouds, all satellites of the MW are dSphs \citep[Leo T, a dwarf galaxy outside the MW but falling into it, is a dTrans object;][]{weisz2014b}.  In between these two morphological classes lie the dwarf transitionals (dTrans), which are generally found as satellites and are gas-rich but are not star forming \citep{Grebel2003}. The dSphs and dTrans follow a density-morphology relationship \citep{weisz2011}, which is thought to be the result of environmental mechanisms such as ram-pressure stripping, starvation, tidal stripping, and tidal stirring. In the Local Volume, dTrans galaxies have tidal indices intermediate between dIrr and dSph \citep{weisz2011}.

We categorize the satellite candidates as dSphs, dTrans, or dIrrs to gain insight into environmental processes they may have undergone. To do this, we divide the satellite candidates into four quadrants, as shown in Figure \ref{fig:SFverGas}.  On the vertical axis, we classify galaxies as quenched or star-forming, and on the horizontal axis, we categorize galaxies as gas-rich or gas-poor.  The quadrants are labeled by morphological type.  The one quadrant without a label is the star-forming, gas-poor quadrant, which does not neatly map into traditional galaxy morphologies.  A galaxy is considered quenched if it lacks an FUV detection and its specific star formation rate (sSFR, calculated by dividing the SFR by the estimated stellar masses) is $<10^{-10}\text{yr}^{-1}$.  It is otherwise considered star-forming. The sSFR benchmark of $10^{-10} \text{yr}^{-1}$ is the average rate of star formation over the age of the Universe. A galaxy is considered gas-rich if $\text{M}_{\text{HI}}/\text{L}_{\text{V}}>1\,\text{M}_\odot/\text{L}_\odot$, which is the benchmark applied to field dwarfs to categorize their gas content \citep{Spekkens2014,Brandford2015,Putman2021}, and gas-poor otherwise. 

Notably, we find that \emph{none} of the 8 low-mass satellite candidates populate the upper right quadrant of Figure \ref{fig:SFverGas}, which corresponds to star-forming and gas-rich candidates. Thus, none of the 8 candidates are members of the dIrr morphological type that dominates the field dwarf population, indicating that environmental affects have likely shaped their morphology and SFH. 

Candidates with signs of recent or on-going star formation and little gas are found in the top left quadrant. These are LEDA 87258 and NGC 672 dwA. Both candidates have NUV and FUV detections in their central regions, indicative of star formation within the last 100 Myr \citep{Leroy2019}. The sSFRs for these candidates are $\sim 6.8\times 10^{-10}- 1.4 \times 10^{-9}\text{yr}^{-1}$ and $\sim 4.8 \times 10^{-11}- 7.1\times 10^{-10}\text{yr}^{-1}$, respectively. The upper limit of $\text{M}_{\text{HI}}/\text{L}_{\text{V}}$ $<0.34 \text{M}_\odot/\text{L}_\odot$ places LEDA 87258 in the gas-poor category.  

NGC 672 dwA's presence in this category is intriguingly debatable.  The HI gas mass $3.0 \pm 0.6 \times 10^5$ M$_\odot$ at the position of the candidate makes $\text{M}_{\text{HI}}/\text{L}_{\text{V}} <1 \,\text{M}_\odot/\text{L}_\odot$,  considerably lower than field galaxies with similar luminosity \citep[e.g.,][]{Brandford2015,sardone2024}.  If the gas we found in a broader search radius of NGC 672 dwA were part of the dwarf candidate, the implications would be exciting.  First, as we showed in the last section, the total HI mass of this system would be nearly an order of magnitude higher, placing this galaxy's $\text{M}_{\text{HI}}/\text{L}_{\text{V}}$ in line with field dwarfs.  More interestingly, though, it could be a sign of ongoing gas stripping.  Satellite galaxies falling into cluster environments undergoing ram pressure stripping are known to be quenched inefficiently and therefore able to continue forming stars while losing their gas content \citep{Oman2021}. Dwarf galaxies falling into cluster environments can also have star formation triggered
by ram pressure stripping depending on the intra-cluster medium density and the dwarf's mass \citep{Steyrleithner2020}. While the hosts of NGC 672 dwA and LEDA 87258 are much smaller than their cluster counterparts, their lack of gas and recent SF indicates similar quenching mechanisms may be at play in their environments.  If we are indeed seeing HI being actively stripped from NGC 672 dwA, it would be one of the few rare observations of such a phenomenon in a low-density environment \citep{pearson2016}.

The quenched and gas-poor dSphs are found in the lower left quadrant and consist of NGC 628 dwA, NGC 628 dwB, and UGC 5086. NGC 628 dwA has a tight upper limit on $\text{M}_{\text{HI}}$, with  $\text{M}_{\text{HI}}/\text{L}_{\text{V}}< 0.08\, \text{M}_\odot/\text{L}_\odot$, making it gas-poor. Although it has an NUV detection, there is no FUV detection. With a sSFR in the range of $\sim (1.0 - 6.7) \times 10^{-11}\text{yr}^{-1}$, we place it in the quenched category.  Its colors indicate an intermediate ($\sim 1 - 2$ Gyr) population.  Thus, quenching may have been recent. NGC 628 dwB has upper limits on its sSFR ranging from $\sim (1.9 - 23) \times 10^{-10}\text{yr}^{-1}$ and has no UV detection, so we consider it to be quenched. It also has an upper limit on $\text{M}_{\text{HI}}/\text{L}_{\text{V}} <5.3\, \text{M}_\odot/\text{L}_\odot$. We place it in the gas-poor category as there is only a suggestive upper limit on its gas mass, but near the gas-rich boundary because the upper limit straddles the two categories.  Finally, UGC 5086 is detected in NUV but not FUV, with an estimated sSFR of $\sim 10^{-11} \text{yr}^{-1}$ based on its WISE W1-estimated stellar mass.  Because of the stringent upper limit on HI emission by \citet{Irwin2009}, it has an upper limit of $\text{M}_{\text{HI}}/\text{L}_{\text{V}} < 0.03 \, \text{M}_\odot/\text{L}_\odot$ on its HI mass to optical luminosity ratio.  Thus, it is definitively dSph-like in its broad-band properties.

A large fraction of the candidates are plausibly dTrans objects and lie in the lower right quenched and gas-rich quadrant and have no FUV detections. Starting with the intrinsically brightest, PWE98 has a low sSFR of between $(1.0 - 8.3) \times 10^{-11}\text{yr}^{-1}$. Its relatively low $\text{M}_{\text{HI}}/\text{L}_{\text{V}} = 2.53 \,\text{M}_\odot/\text{L}_\odot$ is still sufficient to meet the criterion to be considered gas-rich. We note that PWE98 was originally detected in HI, and its gas appears to be undergoing tidal stripping \citep{Pisano1998}.  NGC 3344 dw1 and NGC 672 dwD both have upper limits on their SFRs. Based on W1-based mass estimates for NGC 3344 dw1 and the smallest optical mass estimate for NGC 672 dwD (yielding the most conservative constraints), the upper limits on their sSFR are $9 \times 10^{-12}\text{yr}^{-1}$ and $2 \times 10^{-10}\text{yr}^{-1}$, respectively.   
Based on the ALFALFA and HALOGAS measurements, the implied HI-to-optical-luminosity ratios are a high $\text{M}_{\text{HI}}/\text{L}_{\text{V}}=120 \, \text{M}_\odot/\text{L}_\odot$ and $\text{M}_{\text{HI}}/\text{L}_{\text{V}}\sim \mathcal{O}(1-10) \, \text{M}_\odot/\text{L}_\odot$ for these galaxies,  respectively. 

NGC 672 dwD is a particularly notable dTrans candidate. If its association with NGC 672 is confirmed, its optical luminosity is similar to ultrafaint dwarf galaxies in the Local Group.  Based on high-resolution hydrodynamic simulations, ultrafaint dwarf galaxies with large gas reservoirs are predicted to exist \citep{Jeon2017,wright2019,applebaum2021,gutcke2022,Martin2022}. If NGC 672 dwD is confirmed and the HI mass is indeed associated with it, it would be the first gas-rich ultrafaint dwarf observed. The closest Local Group analog, Leo T, is a low-mass dTrans galaxy, but significantly more luminous than NGC 672 dwD, and has only recently been quenched of star formation \citep{adams2018}.  
Higher resolution imaging is required to confirm the gas's existence and association with NGC 672 dwD or with its host, and NGC 672 dwD's association with the host.  

Finally, we compare the candidates to other satellite systems in terms of their SFH. Studies of the satellite populations of the MW \citep{slater2014,simon2019}, and MW-size hosts such as: M31 \citep{McConnachieM31}, M81 \citep{chiboucas2013confirmation} CenA \citep{Muller2018,crnojevic2019faint}, and M101 \citep{DanieliM101,bennet2019m101,Carlsten2020}, as well as the ELVES survey \citep{Carlsten2022Elves} have found mostly quenched satellites. In contrast, the two satellites of M94 \citep{smercina2018lonely} appear to be star-forming, while the SAGA survey has found dwarf satellites brighter than their limit of $\text{M}_{\text{V}}=-12.1$ to be mostly star forming \citep{geha2017saga,Geha2024,Mao2021}. While there are some suggestions that environment (quantified by the tidal index) plays a role in the properties of satellite populations \citep{bennet2019m101}, there are also indications that quenching times are a strong function of satellite stellar mass \citep{wheeler2014,wetzel2015}.  \citet{karunakaran2019} suggest that the SAGA survey luminosity limit defines a dividing line for satellite galaxies where galaxies fainter than the limit are largely quenched, and galaxies brighter are largely star-forming. 

\begin{figure}
 \includegraphics[width=\columnwidth]{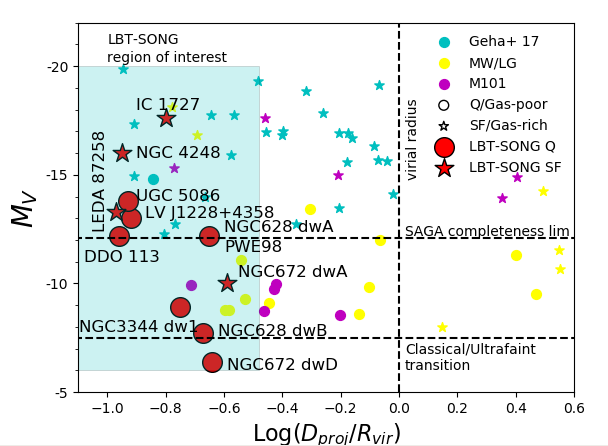}
 \caption{Satellite galaxies of nearby MW analogs as a function of projected host separation $\text{D}_{\text{proj}}$ relative to the virial radius $\text{R}_{\text{vir}}$ and satellite absolute magnitude $\text{M}_{\text{V}}$, based on  \citet{karunakaran2019}. The vertical line separates the virial volume
of the host from the field, and the horizontal lines show the SAGA completeness limit \citep{geha2017saga,Mao2021}, and another line to show the approximate division between classical dwarf galaxies that continue to form stars at late times and reionization fossils (ultrafaint dwarfs). Galaxies are color-coded by survey: MW and Local Group \citep[][with the 2015 update]{McConnachie12}, M101 \citep{Tikhonov2015,DanieliM101,bennet2019m101,karunakaran2019}, SAGA \citep{geha2017saga}, and LBT-SONG \citep{garling2019ddo113,Davis2021,garling2021}. The M101, Local Group, and SAGA satellite galaxies are shown as stars if they are gas-rich or undergoing star formation, and as circles if they are gas-poor or quenched. The LBT-SONG candidates are categorized based on the presence of star formation (SF for star-forming, Q for quenched). The LBT-SONG sensitivity region is shown as the shaded blue region. The circle symbols for PWE98 and NGC 628 dwA overlap.}
 \label{fig:SFQMV}
\end{figure}

Figure \ref{fig:SFQMV} shows the distribution of dwarf galaxies, including our sample, in the space of the projected separation from the host and the V-band absolute magnitude of the dwarf (candidate) galaxies.  
The LBT-SONG search region is shaded in blue. While the satellites from other surveys of MW-size hosts are categorized as star forming (SF) or gas-rich or else quenched (Q) or gas-poor, our candidates are classified in terms of star formation only, as they are a heterogeneous sample in terms of their gas mass measurements. The LBT-SONG candidates have a higher quenched fraction at a brighter magnitude limit, which is more in line with \citet{Carlsten2022Elves}.  
NGC 672 dwA stands out as a star-forming candidate in the quenched region, although it appears to also be gas-poor (but see the previous discussion of its potentially much larger HI reservoir). 

We also compare to satellites of lower-mass hosts. DDO 113 and LVJ1228+4358, the dwarf satellites studied in the LBT-SONG Near Sample \citep{garling2019ddo113,garling2021} are both quenched. DDO 113 shows signs of environmental quenching via strangulation while LV J1228+4358 is tidally disrupting. The MADCASH survey \citep{Carlin16,Carlin_2019,hargis2019,carlin2021}, which studied the satellites of LMC-mass hosts, found five satellites, of which two are quenched dwarf spheroidals. Of the remaining three: DDO 44 is quenched and tidally disrupting, and Antlia B and the Antlia dwarf are quenched dTrans \citep{sand2015,hargis2019,garling2022}. Similarly, many of the LBT-SONG Far Sample candidates are dTrans objects. Thus, compared to other satellites of more massive systems, our galaxies and those dwarf galaxies found in MADCASH have a high quenched fraction, even at fairly bright magnitudes.

Overall, the LBT-SONG Far Sample is a heterogeneous mix of star-forming, quenched, gas-poor and gas-rich candidates. Most of the candidates, which are by design found well within the virial radius of their hosts, are dSph (NGC 628 dwA, NGC 628 dwB, UGC 5086) or dTrans (PWE98, NGC 3344 dw1, NGC 672 dwD), with two galaxies that are star-forming but potentially gas-poor (LEDA 87258 and NGC 672 dwA) and being observed in the process of gas loss. The dSphs and dTrans follow the density-morphology relationship found in the ANGST survey \citep{weisz2011}, which is thought to be the result of a combination of environmental quenching mechanisms. Notably, none of the 8 low-mass dwarf satellites candidates can be classified as dIrr, although dIrr is the characteristic morphology of field dwarf galaxies.  The heterogenous mix of LBT-SONG Far Sample candidates further indicates that strangulation, tidal stripping, and ram-pressure stripping may all have a role to play in quenching star formation in lower mass hosts. Evolutionary pathways similar to those acting within MW-sized hosts may exist in quenching these satellites of intermediate-mass hosts and below.

\section{Satellite counts} \label{sec:satelliteCounts}

\begin{figure}[t]
    \centering
    \includegraphics[width=0.5\textwidth]{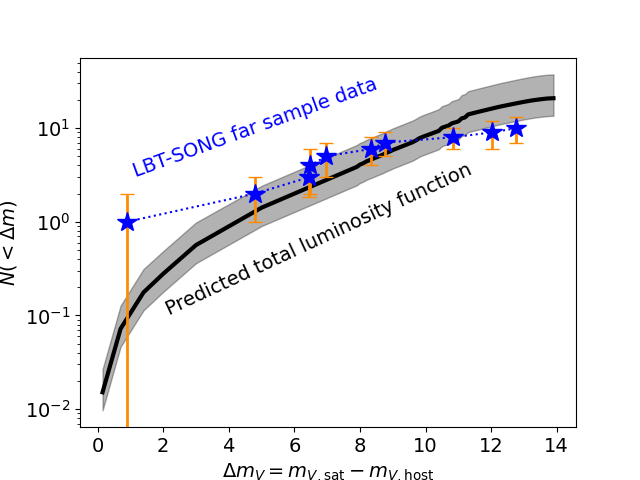}
    \caption{Predicted and observed satellite luminosity function for the LBT-SONG Far Sample.  The luminosity function is cumulative, as a function of the relative magnitude of the satellite with respect to the host, $\Delta m_V = m_\mathrm{V, sat} - m_\mathrm{V, host}$.  The satellite candidates presented in this work are indicated in stars, with uncertainties measured by boot resampling of the hosts. 
    The prediction for the LBT-SONG Far Sample is based on \citet{Sales13}, taking into account observational completeness (Sec.~\ref{sec:completeness}) and the size-luminosity relation \citep{danieli2018,Carlsten2021size}, and assumes the centrally concentrated radial distribution of \citet{Dooley2017a} (solid line).  The band shows the range of predictions for different models of the satellite radial distribution, from satellites following the halos' smooth dark matter distribution (top edge, disfavored), to following the same distribution of the SAGA MW analog sample \citep[lower edge;][]{Mao2024}.  }
    \label{fig:lf}
\end{figure}

In Figure~\ref{fig:lf}, we compare the cumulative luminosity function of observed satellites (points) to predictions (line and shaded band), as a function of the magnitude difference between satellite and host $\Delta m_V = m_\mathrm{V, sat} - m_\mathrm{V, host}$.  Presenting the luminosity function in this way is a common choice for data sets with a heterogeneous set of host stellar masses and distances \citep[e.g.,][]{Nierenberg2012,Sales13,Carlsten2021,Roberts2021}.  We estimate uncertainties in the luminosity function by bootstrap resampling the hosts.  The error bars on the plot represent the 68\% confidence level on the luminosity function based on bootstrapping alone, and neglect uncertainties in the magnitudes. 

The black line represents the prediction for the total luminosity function of the survey based on empirically validated scaling relations, and the grey band represents the uncertainty on the prediction from the uncertainty in the radial distribution of satellite galaxies alone.  The true theory uncertainty is larger than this band.  The ingredients in our prediction are: (1) A prediction for the whole-halo luminosity function for each host.  (2) A prediction for the fraction of the host's satellite population that lies within the LBC footprint.  As we discuss below, there is significant theoretical and observational uncertainty in the radial distribution of satellites, which leads to a significant uncertainty in our prediction for the fraction of a host's satellites that are visible in LBT-SONG.  (3) A well-quantified completeness function (from Sec.~\ref{sec:completeness}) of satellites as a function of magnitude and surface brightness. (4) A convolution of the completeness function with predictions for the size-luminosity relationship of galaxies, to determine what fraction of galaxies at fixed luminosity are observable in our survey.  We then sum the luminosity functions for all hosts to make the cumulative luminosity function expected for the whole LBT-SONG Far Sample.  

In the present study, we make a prediction of satellite populations based on simple scaling relations.  A more rigorous comparison of the candidate satellite population with $\Lambda$CDM predictions, including a likelihood function analysis, will appear in a future analysis that includes the LBT-SONG Near Sample \citep[including hosts from][]{garling2021}. 

We base our predictions for the LBT-SONG Far Sample on the whole-halo luminosity functions from \citet{Sales13}. In this analysis, the full virial volume prediction for each host is constructed with three stellar mass bins, based on the Millennium-II semi-analytic model by \citet{Guo2011} and confirmed with SDSS satellites \citep{Sales13}. Any host with $\log_{10}(\text{M}_*/\text{M}_\odot) < 10$ is assigned the luminosity function of the $\log_{10}(\text{M}_*/\text{M}_\odot) = 9.75$ stellar mass sample, as we expect the luminosity functions to be self-similar in this regime where the stellar-mass--halo-mass relation is a single power law for the hosts and satellites.  Hosts with $10 < \log_{10}(\text{M}_*/\text{M}_\odot) < 10.5$ are assigned the $\log_{10}(\text{M}_*/\text{M}_\odot) = 10.25$ luminosity function, and $10.5 < \log_{10}(\text{M}_*/\text{M}_\odot) < 11$ are assigned the $\log_{10}(\text{M}_*/\text{M}_\odot) = 10.75$ luminosity function from \citet{Sales13}. Luminosity functions are truncated for each host at a $\Delta m_V$ that corresponds to a satellite absolute magnitude $M_V > -6$, since we cannot reliably identify less luminous satellites. 

Next, we determine what fraction of satellites may lie within the LBC footprint for each host.  Our fiducial assumption is that the satellites have a radial distribution similar to that of the MW's. This distribution is well-matched to a model in which satellites live in the most massive halos at infall, subject to a reionization-based criterion as to which halos form stars or remain non-luminous \citep{Dooley2017a}.  There are hints of similarly concentrated satellite populations in other MW-mass Local Volume hosts \citep{Carlsten2020Rad}, although the radial distribution of satellites for the 101 SAGA MW analogs is significantly less concentrated \citep{Mao2024}.  Hence, the satellite radial distribution is still uncertain for MW-mass galaxies, with even greater uncertainty for the lower-mass hosts in the LBT-SONG Far Sample.  Thus, we explore a range of radial distributions, illustrated with the grey band in Fig.~\ref{fig:lf}.  The upper edge of our prediction band assumes that satellite galaxies follow the smooth dark-matter profile of the host, here assumed to be a Navarro-Frenk-White cusped profile with a concentration fixed to $c=10$ \citep{NFW1997,Han16,green2022}, which is an unlikely scenario.  The bottom edge of the band closely corresponds to the radial distribution found by \citet{Mao2024}.  The distribution is nearly identical to a theoretical model in which the satellite population experiences additional tidal disruption relative to that in dark-matter-only simulations because of the extra tidal field of the host galaxy.  In hydrodynamic simulations, this effect is shown to reduce the satellite population throughout the host halo, but most significantly near the halo center \citep{donghia2010,Brooks2013,garrison-kimmel2017,richings2020}.  Our specific choice here is to modify the \citet{Dooley2017a} distribution according to the disk stripping radial distribution of the Auriga simulations, as reported in \citet{richings2020}.  

Finally, we include our observational completeness functions from Sec. \ref{sec:completeness}, convolved with a galaxy size-luminosity relation.  Our completeness measurement tells us what fraction of artificial satellites with fixed absolute magnitude, surface brightness, and stellar populations are recovered in the LBC footprint, including the effects of masking, for each host.  A galaxy luminosity-size relation tells us the distribution of surface brightnesses of galaxies at fixed magnitude. By convolving the completeness function for each host with the luminosity-size relation, we are able to determine what fraction of satellites at fixed magnitude for each host are recoverable in our survey.  We use the luminosity-size relation from \citet{danieli2018} to best model Local Group galaxies, with a log-normal model and a scatter of $\sigma = 0.15$ dex in effective size with $L_V$ in the power-law scaling relation.  The best-fit power law and scatter are similar to that found by \citet{Carlsten2021size} for satellites of MW-mass hosts in the Local Volume.  However, the theoretical model of \citet{kravtsov2021grumpy} suggests the true scatter of galaxy sizes may be even greater, especially for galaxies fainter than the $M_V = -9$ survey limit of \citet{Carlsten2021size}.  The effect of the larger scatter in the luminosity-size relation on our luminosity function is to slightly flatten the predicted luminosity function at large $\Delta m_V$.   

Our prediction for the observed luminosity function for the Far Sample of LBT-SONG is shown with the solid line and shaded band in Fig.~\ref{fig:lf}.  As shown by the data points in Fig.~\ref{fig:lf}, the luminosity function of our candidates is broadly consistent with $\Lambda$CDM predictions, especially when we consider that we have not included all sources of scatter (e.g., the halo-to-halo scatter in the luminosity function for each host, uncertainty in host distance).  However, we do find relatively more bright satellites, and fewer faint satellites, compared to the model.  Although this is within the statistical uncertainties of the measurement and prediction, if it reflects a true difference, this feature has several plausible explanations.  First, there are hints in the Local Volume that massive satellites are more centrally concentrated about their hosts than the satellite population as a whole, potentially on account of dynamical friction \citep{Carlsten2022Elves}.  Second, \citet{Carlsten2022Elves} also found hints that satellites close in projection to their hosts have lower surface brightness than satellites of similar magnitude but more distant from the hosts.  If true, our predictions would flatten for $\Delta m_V > 10$, especially if the scatter in the luminosity-size relation is higher \citep[as predicted by][]{kravtsov2021grumpy}, since the low-luminosity galaxies would scatter to lower surface brightness, where LBT-SONG is less complete.  Finally, the measurements from \citet{Sales13} are made for $\Delta m < 5$.  The extrapolation of their power-law tail to higher $\Delta m$ makes sense in the context of the power-law behavior of the stellar-to-halo mass relation and a power-law subhalo mass function if the average mass-to-light ratio does not vary as a function of $\Delta m$.  However, we expect that the mass-to-light ratio should increase for increasing $\Delta m$, because the mix of SFH at fixed $\Delta m$ will change.  At the largest $\Delta m$, we expect to see mainly reionization fossils with large optical mass-to-light ratios, but at modest $\Delta m$ we find a heterogeneous population in terms of SFH and potentially lower mass-to-light ratios.  Thus, at large $\Delta m$, the shape of the luminosity function may change.  In future work, we will explore these issues in more quantitative detail.  For this study, though, we find that the satellite number counts are consistent with $\Lambda$CDM expectations.  

\section{Summary and Conclusions} \label{sec:summary}
The LBT-SONG is the first survey to systematically study the satellites and candidate satellites of nearby intermediate-mass galaxy systems outside of the Local Group. These satellites are found in the LBT-SONG Near Sample \citep{garling2019ddo113,garling2021} and Far Sample \citep[][and in this work]{Davis2021}. In this work we presented the candidate satellite populations and completeness results for 13 intermediate-mass Far Sample host galaxies with stellar masses from $4.7 \times 10^{8} \,\text{M}_\odot$, roughly that of the Small Magellanic Cloud, to $5.4 \times 10^{10} \,\text{M}_\odot$, roughly that of the MW. In distance, they range from 4.7 Mpc to 10.7 Mpc. 

Our key takeaways are: 
\begin{itemize}
	\item In the LBT-SONG Far Sample we discover 3 new satellites candidates: NGC 628 dwA and NGC 628 dwB \citep{Davis2021}, and NGC 672 dwD. NGC 672 dwD, with $\text{M}_{\text{V}} = -6.6$ and an estimated WISE1 stellar mass upper limit of $5.6 \times 10^4 M_\odot$, is notable as a potentially quenched but gas-rich ultrafaint dwarf, although deeper radio  maps and higher-resolution optical data are needed to confirm its association with the HI and with the putative host.
	\item We present photometric and morphological properties of the heterogeneous set of 8 dwarf spheroidal, dwarf transitional and star-forming satellite candidates found in the Far Sample, and note the presence of 2 larger well-known satellites. 
	\item The satellite candidates are heterogeneous in gas content, present-day star-formation activity, and morphology.  The sample properties indicate a variety of environmental quenching mechanisms such as strangulation, tidal stripping, and ram-pressure stripping may be at work in hosts less massive than the MW.  Importantly, none of the 8 low-mass dwarf candidates share the gas-rich, star-forming characteristics that dominate field dwarf populations of similar luminosities.  The predominantly quenched population indicates there are galaxy evolutionary pathways for suppressing star formation in satellites of intermediate-mass hosts similar to those at play in satellites of the larger MW-mass galaxies. 
	\item We use our satellite counts and completeness estimates to make luminosity predictions for the LBT-SONG Far Sample and find that it is in agreement with predictions of $\Lambda$CDM.
\end{itemize}

With the LBT-SONG survey, we now have a statistical sample of hosts and their close satellites in the intermediate-mass regime outside the Local Group. We plan future theoretical work based on this sample and completeness results using semi-analytic models \citep[as in][]{garling2021}. More broadly, the LBT-SONG survey provides an additional data set for simulators to consider. These data and theoretical methods combined will help to disentangle the many physical effects governing star formation and quenching of low-mass dwarf galaxies over a previously understudied range of intermediate host masses, which will allow us to better understand these galaxies, their halos, and the  $\text{M}_*$-$\text{M}_{\text{halo}}$ relation.

\section*{Acknowledgements}
We thank David Stark for his assistance with HI measurements. The computations in this paper were run on the CCAPP condos of the Ruby Cluster and the Pitzer Cluster at the Ohio Supercomputer Center \citep{OhioSupercomputerCenter1987}. This material is based upon work supported by the National Science Foundation under Grant No. AST-1615838. AHGP and CTG were also supported by National Science Foundation Grant No. AST-1813628. CTG is supported by the Owens Family Foundation.  AMN was supported by a NASA Postdoctoral Program Fellowship. CSK is supported by AST-1907570, 2307385, and 2407206. KJC, DMR, AS, and AHGP are additionally supported by National Science Foundation Grant No. AST-2008110. DJS acknowledges support from NSF grant AST-2205863. AS is supported by an NSF Astronomy and Astrophysics Postdoctoral Fellowship under award AST-1903834.

The LBT is an international collaboration among institutions in the
United States, Italy and Germany. LBT Corporation partners are: The
University of Arizona on behalf of the Arizona Board of Regents;
Istituto Nazionale di Astrofisica, Italy; LBT Beteiligungsgesellschaft,
Germany, representing the Max-Planck Society, The Leibniz Institute
for Astrophysics Potsdam, and Heidelberg University; The Ohio State
University, representing OSU, University of Notre Dame, University
of Minnesota and University of Virginia.





\begin{thebibliography}{}
\makeatletter
\relax
\def\mn@urlcharsother{\let\do\@makeother \do\$\do\&\do\#\do\^\do\_\do\%\do\~}
\def\mn@doi{\begingroup\mn@urlcharsother \@ifnextchar [ {\mn@doi@}
  {\mn@doi@[]}}
\def\mn@doi@[#1]#2{\def\@tempa{#1}\ifx\@tempa\@empty \href
  {http://dx.doi.org/#2} {doi:#2}\else \href {http://dx.doi.org/#2} {#1}\fi
  \endgroup}
\def\mn@eprint#1#2{\mn@eprint@#1:#2::\@nil}
\def\mn@eprint@arXiv#1{\href {http://arxiv.org/abs/#1} {{\tt arXiv:#1}}}
\def\mn@eprint@dblp#1{\href {http://dblp.uni-trier.de/rec/bibtex/#1.xml}
  {dblp:#1}}
\def\mn@eprint@#1:#2:#3:#4\@nil{\def\@tempa {#1}\def\@tempb {#2}\def\@tempc
  {#3}\ifx \@tempc \@empty \let \@tempc \@tempb \let \@tempb \@tempa \fi \ifx
  \@tempb \@empty \def\@tempb {arXiv}\fi \@ifundefined
  {mn@eprint@\@tempb}{\@tempb:\@tempc}{\expandafter \expandafter \csname
  mn@eprint@\@tempb\endcsname \expandafter{\@tempc}}}

\bibitem[\protect\citeauthoryear{{Abazajian} et~al.,}{{Abazajian}
  et~al.}{2009}]{sdss2009}
{Abazajian} K.~N.,  et~al., 2009, \mn@doi [\apjs]
  {10.1088/0067-0049/182/2/543}, \href
  {https://ui.adsabs.harvard.edu/abs/2009ApJS..182..543A} {182, 543}

\bibitem[\protect\citeauthoryear{{Adams} \& {Oosterloo}}{{Adams} \&
  {Oosterloo}}{2018}]{adams2018}
{Adams} E. A.~K.,  {Oosterloo} T.~A.,  2018, \mn@doi [\aap]
  {10.1051/0004-6361/201732017}, \href
  {https://ui.adsabs.harvard.edu/abs/2018A&A...612A..26A} {612, A26}

\bibitem[\protect\citeauthoryear{{Adams}, {Kochanek}, {Gerke}  \&
  {Stanek}}{{Adams} et~al.}{2017}]{Adams2017}
{Adams} S.~M.,  {Kochanek} C.~S.,  {Gerke} J.~R.,   {Stanek} K.~Z.,  2017,
  \mn@doi [\mnras] {10.1093/mnras/stx898}, \href
  {http://adsabs.harvard.edu/abs/2017MNRAS.469.1445A} {469, 1445}

\bibitem[\protect\citeauthoryear{{Akins}, {Christensen}, {Brooks}, {Munshi},
  {Applebaum}, {Engelhardt}  \& {Chamberland}}{{Akins}
  et~al.}{2021}]{akins2021}
{Akins} H.~B.,  {Christensen} C.~R.,  {Brooks} A.~M.,  {Munshi} F.,
  {Applebaum} E.,  {Engelhardt} A.,   {Chamberland} L.,  2021, \mn@doi [\apj]
  {10.3847/1538-4357/abe2ab}, \href
  {https://ui.adsabs.harvard.edu/abs/2021ApJ...909..139A} {909, 139}

\bibitem[\protect\citeauthoryear{{Albareti} et~al.,}{{Albareti}
  et~al.}{2017}]{sdssdr13_2017}
{Albareti} F.~D.,  et~al., 2017, \mn@doi [\apjs] {10.3847/1538-4365/aa8992},
  \href {https://ui.adsabs.harvard.edu/abs/2017ApJS..233...25A} {233, 25}

\bibitem[\protect\citeauthoryear{{Applebaum}, {Brooks}, {Christensen},
  {Munshi}, {Quinn}, {Shen}  \& {Tremmel}}{{Applebaum}
  et~al.}{2021}]{applebaum2021}
{Applebaum} E.,  {Brooks} A.~M.,  {Christensen} C.~R.,  {Munshi} F.,  {Quinn}
  T.~R.,  {Shen} S.,   {Tremmel} M.,  2021, \mn@doi [\apj]
  {10.3847/1538-4357/abcafa}, \href
  {https://ui.adsabs.harvard.edu/abs/2021ApJ...906...96A} {906, 96}

\bibitem[\protect\citeauthoryear{{Assef} et~al.,}{{Assef}
  et~al.}{2010}]{assef2010}
{Assef} R.~J.,  et~al., 2010, \mn@doi [\apj] {10.1088/0004-637X/713/2/970},
  \href {https://ui.adsabs.harvard.edu/abs/2010ApJ...713..970A} {713, 970}

\bibitem[\protect\citeauthoryear{{Baldry}, {Glazebrook}  \& {Driver}}{{Baldry}
  et~al.}{2008}]{Baldry08}
{Baldry} I.~K.,  {Glazebrook} K.,   {Driver} S.~P.,  2008, \mn@doi [\mnras]
  {10.1111/j.1365-2966.2008.13348.x}, \href
  {https://ui.adsabs.harvard.edu/abs/2008MNRAS.388..945B} {388, 945}

\bibitem[\protect\citeauthoryear{{Baldry} et~al.,}{{Baldry}
  et~al.}{2012}]{Baldry2012}
{Baldry} I.~K.,  et~al., 2012, \mn@doi [\mnras]
  {10.1111/j.1365-2966.2012.20340.x}, \href
  {https://ui.adsabs.harvard.edu/abs/2012MNRAS.421..621B} {421, 621}

\bibitem[\protect\citeauthoryear{{Bechtol} et~al.,}{{Bechtol}
  et~al.}{2015}]{Bechtol15}
{Bechtol} K.,  et~al., 2015, \mn@doi [\apj] {10.1088/0004-637X/807/1/50}, \href
  {http://adsabs.harvard.edu/abs/2015ApJ...807...50B} {807, 50}

\bibitem[\protect\citeauthoryear{{Behroozi}, {Wechsler}  \&
  {Conroy}}{{Behroozi} et~al.}{2013}]{Behroozi13}
{Behroozi} P.~S.,  {Wechsler} R.~H.,   {Conroy} C.,  2013, \mn@doi [\apj]
  {10.1088/0004-637X/770/1/57}, \href
  {http://adsabs.harvard.edu/abs/2013ApJ...770...57B} {770, 57}

\bibitem[\protect\citeauthoryear{{Behroozi}, {Wechsler}, {Hearin}  \&
  {Conroy}}{{Behroozi} et~al.}{2019}]{behroozi2019}
{Behroozi} P.,  {Wechsler} R.~H.,  {Hearin} A.~P.,   {Conroy} C.,  2019,
  \mn@doi [\mnras] {10.1093/mnras/stz1182}, \href
  {https://ui.adsabs.harvard.edu/abs/2019MNRAS.488.3143B} {488, 3143}

\bibitem[\protect\citeauthoryear{{Belokurov} et~al.,}{{Belokurov}
  et~al.}{2008}]{belokurov2008}
{Belokurov} V.,  et~al., 2008, \mn@doi [\apjl] {10.1086/592962}, \href
  {http://adsabs.harvard.edu/abs/2008ApJ...686L..83B} {686, L83}

\bibitem[\protect\citeauthoryear{{Belokurov} et~al.,}{{Belokurov}
  et~al.}{2010}]{belokurov2010}
{Belokurov} V.,  et~al., 2010, \mn@doi [\apj] {10.1088/2041-8205/712/1/L103},
  \href {http://adsabs.harvard.edu/abs/2010ApJ...712L.103B} {712, L103}

\bibitem[\protect\citeauthoryear{{Bennet}, {Sand}, {Crnojevi{\'c}}, {Spekkens},
  {Karunakaran}, {Zaritsky}  \& {Mutlu-Pakdil}}{{Bennet}
  et~al.}{2019}]{bennet2019m101}
{Bennet} P.,  {Sand} D.~J.,  {Crnojevi{\'c}} D.,  {Spekkens} K.,  {Karunakaran}
  A.,  {Zaritsky} D.,   {Mutlu-Pakdil} B.,  2019, \mn@doi [\apj]
  {10.3847/1538-4357/ab46ab}, \href
  {https://ui.adsabs.harvard.edu/abs/2019ApJ...885..153B} {885, 153}

\bibitem[\protect\citeauthoryear{{Bennet}, {Sand}, {Crnojevi{\'c}}, {Spekkens},
  {Karunakaran}, {Zaritsky}  \& {Mutlu-Pakdil}}{{Bennet}
  et~al.}{2020}]{bennet2020}
{Bennet} P.,  {Sand} D.~J.,  {Crnojevi{\'c}} D.,  {Spekkens} K.,  {Karunakaran}
  A.,  {Zaritsky} D.,   {Mutlu-Pakdil} B.,  2020, \mn@doi [\apjl]
  {10.3847/2041-8213/ab80c5}, \href
  {https://ui.adsabs.harvard.edu/abs/2020ApJ...893L...9B} {893, L9}

\bibitem[\protect\citeauthoryear{{Benson}, {Frenk}, {Lacey}, {Baugh}  \&
  {Cole}}{{Benson} et~al.}{2002}]{Benson02}
{Benson} A.~J.,  {Frenk} C.~S.,  {Lacey} C.~G.,  {Baugh} C.~M.,   {Cole} S.,
  2002, \mn@doi [\mnras] {10.1046/j.1365-8711.2002.05388.x}, \href
  {http://adsabs.harvard.edu/abs/2002MNRAS.333..177B} {333, 177}

\bibitem[\protect\citeauthoryear{{Bertin} \& {Arnouts}}{{Bertin} \&
  {Arnouts}}{1996}]{Bertin1996}
{Bertin} E.,  {Arnouts} S.,  1996, \mn@doi [\aaps] {10.1051/aas:1996164}, \href
  {http://adsabs.harvard.edu/abs/1996A%26AS..117..393B} {117, 393}

\bibitem[\protect\citeauthoryear{{Bhattacharyya} et~al.,}{{Bhattacharyya}
  et~al.}{2023}]{Bhattacharyya2023}
{Bhattacharyya} J.,  et~al., 2023, \mn@doi [arXiv e-prints]
  {10.48550/arXiv.2312.00773}, \href
  {https://ui.adsabs.harvard.edu/abs/2023arXiv231200773B} {p. arXiv:2312.00773}

\bibitem[\protect\citeauthoryear{{Bose} et~al.,}{{Bose}
  et~al.}{2017}]{Bose2017}
{Bose} S.,  et~al., 2017, \mn@doi [\mnras] {10.1093/mnras/stw2686}, \href
  {https://ui.adsabs.harvard.edu/abs/2017MNRAS.464.4520B} {464, 4520}

\bibitem[\protect\citeauthoryear{{Bovill} \& {Ricotti}}{{Bovill} \&
  {Ricotti}}{2011}]{Bovill11}
{Bovill} M.~S.,  {Ricotti} M.,  2011, \mn@doi [\apj]
  {10.1088/0004-637X/741/1/17}, \href
  {http://adsabs.harvard.edu/abs/2011ApJ...741...17B} {741, 17}

\bibitem[\protect\citeauthoryear{{Bradford}, {Geha}  \& {Blanton}}{{Bradford}
  et~al.}{2015}]{Brandford2015}
{Bradford} J.~D.,  {Geha} M.~C.,   {Blanton} M.~R.,  2015, \mn@doi [\apj]
  {10.1088/0004-637X/809/2/146}, \href
  {https://ui.adsabs.harvard.edu/abs/2015ApJ...809..146B} {809, 146}

\bibitem[\protect\citeauthoryear{{Brooks}, {Kuhlen}, {Zolotov}  \&
  {Hooper}}{{Brooks} et~al.}{2013}]{Brooks2013}
{Brooks} A.~M.,  {Kuhlen} M.,  {Zolotov} A.,   {Hooper} D.,  2013, \mn@doi
  [\apj] {10.1088/0004-637X/765/1/22}, \href
  {http://adsabs.harvard.edu/abs/2013ApJ...765...22B} {765, 22}

\bibitem[\protect\citeauthoryear{{Brown} et~al.,}{{Brown}
  et~al.}{2014}]{Brown14b}
{Brown} T.~M.,  et~al., 2014, \mn@doi [\apj] {10.1088/0004-637X/796/2/91},
  \href {http://adsabs.harvard.edu/abs/2014ApJ...796...91B} {796, 91}

\bibitem[\protect\citeauthoryear{{Buck}, {Macci{\`o}}, {Dutton}, {Obreja}  \&
  {Frings}}{{Buck} et~al.}{2019}]{Buck2019}
{Buck} T.,  {Macci{\`o}} A.~V.,  {Dutton} A.~A.,  {Obreja} A.,   {Frings} J.,
  2019, \mn@doi [\mnras] {10.1093/mnras/sty2913}, \href
  {https://ui.adsabs.harvard.edu/abs/2019MNRAS.483.1314B} {483, 1314}

\bibitem[\protect\citeauthoryear{{Carlin} et~al.,}{{Carlin}
  et~al.}{2016}]{Carlin16}
{Carlin} J.~L.,  et~al., 2016, \mn@doi [\apjl] {10.3847/2041-8205/828/1/L5},
  \href {https://ui.adsabs.harvard.edu/abs/2016ApJ...828L...5C} {828, L5}

\bibitem[\protect\citeauthoryear{Carlin et~al.,}{Carlin
  et~al.}{2019}]{Carlin_2019}
Carlin J.~L.,  et~al., 2019, \mn@doi [The Astrophysical Journal]
  {10.3847/1538-4357/ab4c32}, 886, 109

\bibitem[\protect\citeauthoryear{{Carlin} et~al.,}{{Carlin}
  et~al.}{2021}]{carlin2021}
{Carlin} J.~L.,  et~al., 2021, \mn@doi [\apj] {10.3847/1538-4357/abe040}, \href
  {https://ui.adsabs.harvard.edu/abs/2021ApJ...909..211C} {909, 211}

\bibitem[\protect\citeauthoryear{{Carlsten}, {Greco}, {Beaton}  \&
  {Greene}}{{Carlsten} et~al.}{2020a}]{Carlsten2020}
{Carlsten} S.~G.,  {Greco} J.~P.,  {Beaton} R.~L.,   {Greene} J.~E.,  2020a,
  \mn@doi [\apj] {10.3847/1538-4357/ab7758}, \href
  {https://ui.adsabs.harvard.edu/abs/2020ApJ...891..144C} {891, 144}

\bibitem[\protect\citeauthoryear{{Carlsten}, {Greene}, {Peter}, {Greco}  \&
  {Beaton}}{{Carlsten} et~al.}{2020b}]{Carlsten2020Rad}
{Carlsten} S.~G.,  {Greene} J.~E.,  {Peter} A. H.~G.,  {Greco} J.~P.,
  {Beaton} R.~L.,  2020b, \mn@doi [\apj] {10.3847/1538-4357/abb60b}, \href
  {https://ui.adsabs.harvard.edu/abs/2020ApJ...902..124C} {902, 124}

\bibitem[\protect\citeauthoryear{{Carlsten}, {Greene}, {Peter}, {Beaton}  \&
  {Greco}}{{Carlsten} et~al.}{2021a}]{Carlsten2021}
{Carlsten} S.~G.,  {Greene} J.~E.,  {Peter} A. H.~G.,  {Beaton} R.~L.,
  {Greco} J.~P.,  2021a, \mn@doi [\apj] {10.3847/1538-4357/abd039}, \href
  {https://ui.adsabs.harvard.edu/abs/2021ApJ...908..109C} {908, 109}

\bibitem[\protect\citeauthoryear{{Carlsten}, {Greene}, {Greco}, {Beaton}  \&
  {Kado-Fong}}{{Carlsten} et~al.}{2021b}]{Carlsten2021size}
{Carlsten} S.~G.,  {Greene} J.~E.,  {Greco} J.~P.,  {Beaton} R.~L.,
  {Kado-Fong} E.,  2021b, \mn@doi [\apj] {10.3847/1538-4357/ac2581}, \href
  {https://ui.adsabs.harvard.edu/abs/2021ApJ...922..267C} {922, 267}

\bibitem[\protect\citeauthoryear{{Carlsten}, {Greene}, {Beaton}, {Danieli}  \&
  {Greco}}{{Carlsten} et~al.}{2022}]{Carlsten2022Elves}
{Carlsten} S.~G.,  {Greene} J.~E.,  {Beaton} R.~L.,  {Danieli} S.,   {Greco}
  J.~P.,  2022, \mn@doi [\apj] {10.3847/1538-4357/ac6fd7}, \href
  {https://ui.adsabs.harvard.edu/abs/2022ApJ...933...47C} {933, 47}

\bibitem[\protect\citeauthoryear{{Casey}, {Greco}, {Peter}  \& {Davis}}{{Casey}
  et~al.}{2023}]{casey2023}
{Casey} K.~J.,  {Greco} J.~P.,  {Peter} A. H.~G.,   {Davis} A.~B.,  2023,
  \mn@doi [\mnras] {10.1093/mnras/stad352}, \href
  {https://ui.adsabs.harvard.edu/abs/2023MNRAS.520.4715C} {520, 4715}

\bibitem[\protect\citeauthoryear{Center}{Center}{1987}]{OhioSupercomputerCenter1987}
Center O.~S.,  1987, Ohio Supercomputer Center, \url
  {http://osc.edu/ark:/19495/f5s1ph73}

\bibitem[\protect\citeauthoryear{Chabrier}{Chabrier}{2003}]{Chabrier2001}
Chabrier G.,  2003, Publications of the Astronomical Society of the Pacific,
  115, 763

\bibitem[\protect\citeauthoryear{{Chau}, {Mayer}  \& {Governato}}{{Chau}
  et~al.}{2017}]{Chau2017}
{Chau} A.,  {Mayer} L.,   {Governato} F.,  2017, \mn@doi [\apj]
  {10.3847/1538-4357/aa7e74}, \href
  {https://ui.adsabs.harvard.edu/abs/2017ApJ...845...17C} {845, 17}

\bibitem[\protect\citeauthoryear{Chiboucas, Jacobs, Tully  \&
  Karachentsev}{Chiboucas et~al.}{2013}]{chiboucas2013confirmation}
Chiboucas K.,  Jacobs B.~A.,  Tully R.~B.,   Karachentsev I.~D.,  2013, The
  Astronomical Journal, 146, 126

\bibitem[\protect\citeauthoryear{{Christensen}, {Brooks}, {Munshi}, {Riggs},
  {Van Nest}, {Akins}, {Quinn}  \& {Chamberland}}{{Christensen}
  et~al.}{2024}]{Christensen2024}
{Christensen} C.~R.,  {Brooks} A.~M.,  {Munshi} F.,  {Riggs} C.,  {Van Nest}
  J.,  {Akins} H.,  {Quinn} T.~R.,   {Chamberland} L.,  2024, \mn@doi [\apj]
  {10.3847/1538-4357/ad0c5a}, \href
  {https://ui.adsabs.harvard.edu/abs/2024ApJ...961..236C} {961, 236}

\bibitem[\protect\citeauthoryear{{Conroy}, {Wechsler}  \& {Kravtsov}}{{Conroy}
  et~al.}{2006}]{Conroy_2006}
{Conroy} C.,  {Wechsler} R.~H.,   {Kravtsov} A.~V.,  2006, \mn@doi [\apj]
  {10.1086/503602}, \href
  {https://ui.adsabs.harvard.edu/abs/2006ApJ...647..201C} {647, 201}

\bibitem[\protect\citeauthoryear{Crnojevi{\'c} et~al.,}{Crnojevi{\'c}
  et~al.}{2019}]{crnojevic2019faint}
Crnojevi{\'c} D.,  et~al., 2019, The Astrophysical Journal, 872, 80

\bibitem[\protect\citeauthoryear{Cutri et~al.,}{Cutri et~al.}{2013}]{ALLWISE}
Cutri R.,  et~al., 2013, Explanatory Supplement to the AllWISE Data Release
  Products, by RM Cutri et al.

\bibitem[\protect\citeauthoryear{{D'Onghia}, {Springel}, {Hernquist}  \&
  {Keres}}{{D'Onghia} et~al.}{2010}]{donghia2010}
{D'Onghia} E.,  {Springel} V.,  {Hernquist} L.,   {Keres} D.,  2010, \mn@doi
  [\apj] {10.1088/0004-637X/709/2/1138}, \href
  {https://ui.adsabs.harvard.edu/abs/2010ApJ...709.1138D} {709, 1138}

\bibitem[\protect\citeauthoryear{{Danieli}, {van Dokkum}, {Merritt}, {Abraham},
  {Zhang}, {Karachentsev}  \& {Makarova}}{{Danieli} et~al.}{2017}]{DanieliM101}
{Danieli} S.,  {van Dokkum} P.,  {Merritt} A.,  {Abraham} R.,  {Zhang} J.,
  {Karachentsev} I.~D.,   {Makarova} L.~N.,  2017, \mn@doi [\apj]
  {10.3847/1538-4357/aa615b}, \href
  {https://ui.adsabs.harvard.edu/abs/2017ApJ...837..136D} {837, 136}

\bibitem[\protect\citeauthoryear{{Danieli}, {van Dokkum}  \&
  {Conroy}}{{Danieli} et~al.}{2018}]{danieli2018}
{Danieli} S.,  {van Dokkum} P.,   {Conroy} C.,  2018, \mn@doi [\apj]
  {10.3847/1538-4357/aaadfb}, \href
  {http://adsabs.harvard.edu/abs/2018ApJ...856...69D} {856, 69}

\bibitem[\protect\citeauthoryear{{Davis} et~al.,}{{Davis}
  et~al.}{2021}]{Davis2021}
{Davis} A.~B.,  et~al., 2021, \mn@doi [\mnras] {10.1093/mnras/staa3246}, \href
  {https://ui.adsabs.harvard.edu/abs/2021MNRAS.500.3854D} {500, 3854}

\bibitem[\protect\citeauthoryear{{Dekel} \& {Birnboim}}{{Dekel} \&
  {Birnboim}}{2006}]{dekel2006}
{Dekel} A.,  {Birnboim} Y.,  2006, \mn@doi [\mnras]
  {10.1111/j.1365-2966.2006.10145.x}, \href
  {https://ui.adsabs.harvard.edu/abs/2006MNRAS.368....2D} {368, 2}

\bibitem[\protect\citeauthoryear{{Dekker} \& {Kravtsov}}{{Dekker} \&
  {Kravtsov}}{2024}]{Dekker2024}
{Dekker} A.,  {Kravtsov} A.,  2024, \mn@doi [arXiv e-prints]
  {10.48550/arXiv.2407.04198}, \href
  {https://ui.adsabs.harvard.edu/abs/2024arXiv240704198D} {p. arXiv:2407.04198}

\bibitem[\protect\citeauthoryear{{Dekker}, {Ando}, {Correa}  \& {Ng}}{{Dekker}
  et~al.}{2022}]{Dekker2022}
{Dekker} A.,  {Ando} S.,  {Correa} C.~A.,   {Ng} K. C.~Y.,  2022, \mn@doi
  [\prd] {10.1103/PhysRevD.106.123026}, \href
  {https://ui.adsabs.harvard.edu/abs/2022PhRvD.106l3026D} {106, 123026}

\bibitem[\protect\citeauthoryear{{Digby} et~al.,}{{Digby}
  et~al.}{2019}]{digby2019}
{Digby} R.,  et~al., 2019, \mn@doi [\mnras] {10.1093/mnras/stz745}, \href
  {https://ui.adsabs.harvard.edu/abs/2019MNRAS.485.5423D} {485, 5423}

\bibitem[\protect\citeauthoryear{{Dooley}, {Peter}, {Yang}, {Willman},
  {Griffen}  \& {Frebel}}{{Dooley} et~al.}{2017a}]{Dooley2017a}
{Dooley} G.~A.,  {Peter} A. H.~G.,  {Yang} T.,  {Willman} B.,  {Griffen} B.~F.,
    {Frebel} A.,  2017a, \mn@doi [\mnras] {10.1093/mnras/stx1900}, \href
  {https://ui.adsabs.harvard.edu/abs/2017MNRAS.471.4894D} {471, 4894}

\bibitem[\protect\citeauthoryear{{Dooley}, {Peter}, {Carlin}, {Frebel},
  {Bechtol}  \& {Willman}}{{Dooley} et~al.}{2017b}]{Dooley17b}
{Dooley} G.~A.,  {Peter} A. H.~G.,  {Carlin} J.~L.,  {Frebel} A.,  {Bechtol}
  K.,   {Willman} B.,  2017b, \mn@doi [\mnras] {10.1093/mnras/stx2001}, \href
  {https://ui.adsabs.harvard.edu/abs/2017MNRAS.472.1060D} {472, 1060}

\bibitem[\protect\citeauthoryear{{Driver} et~al.,}{{Driver}
  et~al.}{2022}]{driver2022}
{Driver} S.~P.,  et~al., 2022, \mn@doi [\mnras] {10.1093/mnras/stac472}, \href
  {https://ui.adsabs.harvard.edu/abs/2022MNRAS.tmp..552D} {}

\bibitem[\protect\citeauthoryear{{Drlica-Wagner} et~al.,}{{Drlica-Wagner}
  et~al.}{2015}]{DrlicaWagner2015}
{Drlica-Wagner} A.,  et~al., 2015, \mn@doi [\apj]
  {10.1088/0004-637X/813/2/109}, \href
  {http://adsabs.harvard.edu/abs/2015ApJ...813..109D} {813, 109}

\bibitem[\protect\citeauthoryear{{Drlica-Wagner} et~al.,}{{Drlica-Wagner}
  et~al.}{2020}]{drlica-wagner2020}
{Drlica-Wagner} A.,  et~al., 2020, \mn@doi [\apj] {10.3847/1538-4357/ab7eb9},
  \href {https://ui.adsabs.harvard.edu/abs/2020ApJ...893...47D} {893, 47}

\bibitem[\protect\citeauthoryear{{Drlica-Wagner} et~al.,}{{Drlica-Wagner}
  et~al.}{2021}]{delve2021}
{Drlica-Wagner} A.,  et~al., 2021, \mn@doi [\apjs] {10.3847/1538-4365/ac079d},
  \href {https://ui.adsabs.harvard.edu/abs/2021ApJS..256....2D} {256, 2}

\bibitem[\protect\citeauthoryear{{Erwin}}{{Erwin}}{2015}]{Erwin2015}
{Erwin} P.,  2015, \mn@doi [\apj] {10.1088/0004-637X/799/2/226}, \href
  {https://ui.adsabs.harvard.edu/abs/2015ApJ...799..226E} {799, 226}

\bibitem[\protect\citeauthoryear{{Esteban}, {Peter}  \& {Kim}}{{Esteban}
  et~al.}{2023}]{Esteban2023}
{Esteban} I.,  {Peter} A. H.~G.,   {Kim} S.~Y.,  2023, \mn@doi [arXiv e-prints]
  {10.48550/arXiv.2306.04674}, \href
  {https://ui.adsabs.harvard.edu/abs/2023arXiv230604674E} {p. arXiv:2306.04674}

\bibitem[\protect\citeauthoryear{{Fillingham}, {Cooper}, {Pace},
  {Boylan-Kolchin}, {Bullock}, {Garrison-Kimmel}  \& {Wheeler}}{{Fillingham}
  et~al.}{2016}]{fillingham2016}
{Fillingham} S.~P.,  {Cooper} M.~C.,  {Pace} A.~B.,  {Boylan-Kolchin} M.,
  {Bullock} J.~S.,  {Garrison-Kimmel} S.,   {Wheeler} C.,  2016, \mn@doi
  [\mnras] {10.1093/mnras/stw2131}, \href
  {https://ui.adsabs.harvard.edu/abs/2016MNRAS.463.1916F} {463, 1916}

\bibitem[\protect\citeauthoryear{{Font}, {McCarthy}, {Belokurov}, {Brown}  \&
  {Stafford}}{{Font} et~al.}{2022}]{font2022}
{Font} A.~S.,  {McCarthy} I.~G.,  {Belokurov} V.,  {Brown} S.~T.,   {Stafford}
  S.~G.,  2022, \mn@doi [\mnras] {10.1093/mnras/stac183}, \href
  {https://ui.adsabs.harvard.edu/abs/2022MNRAS.511.1544F} {511, 1544}

\bibitem[\protect\citeauthoryear{{Garling}, {Peter}, {Kochanek}, {Sand}  \&
  {Crnojevi{\'c}}}{{Garling} et~al.}{2020}]{garling2019ddo113}
{Garling} C.~T.,  {Peter} A. H.~G.,  {Kochanek} C.~S.,  {Sand} D.~J.,
  {Crnojevi{\'c}} D.,  2020, \mn@doi [\mnras] {10.1093/mnras/stz3526}, \href
  {https://ui.adsabs.harvard.edu/abs/2020MNRAS.492.1713G} {492, 1713}

\bibitem[\protect\citeauthoryear{{Garling}, {Peter}, {Kochanek}, {Sand}  \&
  {Crnojevi{\'c}}}{{Garling} et~al.}{2021}]{garling2021}
{Garling} C.~T.,  {Peter} A. H.~G.,  {Kochanek} C.~S.,  {Sand} D.~J.,
  {Crnojevi{\'c}} D.,  2021, \mn@doi [\mnras] {10.1093/mnras/stab2447}, \href
  {https://ui.adsabs.harvard.edu/abs/2021MNRAS.507.4764G} {507, 4764}

\bibitem[\protect\citeauthoryear{{Garling}, {Peter}, {Spekkens}, {Sand},
  {Hargis}, {Crnojevi{\'c}}  \& {Carlin}}{{Garling} et~al.}{2022}]{garling2022}
{Garling} C.~T.,  {Peter} A. H.~G.,  {Spekkens} K.,  {Sand} D.~J.,  {Hargis}
  J.,  {Crnojevi{\'c}} D.,   {Carlin} J.~L.,  2022, arXiv e-prints, \href
  {https://ui.adsabs.harvard.edu/abs/2022arXiv220909262G} {p. arXiv:2209.09262}

\bibitem[\protect\citeauthoryear{{Garling}, {Peter}, {Spekkens}, {Sand},
  {Hargis}, {Crnojevi{\'c}}  \& {Carlin}}{{Garling} et~al.}{2024}]{Garling2024}
{Garling} C.~T.,  {Peter} A. H.~G.,  {Spekkens} K.,  {Sand} D.~J.,  {Hargis}
  J.,  {Crnojevi{\'c}} D.,   {Carlin} J.~L.,  2024, \mn@doi [\mnras]
  {10.1093/mnras/stae014}, \href
  {https://ui.adsabs.harvard.edu/abs/2024MNRAS.528..365G} {528, 365}

\bibitem[\protect\citeauthoryear{{Garrison-Kimmel} et~al.,}{{Garrison-Kimmel}
  et~al.}{2017}]{garrison-kimmel2017}
{Garrison-Kimmel} S.,  et~al., 2017, \mn@doi [\mnras] {10.1093/mnras/stx1710},
  \href {https://ui.adsabs.harvard.edu/abs/2017MNRAS.471.1709G} {471, 1709}

\bibitem[\protect\citeauthoryear{{Garrison-Kimmel} et~al.,}{{Garrison-Kimmel}
  et~al.}{2019}]{garrison-kimmel2019}
{Garrison-Kimmel} S.,  et~al., 2019, \mn@doi [\mnras] {10.1093/mnras/stz2507},
  \href {https://ui.adsabs.harvard.edu/abs/2019MNRAS.489.4574G} {489, 4574}

\bibitem[\protect\citeauthoryear{{Gatto}, {Fraternali}, {Read}, {Marinacci},
  {Lux}  \& {Walch}}{{Gatto} et~al.}{2013}]{gatto2013}
{Gatto} A.,  {Fraternali} F.,  {Read} J.~I.,  {Marinacci} F.,  {Lux} H.,
  {Walch} S.,  2013, \mn@doi [\mnras] {10.1093/mnras/stt896}, \href
  {https://ui.adsabs.harvard.edu/abs/2013MNRAS.433.2749G} {433, 2749}

\bibitem[\protect\citeauthoryear{{Geha}, {Blanton}, {Yan}  \& {Tinker}}{{Geha}
  et~al.}{2012}]{geha2012}
{Geha} M.,  {Blanton} M.~R.,  {Yan} R.,   {Tinker} J.~L.,  2012, \mn@doi [\apj]
  {10.1088/0004-637X/757/1/85}, \href
  {http://adsabs.harvard.edu/abs/2012ApJ...757...85G} {757, 85}

\bibitem[\protect\citeauthoryear{Geha et~al.,}{Geha
  et~al.}{2017}]{geha2017saga}
Geha M.,  et~al., 2017, The Astrophysical Journal, 847, 4

\bibitem[\protect\citeauthoryear{{Geha} et~al.,}{{Geha}
  et~al.}{2024}]{Geha2024}
{Geha} M.,  et~al., 2024, \mn@doi [arXiv e-prints] {10.48550/arXiv.2404.14499},
  \href {https://ui.adsabs.harvard.edu/abs/2024arXiv240414499G} {p.
  arXiv:2404.14499}

\bibitem[\protect\citeauthoryear{Gerke, Kochanek  \& Stanek}{Gerke
  et~al.}{2015}]{Gerke2015}
Gerke J.~R.,  Kochanek C.~S.,   Stanek K.~Z.,  2015, \mn@doi [MNRAS]
  {10.1093/mnras/stv776}, 450, 3289

\bibitem[\protect\citeauthoryear{Giallongo et~al.,}{Giallongo
  et~al.}{2008}]{Giallongo2008}
Giallongo E.,  et~al., 2008, Astronomy \& Astrophysics, 482, 349

\bibitem[\protect\citeauthoryear{{Grebel}, {Gallagher}  \& {Harbeck}}{{Grebel}
  et~al.}{2003}]{Grebel2003}
{Grebel} E.~K.,  {Gallagher} John~S. I.,   {Harbeck} D.,  2003, \mn@doi [\aj]
  {10.1086/368363}, \href
  {https://ui.adsabs.harvard.edu/abs/2003AJ....125.1926G} {125, 1926}

\bibitem[\protect\citeauthoryear{{Green}, {van den Bosch}  \& {Jiang}}{{Green}
  et~al.}{2022}]{green2022}
{Green} S.~B.,  {van den Bosch} F.~C.,   {Jiang} F.,  2022, \mn@doi [\mnras]
  {10.1093/mnras/stab3130}, \href
  {https://ui.adsabs.harvard.edu/abs/2022MNRAS.509.2624G} {509, 2624}

\bibitem[\protect\citeauthoryear{{Greene}, {Danieli}, {Carlsten}, {Beaton},
  {Jiang}  \& {Li}}{{Greene} et~al.}{2023}]{Greene2023}
{Greene} J.~E.,  {Danieli} S.,  {Carlsten} S.,  {Beaton} R.,  {Jiang} F.,
  {Li} J.,  2023, \mn@doi [\apj] {10.3847/1538-4357/acc58c}, \href
  {https://ui.adsabs.harvard.edu/abs/2023ApJ...949...94G} {949, 94}

\bibitem[\protect\citeauthoryear{{Gunn} \& {Gott}}{{Gunn} \&
  {Gott}}{1972}]{gunn1972}
{Gunn} J.~E.,  {Gott} J.~R.~I.,  1972, \mn@doi [\apj] {10.1086/151605}, \href
  {http://adsabs.harvard.edu/abs/1972ApJ...176....1G} {176, 1}

\bibitem[\protect\citeauthoryear{{Guo} et~al.,}{{Guo} et~al.}{2011}]{Guo2011}
{Guo} Q.,  et~al., 2011, \mn@doi [\mnras] {10.1111/j.1365-2966.2010.18114.x},
  \href {https://ui.adsabs.harvard.edu/abs/2011MNRAS.413..101G} {413, 101}

\bibitem[\protect\citeauthoryear{{Gutcke}, {Pfrommer}, {Bryan}, {Pakmor},
  {Springel}  \& {Naab}}{{Gutcke} et~al.}{2022}]{gutcke2022}
{Gutcke} T.~A.,  {Pfrommer} C.,  {Bryan} G.~L.,  {Pakmor} R.,  {Springel} V.,
  {Naab} T.,  2022, \mn@doi [\apj] {10.3847/1538-4357/aca1b4}, \href
  {https://ui.adsabs.harvard.edu/abs/2022ApJ...941..120G} {941, 120}

\bibitem[\protect\citeauthoryear{{Haines}, {Gargiulo}, {La Barbera},
  {Mercurio}, {Merluzzi}  \& {Busarello}}{{Haines} et~al.}{2007}]{haines2007}
{Haines} C.~P.,  {Gargiulo} A.,  {La Barbera} F.,  {Mercurio} A.,  {Merluzzi}
  P.,   {Busarello} G.,  2007, \mn@doi [\mnras]
  {10.1111/j.1365-2966.2007.12189.x}, \href
  {https://ui.adsabs.harvard.edu/abs/2007MNRAS.381....7H} {381, 7}

\bibitem[\protect\citeauthoryear{{Han}, {Cole}, {Frenk}  \& {Jing}}{{Han}
  et~al.}{2016}]{Han16}
{Han} J.,  {Cole} S.,  {Frenk} C.~S.,   {Jing} Y.,  2016, \mn@doi [\mnras]
  {10.1093/mnras/stv2900}, \href
  {http://adsabs.harvard.edu/abs/2016MNRAS.457.1208H} {457, 1208}

\bibitem[\protect\citeauthoryear{{Hargis}, {Willman}  \& {Peter}}{{Hargis}
  et~al.}{2014}]{Hargis14}
{Hargis} J.~R.,  {Willman} B.,   {Peter} A.~H.~G.,  2014, \mn@doi [\apjl]
  {10.1088/2041-8205/795/1/L13}, \href
  {http://adsabs.harvard.edu/abs/2014ApJ...795L..13H} {795, L13}

\bibitem[\protect\citeauthoryear{{Hargis} et~al.,}{{Hargis}
  et~al.}{2020}]{hargis2019}
{Hargis} J.~R.,  et~al., 2020, \mn@doi [\apj] {10.3847/1538-4357/ab58d2}, \href
  {https://ui.adsabs.harvard.edu/abs/2020ApJ...888...31H} {888, 31}

\bibitem[\protect\citeauthoryear{{Hayashi}, {Ibe}, {Kobayashi}, {Nakayama}  \&
  {Shirai}}{{Hayashi} et~al.}{2021}]{Hayashi2021}
{Hayashi} K.,  {Ibe} M.,  {Kobayashi} S.,  {Nakayama} Y.,   {Shirai} S.,  2021,
  \mn@doi [\prd] {10.1103/PhysRevD.103.023017}, \href
  {https://ui.adsabs.harvard.edu/abs/2021PhRvD.103b3017H} {103, 023017}

\bibitem[\protect\citeauthoryear{{Haynes} et~al.,}{{Haynes}
  et~al.}{2011}]{ALFALFA2011}
{Haynes} M.~P.,  et~al., 2011, \mn@doi [\aj] {10.1088/0004-6256/142/5/170},
  \href {https://ui.adsabs.harvard.edu/abs/2011AJ....142..170H} {142, 170}

\bibitem[\protect\citeauthoryear{{Heald} et~al.,}{{Heald}
  et~al.}{2011}]{Heald2011}
{Heald} G.,  et~al., 2011, \mn@doi [\aap] {10.1051/0004-6361/201015938}, \href
  {https://ui.adsabs.harvard.edu/abs/2011A&A...526A.118H} {526, A118}

\bibitem[\protect\citeauthoryear{{Hill}, {Green}, {Ashby}, {Brynnel},
  {Cushing}, {Little}, {Slagle}  \& {Wagner}}{{Hill}
  et~al.}{2010}]{Hill2010LBT}
{Hill} J.~M.,  {Green} R.~F.,  {Ashby} D.~S.,  {Brynnel} J.~G.,  {Cushing}
  N.~J.,  {Little} J.,  {Slagle} J.~H.,   {Wagner} R.~M.,  2010, in {Stepp}
  L.~M.,  {Gilmozzi} R.,   {Hall} H.~J.,  eds,  Society of Photo-Optical
  Instrumentation Engineers (SPIE) Conference Series Vol. 7733, Ground-based
  and Airborne Telescopes III. p. 77330C, \mn@doi{10.1117/12.856479}

\bibitem[\protect\citeauthoryear{{Homma} et~al.,}{{Homma}
  et~al.}{2018}]{Homma2018}
{Homma} D.,  et~al., 2018, \mn@doi [\pasj] {10.1093/pasj/psx050}, \href
  {http://adsabs.harvard.edu/abs/2018PASJ...70S..18H} {70, S18}

\bibitem[\protect\citeauthoryear{{Homma} et~al.,}{{Homma}
  et~al.}{2019}]{Homma2019}
{Homma} D.,  et~al., 2019, \mn@doi [\pasj] {10.1093/pasj/psz076}, \href
  {https://ui.adsabs.harvard.edu/abs/2019PASJ...71...94H} {71, 94}

\bibitem[\protect\citeauthoryear{{Horiuchi}, {Humphrey}, {O{\~n}orbe},
  {Abazajian}, {Kaplinghat}  \& {Garrison-Kimmel}}{{Horiuchi}
  et~al.}{2014}]{Horiuchi2014}
{Horiuchi} S.,  {Humphrey} P.~J.,  {O{\~n}orbe} J.,  {Abazajian} K.~N.,
  {Kaplinghat} M.,   {Garrison-Kimmel} S.,  2014, \mn@doi [\prd]
  {10.1103/PhysRevD.89.025017}, \href
  {https://ui.adsabs.harvard.edu/abs/2014PhRvD..89b5017H} {89, 025017}

\bibitem[\protect\citeauthoryear{{Irwin} et~al.,}{{Irwin}
  et~al.}{2009}]{Irwin2009}
{Irwin} J.~A.,  et~al., 2009, \mn@doi [\apj] {10.1088/0004-637X/692/2/1447},
  \href {https://ui.adsabs.harvard.edu/abs/2009ApJ...692.1447I} {692, 1447}

\bibitem[\protect\citeauthoryear{{Jacobs}, {Rizzi}, {Tully}, {Shaya}, {Makarov}
   \& {Makarova}}{{Jacobs} et~al.}{2009}]{Jacobs2009}
{Jacobs} B.~A.,  {Rizzi} L.,  {Tully} R.~B.,  {Shaya} E.~J.,  {Makarov} D.~I.,
   {Makarova} L.,  2009, \mn@doi [\aj] {10.1088/0004-6256/138/2/332}, \href
  {https://ui.adsabs.harvard.edu/abs/2009AJ....138..332J} {138, 332}

\bibitem[\protect\citeauthoryear{{Jahn}, {Sales}, {Wetzel}, {Samuel},
  {El-Badry}, {Boylan-Kolchin}  \& {Bullock}}{{Jahn} et~al.}{2022}]{jahn2021}
{Jahn} E.~D.,  {Sales} L.~V.,  {Wetzel} A.,  {Samuel} J.,  {El-Badry} K.,
  {Boylan-Kolchin} M.,   {Bullock} J.~S.,  2022, \mn@doi [\mnras]
  {10.1093/mnras/stac811}, \href
  {https://ui.adsabs.harvard.edu/abs/2022MNRAS.513.2673J} {513, 2673}

\bibitem[\protect\citeauthoryear{{Jeon}, {Besla}  \& {Bromm}}{{Jeon}
  et~al.}{2017}]{Jeon2017}
{Jeon} M.,  {Besla} G.,   {Bromm} V.,  2017, \mn@doi [\apj]
  {10.3847/1538-4357/aa8c80}, \href
  {https://ui.adsabs.harvard.edu/abs/2017ApJ...848...85J} {848, 85}

\bibitem[\protect\citeauthoryear{{Jethwa}, {Erkal}  \& {Belokurov}}{{Jethwa}
  et~al.}{2018}]{Jethwa_2017}
{Jethwa} P.,  {Erkal} D.,   {Belokurov} V.,  2018, \mn@doi [\mnras]
  {10.1093/mnras/stx2330}, \href
  {https://ui.adsabs.harvard.edu/abs/2018MNRAS.473.2060J} {473, 2060}

\bibitem[\protect\citeauthoryear{{Jones} et~al.,}{{Jones}
  et~al.}{2024}]{jones2024}
{Jones} M.~G.,  et~al., 2024, \mn@doi [\apj] {10.3847/1538-4357/ad3076}, \href
  {https://ui.adsabs.harvard.edu/abs/2024ApJ...966...93J} {966, 93}

\bibitem[\protect\citeauthoryear{Jordi, Grebel  \& Ammon}{Jordi
  et~al.}{2006}]{Jordi2006}
Jordi K.,  Grebel E.~K.,   Ammon K.,  2006, \mn@doi [A{\&}A]
  {10.1051/0004-6361:20066082}, 460, 339

\bibitem[\protect\citeauthoryear{{Karachentsev} \& {Kaisina}}{{Karachentsev} \&
  {Kaisina}}{2019}]{Karachentsev2019}
{Karachentsev} I.~D.,  {Kaisina} E.~I.,  2019, in {McQuinn} K. B.~W.,
  {Stierwalt} S.,  eds,  Vol. 344, Dwarf Galaxies: From the Deep Universe to
  the Present. pp 381--383, \mn@doi{10.1017/S1743921318005975}

\bibitem[\protect\citeauthoryear{{Karachentsev} et~al.,}{{Karachentsev}
  et~al.}{2003}]{Karachentsev2003}
{Karachentsev} I.~D.,  et~al., 2003, \mn@doi [\aap]
  {10.1051/0004-6361:20021566}, \href
  {https://ui.adsabs.harvard.edu/abs/2003A&A...398..479K} {398, 479}

\bibitem[\protect\citeauthoryear{{Karachentsev} et~al.,}{{Karachentsev}
  et~al.}{2015}]{karachentsev2015spiral}
{Karachentsev} I.~D.,  et~al., 2015, \mn@doi [Astrophysical Bulletin]
  {10.1134/S199034131504001X}, \href
  {https://ui.adsabs.harvard.edu/abs/2015AstBu..70..379K} {70, 379}

\bibitem[\protect\citeauthoryear{{Karachentsev}, {Kaisina}  \&
  {Makarov}}{{Karachentsev} et~al.}{2018}]{Karachentsev2018}
{Karachentsev} I.~D.,  {Kaisina} E.~I.,   {Makarov} D.~I.,  2018, \mn@doi
  [\mnras] {10.1093/mnras/sty1774}, \href
  {https://ui.adsabs.harvard.edu/abs/2018MNRAS.479.4136K} {479, 4136}

\bibitem[\protect\citeauthoryear{{Karunakaran}, {Spekkens}, {Bennet}, {Sand},
  {Crnojevi{\'c}}  \& {Zaritsky}}{{Karunakaran} et~al.}{2020}]{karunakaran2019}
{Karunakaran} A.,  {Spekkens} K.,  {Bennet} P.,  {Sand} D.~J.,  {Crnojevi{\'c}}
  D.,   {Zaritsky} D.,  2020, \mn@doi [\aj] {10.3847/1538-3881/ab5af1}, \href
  {https://ui.adsabs.harvard.edu/abs/2020AJ....159...37K} {159, 37}

\bibitem[\protect\citeauthoryear{{Karunakaran} et~al.,}{{Karunakaran}
  et~al.}{2021}]{karunakaran2021}
{Karunakaran} A.,  et~al., 2021, \mn@doi [\apjl] {10.3847/2041-8213/ac0e3a},
  \href {https://ui.adsabs.harvard.edu/abs/2021ApJ...916L..19K} {916, L19}

\bibitem[\protect\citeauthoryear{{Karunakaran}, {Spekkens}, {Carroll}, {Sand},
  {Bennet}, {Crnojevi{\'c}}, {Jones}  \& {Mutlu-Pakdil}}{{Karunakaran}
  et~al.}{2022a}]{karunakaran2022}
{Karunakaran} A.,  {Spekkens} K.,  {Carroll} R.,  {Sand} D.~J.,  {Bennet} P.,
  {Crnojevi{\'c}} D.,  {Jones} M.~G.,   {Mutlu-Pakdil} B.,  2022a, arXiv
  e-prints, \href {https://ui.adsabs.harvard.edu/abs/2022arXiv220611907K} {p.
  arXiv:2206.11907}

\bibitem[\protect\citeauthoryear{{Karunakaran}, {Sand}, {Jones}, {Spekkens},
  {Bennet}, {Crnojevi{\'c}}, {Mutlu-Pakdil}  \& {Zaritsky}}{{Karunakaran}
  et~al.}{2022b}]{karunakaran2022b}
{Karunakaran} A.,  {Sand} D.~J.,  {Jones} M.~G.,  {Spekkens} K.,  {Bennet} P.,
  {Crnojevi{\'c}} D.,  {Mutlu-Pakdil} B.,   {Zaritsky} D.,  2022b, \mn@doi
  [arXiv e-prints] {10.48550/arXiv.2210.03748}, \href
  {https://ui.adsabs.harvard.edu/abs/2022arXiv221003748K} {p. arXiv:2210.03748}

\bibitem[\protect\citeauthoryear{{Kawinwanichakij} et~al.,}{{Kawinwanichakij}
  et~al.}{2017}]{kawinwanichakij2017}
{Kawinwanichakij} L.,  et~al., 2017, \mn@doi [\apj] {10.3847/1538-4357/aa8b75},
  \href {https://ui.adsabs.harvard.edu/abs/2017ApJ...847..134K} {847, 134}

\bibitem[\protect\citeauthoryear{{Kennedy}, {Frenk}, {Cole}  \&
  {Benson}}{{Kennedy} et~al.}{2014}]{Kennedy2014}
{Kennedy} R.,  {Frenk} C.,  {Cole} S.,   {Benson} A.,  2014, \mn@doi [\mnras]
  {10.1093/mnras/stu719}, \href
  {https://ui.adsabs.harvard.edu/abs/2014MNRAS.442.2487K} {442, 2487}

\bibitem[\protect\citeauthoryear{{Kim} \& {Jerjen}}{{Kim} \&
  {Jerjen}}{2015}]{Kim15b}
{Kim} D.,  {Jerjen} H.,  2015, \mn@doi [\apjl] {10.1088/2041-8205/808/2/L39},
  \href {http://adsabs.harvard.edu/abs/2015ApJ...808L..39K} {808, L39}

\bibitem[\protect\citeauthoryear{{Kim} \& {Peter}}{{Kim} \&
  {Peter}}{2021}]{Kim2021}
{Kim} S.~Y.,  {Peter} A. H.~G.,  2021, arXiv e-prints, \href
  {https://ui.adsabs.harvard.edu/abs/2021arXiv210609050K} {p. arXiv:2106.09050}

\bibitem[\protect\citeauthoryear{{Kim}, {Jerjen}, {Mackey}, {Da Costa}  \&
  {Milone}}{{Kim} et~al.}{2015}]{Kim15a}
{Kim} D.,  {Jerjen} H.,  {Mackey} D.,  {Da Costa} G.~S.,   {Milone} A.~P.,
  2015, \mn@doi [\apjl] {10.1088/2041-8205/804/2/L44}, \href
  {http://adsabs.harvard.edu/abs/2015ApJ...804L..44K} {804, L44}

\bibitem[\protect\citeauthoryear{{Kim}, {Peter}  \& {Hargis}}{{Kim}
  et~al.}{2018}]{Kim2018}
{Kim} S.~Y.,  {Peter} A.~H.~G.,   {Hargis} J.~R.,  2018, \mn@doi [Physical
  Review Letters] {10.1103/PhysRevLett.121.211302}, \href
  {http://adsabs.harvard.edu/abs/2018PhRvL.121u1302K} {121, 211302}

\bibitem[\protect\citeauthoryear{{Kirby}, {Cohen}, {Guhathakurta}, {Cheng},
  {Bullock}  \& {Gallazzi}}{{Kirby} et~al.}{2013}]{Kirby13}
{Kirby} E.~N.,  {Cohen} J.~G.,  {Guhathakurta} P.,  {Cheng} L.,  {Bullock}
  J.~S.,   {Gallazzi} A.,  2013, \mn@doi [\apj] {10.1088/0004-637X/779/2/102},
  \href {http://adsabs.harvard.edu/abs/2013ApJ...779..102K} {779, 102}

\bibitem[\protect\citeauthoryear{{Klypin}, {Kravtsov}, {Valenzuela}  \&
  {Prada}}{{Klypin} et~al.}{1999}]{Klypin99}
{Klypin} A.,  {Kravtsov} A.~V.,  {Valenzuela} O.,   {Prada} F.,  1999, \mn@doi
  [\apj] {10.1086/307643}, \href
  {http://adsabs.harvard.edu/abs/1999ApJ...522...82K} {522, 82}

\bibitem[\protect\citeauthoryear{{Kochanek}, {Beacom}, {Kistler}, {Prieto},
  {Stanek}, {Thompson}  \& {Y{\"u}ksel}}{{Kochanek} et~al.}{2008}]{Kochanek08}
{Kochanek} C.~S.,  {Beacom} J.~F.,  {Kistler} M.~D.,  {Prieto} J.~L.,  {Stanek}
  K.~Z.,  {Thompson} T.~A.,   {Y{\"u}ksel} H.,  2008, \mn@doi [\apj]
  {10.1086/590053}, \href {http://adsabs.harvard.edu/abs/2008ApJ...684.1336K}
  {684, 1336}

\bibitem[\protect\citeauthoryear{{Kondapally}, {Russell}, {Conselice}  \&
  {Penny}}{{Kondapally} et~al.}{2018}]{Kondapally2018}
{Kondapally} R.,  {Russell} G.~A.,  {Conselice} C.~J.,   {Penny} S.~J.,  2018,
  \mn@doi [\mnras] {10.1093/mnras/sty2333}, \href
  {https://ui.adsabs.harvard.edu/abs/2018MNRAS.481.1759K} {481, 1759}

\bibitem[\protect\citeauthoryear{{Koposov}, {Yoo}, {Rix}, {Weinberg},
  {Macci{\`o}}  \& {Escud{\'e}}}{{Koposov} et~al.}{2009}]{Koposov2009}
{Koposov} S.~E.,  {Yoo} J.,  {Rix} H.-W.,  {Weinberg} D.~H.,  {Macci{\`o}}
  A.~V.,   {Escud{\'e}} J.~M.,  2009, \mn@doi [\apj]
  {10.1088/0004-637X/696/2/2179}, \href
  {https://ui.adsabs.harvard.edu/abs/2009ApJ...696.2179K} {696, 2179}

\bibitem[\protect\citeauthoryear{{Koposov}, {Belokurov}, {Torrealba}  \&
  {Evans}}{{Koposov} et~al.}{2015}]{Koposov15}
{Koposov} S.~E.,  {Belokurov} V.,  {Torrealba} G.,   {Evans} N.~W.,  2015,
  \mn@doi [\apj] {10.1088/0004-637X/805/2/130}, \href
  {http://adsabs.harvard.edu/abs/2015ApJ...805..130K} {805, 130}

\bibitem[\protect\citeauthoryear{{Koposov} et~al.,}{{Koposov}
  et~al.}{2018}]{Koposov2018}
{Koposov} S.~E.,  et~al., 2018, \mn@doi [\mnras] {10.1093/mnras/sty1772}, \href
  {https://ui.adsabs.harvard.edu/abs/2018MNRAS.479.5343K} {479, 5343}

\bibitem[\protect\citeauthoryear{{Kravtsov} \& {Manwadkar}}{{Kravtsov} \&
  {Manwadkar}}{2022}]{kravtsov2021grumpy}
{Kravtsov} A.,  {Manwadkar} V.,  2022, \mn@doi [\mnras]
  {10.1093/mnras/stac1439}, \href
  {https://ui.adsabs.harvard.edu/abs/2022MNRAS.514.2667K} {514, 2667}

\bibitem[\protect\citeauthoryear{{Laevens} et~al.,}{{Laevens}
  et~al.}{2015}]{Laevens15}
{Laevens} B.~P.~M.,  et~al., 2015, \mn@doi [\apj] {10.1088/0004-637X/813/1/44},
  \href {http://adsabs.harvard.edu/abs/2015ApJ...813...44L} {813, 44}

\bibitem[\protect\citeauthoryear{{Lan}, {M{\'e}nard}  \& {Mo}}{{Lan}
  et~al.}{2016}]{Lan16}
{Lan} T.-W.,  {M{\'e}nard} B.,   {Mo} H.,  2016, \mn@doi [\mnras]
  {10.1093/mnras/stw898}, \href
  {http://adsabs.harvard.edu/abs/2016MNRAS.459.3998L} {459, 3998}

\bibitem[\protect\citeauthoryear{{Larson}, {Tinsley}  \& {Caldwell}}{{Larson}
  et~al.}{1980}]{larson1980}
{Larson} R.~B.,  {Tinsley} B.~M.,   {Caldwell} C.~N.,  1980, \mn@doi [ApJ]
  {10.1086/157917}, \href {http://adsabs.harvard.edu/abs/1980ApJ...237..692L}
  {237, 692}

\bibitem[\protect\citeauthoryear{Leroy, Walter, Brinks, Bigiel, De~Blok, Madore
   \& Thornley}{Leroy et~al.}{2008}]{Leroy2008THINGS}
Leroy A.~K.,  Walter F.,  Brinks E.,  Bigiel F.,  De~Blok W.,  Madore B.,
  Thornley M.,  2008, The Astronomical Journal, 136, 2782

\bibitem[\protect\citeauthoryear{{Leroy} et~al.,}{{Leroy}
  et~al.}{2019}]{Leroy2019}
{Leroy} A.~K.,  et~al., 2019, \mn@doi [\apjs] {10.3847/1538-4365/ab3925}, \href
  {https://ui.adsabs.harvard.edu/abs/2019ApJS..244...24L} {244, 24}

\bibitem[\protect\citeauthoryear{{Li}, {Greene}, {Carlsten}  \& {Danieli}}{{Li}
  et~al.}{2024}]{li2024}
{Li} J.,  {Greene} J.~E.,  {Carlsten} S.~G.,   {Danieli} S.,  2024, \mn@doi
  [arXiv e-prints] {10.48550/arXiv.2406.00101}, \href
  {https://ui.adsabs.harvard.edu/abs/2024arXiv240600101L} {p. arXiv:2406.00101}

\bibitem[\protect\citeauthoryear{{Mao}, {Geha}, {Wechsler}, {Weiner},
  {Tollerud}, {Nadler}  \& {Kallivayalil}}{{Mao} et~al.}{2021}]{Mao2021}
{Mao} Y.-Y.,  {Geha} M.,  {Wechsler} R.~H.,  {Weiner} B.,  {Tollerud} E.~J.,
  {Nadler} E.~O.,   {Kallivayalil} N.,  2021, \mn@doi [\apj]
  {10.3847/1538-4357/abce58}, \href
  {https://ui.adsabs.harvard.edu/abs/2021ApJ...907...85M} {907, 85}

\bibitem[\protect\citeauthoryear{{Mao} et~al.,}{{Mao} et~al.}{2024}]{Mao2024}
{Mao} Y.-Y.,  et~al., 2024, \mn@doi [arXiv e-prints]
  {10.48550/arXiv.2404.14498}, \href
  {https://ui.adsabs.harvard.edu/abs/2024arXiv240414498M} {p. arXiv:2404.14498}

\bibitem[\protect\citeauthoryear{{Marigo} et~al.,}{{Marigo}
  et~al.}{2017}]{Marigo17}
{Marigo} P.,  et~al., 2017, \mn@doi [\apj] {10.3847/1538-4357/835/1/77}, \href
  {http://adsabs.harvard.edu/abs/2017ApJ...835...77M} {835, 77}

\bibitem[\protect\citeauthoryear{{Martin} et~al.,}{{Martin}
  et~al.}{2005}]{GALEX2005}
{Martin} D.~C.,  et~al., 2005, \mn@doi [\apjl] {10.1086/426387}, \href
  {https://ui.adsabs.harvard.edu/abs/2005ApJ...619L...1M} {619, L1}

\bibitem[\protect\citeauthoryear{{Mart{\'\i}nez-Delgado}
  et~al.,}{{Mart{\'\i}nez-Delgado} et~al.}{2021}]{martinez-delgado2021}
{Mart{\'\i}nez-Delgado} D.,  et~al., 2021, \mn@doi [\aap]
  {10.1051/0004-6361/202141242}, \href
  {https://ui.adsabs.harvard.edu/abs/2021A&A...652A..48M} {652, A48}

\bibitem[\protect\citeauthoryear{{Mau} et~al.,}{{Mau} et~al.}{2022}]{Mau2022}
{Mau} S.,  et~al., 2022, \mn@doi [\apj] {10.3847/1538-4357/ac6e65}, \href
  {https://ui.adsabs.harvard.edu/abs/2022ApJ...932..128M} {932, 128}

\bibitem[\protect\citeauthoryear{{Mayer}, {Mastropietro}, {Wadsley}, {Stadel}
  \& {Moore}}{{Mayer} et~al.}{2006}]{mayer2006}
{Mayer} L.,  {Mastropietro} C.,  {Wadsley} J.,  {Stadel} J.,   {Moore} B.,
  2006, \mn@doi [\mnras] {10.1111/j.1365-2966.2006.10403.x}, \href
  {http://adsabs.harvard.edu/abs/2006MNRAS.369.1021M} {369, 1021}

\bibitem[\protect\citeauthoryear{{McConnachie}}{{McConnachie}}{2012}]{McConnachie12}
{McConnachie} A.~W.,  2012, \mn@doi [\aj] {10.1088/0004-6256/144/1/4}, \href
  {http://adsabs.harvard.edu/abs/2012AJ....144....4M} {144, 4}

\bibitem[\protect\citeauthoryear{McConnachie \& Irwin}{McConnachie \&
  Irwin}{2006}]{McConnachieM31}
McConnachie A.~W.,  Irwin M.~J.,  2006, \mn@doi [Monthly Notices of the Royal
  Astronomical Society] {10.1111/j.1365-2966.2005.09771.x}, 365, 902

\bibitem[\protect\citeauthoryear{{McNanna} et~al.,}{{McNanna}
  et~al.}{2024}]{mcnanna2024}
{McNanna} M.,  et~al., 2024, \mn@doi [\apj] {10.3847/1538-4357/ad07d0}, \href
  {https://ui.adsabs.harvard.edu/abs/2024ApJ...961..126M} {961, 126}

\bibitem[\protect\citeauthoryear{McQuinn, Skillman, Dolphin, Berg  \&
  Kennicutt}{McQuinn et~al.}{2017}]{Mcquinn2017Distance}
McQuinn K.~B.,  Skillman E.~D.,  Dolphin A.~E.,  Berg D.,   Kennicutt R.,
  2017, The Astronomical Journal, 154, 51

\bibitem[\protect\citeauthoryear{{Menci}, {Grazian}, {Castellano}  \&
  {Sanchez}}{{Menci} et~al.}{2016}]{Menci2016}
{Menci} N.,  {Grazian} A.,  {Castellano} M.,   {Sanchez} N.~G.,  2016, \mn@doi
  [\apjl] {10.3847/2041-8205/825/1/L1}, \href
  {https://ui.adsabs.harvard.edu/abs/2016ApJ...825L...1M} {825, L1}

\bibitem[\protect\citeauthoryear{{Menci}, {Merle}, {Totzauer}, {Schneider},
  {Grazian}, {Castellano}  \& {Sanchez}}{{Menci} et~al.}{2017}]{Menci2017}
{Menci} N.,  {Merle} A.,  {Totzauer} M.,  {Schneider} A.,  {Grazian} A.,
  {Castellano} M.,   {Sanchez} N.~G.,  2017, \mn@doi [\apj]
  {10.3847/1538-4357/836/1/61}, \href
  {https://ui.adsabs.harvard.edu/abs/2017ApJ...836...61M} {836, 61}

\bibitem[\protect\citeauthoryear{{Moore}, {Quinn}, {Governato}, {Stadel}  \&
  {Lake}}{{Moore} et~al.}{1999}]{Moore_1999}
{Moore} B.,  {Quinn} T.,  {Governato} F.,  {Stadel} J.,   {Lake} G.,  1999,
  \mn@doi [\mnras] {10.1046/j.1365-8711.1999.03039.x}, \href
  {https://ui.adsabs.harvard.edu/abs/1999MNRAS.310.1147M} {310, 1147}

\bibitem[\protect\citeauthoryear{{Moster}, {Naab}  \& {White}}{{Moster}
  et~al.}{2013}]{Moster13}
{Moster} B.~P.,  {Naab} T.,   {White} S.~D.~M.,  2013, \mn@doi [\mnras]
  {10.1093/mnras/sts261}, \href
  {http://adsabs.harvard.edu/abs/2013MNRAS.428.3121M} {428, 3121}

\bibitem[\protect\citeauthoryear{{Moster}, {Naab}, {Lindstr{\"o}m}  \&
  {O'Leary}}{{Moster} et~al.}{2021}]{moster2021}
{Moster} B.~P.,  {Naab} T.,  {Lindstr{\"o}m} M.,   {O'Leary} J.~A.,  2021,
  \mn@doi [\mnras] {10.1093/mnras/stab1449}, \href
  {https://ui.adsabs.harvard.edu/abs/2021MNRAS.507.2115M} {507, 2115}

\bibitem[\protect\citeauthoryear{{Mu{\~n}oz} et~al.,}{{Mu{\~n}oz}
  et~al.}{2015}]{Munoz15}
{Mu{\~n}oz} R.~P.,  et~al., 2015, \mn@doi [\apjl]
  {10.1088/2041-8205/813/1/L15}, \href
  {http://adsabs.harvard.edu/abs/2015ApJ...813L..15M} {813, L15}

\bibitem[\protect\citeauthoryear{{M{\"u}ller}, {Pawlowski}, {Jerjen}  \&
  {Lelli}}{{M{\"u}ller} et~al.}{2018}]{Muller2018}
{M{\"u}ller} O.,  {Pawlowski} M.~S.,  {Jerjen} H.,   {Lelli} F.,  2018, \mn@doi
  [Science] {10.1126/science.aao1858}, \href
  {https://ui.adsabs.harvard.edu/abs/2018Sci...359..534M} {359, 534}

\bibitem[\protect\citeauthoryear{{Mutlu-Pakdil} et~al.,}{{Mutlu-Pakdil}
  et~al.}{2021}]{mutlu-pakdil2021}
{Mutlu-Pakdil} B.,  et~al., 2021, \mn@doi [\apj] {10.3847/1538-4357/ac0db8},
  \href {https://ui.adsabs.harvard.edu/abs/2021ApJ...918...88M} {918, 88}

\bibitem[\protect\citeauthoryear{{Mutlu-Pakdil} et~al.,}{{Mutlu-Pakdil}
  et~al.}{2022}]{mutlu-pakdil2022n253}
{Mutlu-Pakdil} B.,  et~al., 2022, \mn@doi [\apj] {10.3847/1538-4357/ac4418},
  \href {https://ui.adsabs.harvard.edu/abs/2022ApJ...926...77M} {926, 77}

\bibitem[\protect\citeauthoryear{{Mutlu-Pakdil} et~al.,}{{Mutlu-Pakdil}
  et~al.}{2024}]{mutlu-pakdil2024}
{Mutlu-Pakdil} B.,  et~al., 2024, \mn@doi [\apj] {10.3847/1538-4357/ad36c4},
  \href {https://ui.adsabs.harvard.edu/abs/2024ApJ...966..188M} {966, 188}

\bibitem[\protect\citeauthoryear{{Nadler}, {Gluscevic}, {Boddy}  \&
  {Wechsler}}{{Nadler} et~al.}{2019}]{Nadler2019microphys}
{Nadler} E.~O.,  {Gluscevic} V.,  {Boddy} K.~K.,   {Wechsler} R.~H.,  2019,
  \mn@doi [\apjl] {10.3847/2041-8213/ab1eb2}, \href
  {https://ui.adsabs.harvard.edu/abs/2019ApJ...878L..32N} {878, L32}

\bibitem[\protect\citeauthoryear{{Nadler} et~al.,}{{Nadler}
  et~al.}{2020}]{Nadler2020}
{Nadler} E.~O.,  et~al., 2020, \mn@doi [\apj] {10.3847/1538-4357/ab846a}, \href
  {https://ui.adsabs.harvard.edu/abs/2020ApJ...893...48N} {893, 48}

\bibitem[\protect\citeauthoryear{{Nadler} et~al.,}{{Nadler}
  et~al.}{2021}]{Nadler2021}
{Nadler} E.~O.,  et~al., 2021, \mn@doi [\prl] {10.1103/PhysRevLett.126.091101},
  \href {https://ui.adsabs.harvard.edu/abs/2021PhRvL.126i1101N} {126, 091101}

\bibitem[\protect\citeauthoryear{{Navarro}, {Frenk}  \& {White}}{{Navarro}
  et~al.}{1997}]{NFW1997}
{Navarro} J.~F.,  {Frenk} C.~S.,   {White} S. D.~M.,  1997, \mn@doi [\apj]
  {10.1086/304888}, \href
  {https://ui.adsabs.harvard.edu/abs/1997ApJ...490..493N} {490, 493}

\bibitem[\protect\citeauthoryear{{Neustadt}, {Kochanek}, {Stanek}, {Basinger},
  {Jayasinghe}, {Garling}, {Adams}  \& {Gerke}}{{Neustadt}
  et~al.}{2021}]{Neustadt2021}
{Neustadt} J.~M.~M.,  {Kochanek} C.~S.,  {Stanek} K.~Z.,  {Basinger} C.,
  {Jayasinghe} T.,  {Garling} C.~T.,  {Adams} S.~M.,   {Gerke} J.,  2021,
  \mn@doi [\mnras] {10.1093/mnras/stab2605}, \href
  {https://ui.adsabs.harvard.edu/abs/2021MNRAS.508..516N} {508, 516}

\bibitem[\protect\citeauthoryear{{Newton}, {Cautun}, {Jenkins}, {Frenk}  \&
  {Helly}}{{Newton} et~al.}{2018}]{Newton2018}
{Newton} O.,  {Cautun} M.,  {Jenkins} A.,  {Frenk} C.~S.,   {Helly} J.~C.,
  2018, \mn@doi [\mnras] {10.1093/mnras/sty1085}, \href
  {http://adsabs.harvard.edu/abs/2018MNRAS.479.2853N} {479, 2853}

\bibitem[\protect\citeauthoryear{{Newton} et~al.,}{{Newton}
  et~al.}{2021}]{Newton2021}
{Newton} O.,  et~al., 2021, \mn@doi [\jcap] {10.1088/1475-7516/2021/08/062},
  \href {https://ui.adsabs.harvard.edu/abs/2021JCAP...08..062N} {2021, 062}

\bibitem[\protect\citeauthoryear{{Ni}, {Wang}, {Feng}  \& {Di Matteo}}{{Ni}
  et~al.}{2019}]{Ni2019}
{Ni} Y.,  {Wang} M.-Y.,  {Feng} Y.,   {Di Matteo} T.,  2019, \mn@doi [\mnras]
  {10.1093/mnras/stz2085}, \href
  {https://ui.adsabs.harvard.edu/abs/2019MNRAS.488.5551N} {488, 5551}

\bibitem[\protect\citeauthoryear{{Nichols} \& {Bland-Hawthorn}}{{Nichols} \&
  {Bland-Hawthorn}}{2011}]{nichols2011}
{Nichols} M.,  {Bland-Hawthorn} J.,  2011, \mn@doi [\apj]
  {10.1088/0004-637X/732/1/17}, \href
  {http://adsabs.harvard.edu/abs/2011ApJ...732...17N} {732, 17}

\bibitem[\protect\citeauthoryear{{Nierenberg}, {Auger}, {Treu}, {Marshall},
  {Fassnacht}  \& {Busha}}{{Nierenberg} et~al.}{2012}]{Nierenberg2012}
{Nierenberg} A.~M.,  {Auger} M.~W.,  {Treu} T.,  {Marshall} P.~J.,  {Fassnacht}
  C.~D.,   {Busha} M.~T.,  2012, \mn@doi [\apj] {10.1088/0004-637X/752/2/99},
  \href {https://ui.adsabs.harvard.edu/abs/2012ApJ...752...99N} {752, 99}

\bibitem[\protect\citeauthoryear{{Nierenberg}, {Treu}, {Menci}, {Lu}, {Torrey}
  \& {Vogelsberger}}{{Nierenberg} et~al.}{2016}]{nierenberg2016}
{Nierenberg} A.~M.,  {Treu} T.,  {Menci} N.,  {Lu} Y.,  {Torrey} P.,
  {Vogelsberger} M.,  2016, \mn@doi [\mnras] {10.1093/mnras/stw1860}, \href
  {https://ui.adsabs.harvard.edu/abs/2016MNRAS.462.4473N} {462, 4473}

\bibitem[\protect\citeauthoryear{{Olsen} \& {Gawiser}}{{Olsen} \&
  {Gawiser}}{2022}]{olsen2022}
{Olsen} C.,  {Gawiser} E.,  2022, arXiv e-prints, \href
  {https://ui.adsabs.harvard.edu/abs/2022arXiv221005637O} {p. arXiv:2210.05637}

\bibitem[\protect\citeauthoryear{{Oman}, {Bah{\'e}}, {Healy}, {Hess}, {Hudson}
  \& {Verheijen}}{{Oman} et~al.}{2021}]{Oman2021}
{Oman} K.~A.,  {Bah{\'e}} Y.~M.,  {Healy} J.,  {Hess} K.~M.,  {Hudson} M.~J.,
  {Verheijen} M. A.~W.,  2021, \mn@doi [\mnras] {10.1093/mnras/staa3845}, \href
  {https://ui.adsabs.harvard.edu/abs/2021MNRAS.501.5073O} {501, 5073}

\bibitem[\protect\citeauthoryear{Papastergis, Cattaneo, Huang, Giovanelli  \&
  Haynes}{Papastergis et~al.}{2012}]{papastergis2012}
Papastergis E.,  Cattaneo A.,  Huang S.,  Giovanelli R.,   Haynes M.~P.,  2012,
  The Astrophysical Journal, 759, 138

\bibitem[\protect\citeauthoryear{{Pearson} et~al.,}{{Pearson}
  et~al.}{2016}]{pearson2016}
{Pearson} S.,  et~al., 2016, \mn@doi [\mnras] {10.1093/mnras/stw757}, \href
  {https://ui.adsabs.harvard.edu/abs/2016MNRAS.459.1827P} {459, 1827}

\bibitem[\protect\citeauthoryear{{Pisano}, {Wilcots}  \& {Elmegreen}}{{Pisano}
  et~al.}{1998}]{Pisano1998}
{Pisano} D.~J.,  {Wilcots} E.~M.,   {Elmegreen} B.~G.,  1998, \mn@doi [\aj]
  {10.1086/300239}, \href
  {https://ui.adsabs.harvard.edu/abs/1998AJ....115..975P} {115, 975}

\bibitem[\protect\citeauthoryear{{Plummer}}{{Plummer}}{1911}]{plummer1911}
{Plummer} H.~C.,  1911, \mn@doi [\mnras] {10.1093/mnras/71.5.460}, \href
  {https://ui.adsabs.harvard.edu/abs/1911MNRAS..71..460P} {71, 460}

\bibitem[\protect\citeauthoryear{{Putman}, {Zheng}, {Price-Whelan}, {Grcevich},
  {Johnson}, {Tollerud}  \& {Peek}}{{Putman} et~al.}{2021}]{Putman2021}
{Putman} M.~E.,  {Zheng} Y.,  {Price-Whelan} A.~M.,  {Grcevich} J.,  {Johnson}
  A.~C.,  {Tollerud} E.,   {Peek} J. E.~G.,  2021, \mn@doi [\apj]
  {10.3847/1538-4357/abe391}, \href
  {https://ui.adsabs.harvard.edu/abs/2021ApJ...913...53P} {913, 53}

\bibitem[\protect\citeauthoryear{{Ragazzoni} et~al.,}{{Ragazzoni}
  et~al.}{2006}]{Ragazzoni2006}
{Ragazzoni} R.,  et~al., 2006, in {Stepp} L.~M.,  ed.,  Society of
  Photo-Optical Instrumentation Engineers (SPIE) Conference Series Vol. 6267,
  Society of Photo-Optical Instrumentation Engineers (SPIE) Conference Series.
  p. 626710, \mn@doi{10.1117/12.673919}

\bibitem[\protect\citeauthoryear{{Read} \& {Erkal}}{{Read} \&
  {Erkal}}{2019}]{Read2019}
{Read} J.~I.,  {Erkal} D.,  2019, \mn@doi [\mnras] {10.1093/mnras/stz1320},
  \href {https://ui.adsabs.harvard.edu/abs/2019MNRAS.487.5799R} {487, 5799}

\bibitem[\protect\citeauthoryear{Reddick, Wechsler, Tinker  \&
  Behroozi}{Reddick et~al.}{2013}]{Reddick_2013}
Reddick R.~M.,  Wechsler R.~H.,  Tinker J.~L.,   Behroozi P.~S.,  2013, \mn@doi
  [The Astrophysical Journal] {10.1088/0004-637x/771/1/30}, 771, 30

\bibitem[\protect\citeauthoryear{{Rey}, {Pontzen}, {Agertz}, {Orkney}, {Read},
  {Saintonge}, {Kim}  \& {Das}}{{Rey} et~al.}{2022}]{Martin2022}
{Rey} M.~P.,  {Pontzen} A.,  {Agertz} O.,  {Orkney} M. D.~A.,  {Read} J.~I.,
  {Saintonge} A.,  {Kim} S.~Y.,   {Das} P.,  2022, \mn@doi [\mnras]
  {10.1093/mnras/stac502}, \href
  {https://ui.adsabs.harvard.edu/abs/2022MNRAS.511.5672R} {511, 5672}

\bibitem[\protect\citeauthoryear{{Richings} et~al.,}{{Richings}
  et~al.}{2020}]{richings2020}
{Richings} J.,  et~al., 2020, \mn@doi [\mnras] {10.1093/mnras/stz3448}, \href
  {https://ui.adsabs.harvard.edu/abs/2020MNRAS.492.5780R} {492, 5780}

\bibitem[\protect\citeauthoryear{{Roberts}, {Nierenberg}  \& {Peter}}{{Roberts}
  et~al.}{2021}]{Roberts2021}
{Roberts} D.~M.,  {Nierenberg} A.~M.,   {Peter} A. H.~G.,  2021, \mn@doi
  [\mnras] {10.1093/mnras/stab069}, \href
  {https://ui.adsabs.harvard.edu/abs/2021MNRAS.502.1205R} {502, 1205}

\bibitem[\protect\citeauthoryear{{Rodr{\'\i}guez-Puebla}, {Primack},
  {Avila-Reese}  \& {Faber}}{{Rodr{\'\i}guez-Puebla}
  et~al.}{2017}]{rodriguez-puebla2017}
{Rodr{\'\i}guez-Puebla} A.,  {Primack} J.~R.,  {Avila-Reese} V.,   {Faber}
  S.~M.,  2017, \mn@doi [\mnras] {10.1093/mnras/stx1172}, \href
  {https://ui.adsabs.harvard.edu/abs/2017MNRAS.470..651R} {470, 651}

\bibitem[\protect\citeauthoryear{{Rodriguez Wimberly}, {Cooper}, {Fillingham},
  {Boylan-Kolchin}, {Bullock}  \& {Garrison-Kimmel}}{{Rodriguez Wimberly}
  et~al.}{2019}]{rodriguezwimberly2019}
{Rodriguez Wimberly} M.~K.,  {Cooper} M.~C.,  {Fillingham} S.~P.,
  {Boylan-Kolchin} M.,  {Bullock} J.~S.,   {Garrison-Kimmel} S.,  2019, \mn@doi
  [\mnras] {10.1093/mnras/sty3357}, \href
  {http://adsabs.harvard.edu/abs/2019MNRAS.483.4031R} {483, 4031}

\bibitem[\protect\citeauthoryear{{Rudakovskyi}, {Mesinger}, {Savchenko}  \&
  {Gillet}}{{Rudakovskyi} et~al.}{2021}]{Rudakovskyi2021}
{Rudakovskyi} A.,  {Mesinger} A.,  {Savchenko} D.,   {Gillet} N.,  2021,
  \mn@doi [\mnras] {10.1093/mnras/stab2333}, \href
  {https://ui.adsabs.harvard.edu/abs/2021MNRAS.507.3046R} {507, 3046}

\bibitem[\protect\citeauthoryear{{Sabbi} et~al.,}{{Sabbi}
  et~al.}{2018}]{Sabbi2018}
{Sabbi} E.,  et~al., 2018, \mn@doi [\apjs] {10.3847/1538-4365/aaa8e5}, \href
  {https://ui.adsabs.harvard.edu/abs/2018ApJS..235...23S} {235, 23}

\bibitem[\protect\citeauthoryear{{Sales}, {Wang}, {White}  \&
  {Navarro}}{{Sales} et~al.}{2013}]{Sales13}
{Sales} L.~V.,  {Wang} W.,  {White} S.~D.~M.,   {Navarro} J.~F.,  2013, \mn@doi
  [\mnras] {10.1093/mnras/sts054}, \href
  {http://adsabs.harvard.edu/abs/2013MNRAS.428..573S} {428, 573}

\bibitem[\protect\citeauthoryear{{Sameie}, {Benson}, {Sales}, {Yu}, {Moustakas}
   \& {Creasey}}{{Sameie} et~al.}{2019}]{Sameie2019}
{Sameie} O.,  {Benson} A.~J.,  {Sales} L.~V.,  {Yu} H.-b.,  {Moustakas} L.~A.,
   {Creasey} P.,  2019, \mn@doi [\apj] {10.3847/1538-4357/ab0824}, \href
  {https://ui.adsabs.harvard.edu/abs/2019ApJ...874..101S} {874, 101}

\bibitem[\protect\citeauthoryear{{Samuel}, {Wetzel}, {Santistevan}, {Tollerud},
  {Moreno}, {Boylan-Kolchin}, {Bailin}  \& {Pardasani}}{{Samuel}
  et~al.}{2022}]{samuel2022}
{Samuel} J.,  {Wetzel} A.,  {Santistevan} I.,  {Tollerud} E.,  {Moreno} J.,
  {Boylan-Kolchin} M.,  {Bailin} J.,   {Pardasani} B.,  2022, \mn@doi [\mnras]
  {10.1093/mnras/stac1706}, \href
  {https://ui.adsabs.harvard.edu/abs/2022MNRAS.514.5276S} {514, 5276}

\bibitem[\protect\citeauthoryear{{Samuel}, {Pardasani}, {Wetzel},
  {Santistevan}, {Boylan-Kolchin}, {Moreno}  \& {Faucher-Gigu{\`e}re}}{{Samuel}
  et~al.}{2023}]{Samuel2023}
{Samuel} J.,  {Pardasani} B.,  {Wetzel} A.,  {Santistevan} I.,
  {Boylan-Kolchin} M.,  {Moreno} J.,   {Faucher-Gigu{\`e}re} C.-A.,  2023,
  \mn@doi [\mnras] {10.1093/mnras/stad2576}, \href
  {https://ui.adsabs.harvard.edu/abs/2023MNRAS.525.3849S} {525, 3849}

\bibitem[\protect\citeauthoryear{{Sand}, {Spekkens}, {Crnojevi{\'c}}, {Hargis},
  {Willman}, {Strader}  \& {Grillmair}}{{Sand} et~al.}{2015}]{sand2015}
{Sand} D.~J.,  {Spekkens} K.,  {Crnojevi{\'c}} D.,  {Hargis} J.~R.,  {Willman}
  B.,  {Strader} J.,   {Grillmair} C.~J.,  2015, \mn@doi [\apjl]
  {10.1088/2041-8205/812/1/L13}, \href
  {http://adsabs.harvard.edu/abs/2015ApJ...812L..13S} {812, L13}

\bibitem[\protect\citeauthoryear{{Sand} et~al.,}{{Sand}
  et~al.}{2022}]{sand2022}
{Sand} D.~J.,  et~al., 2022, arXiv e-prints, \href
  {https://ui.adsabs.harvard.edu/abs/2022arXiv220509129S} {p. arXiv:2205.09129}

\bibitem[\protect\citeauthoryear{{Santos-Santos}, {Sales}, {Fattahi}  \&
  {Navarro}}{{Santos-Santos} et~al.}{2021}]{Santos-Santos2021}
{Santos-Santos} I. M.~E.,  {Sales} L.~V.,  {Fattahi} A.,   {Navarro} J.~F.,
  2021, arXiv e-prints, \href
  {https://ui.adsabs.harvard.edu/abs/2021arXiv211101158S} {p. arXiv:2111.01158}

\bibitem[\protect\citeauthoryear{{Sardone}, {Peter}, {Brooks}  \&
  {Kaczmarek}}{{Sardone} et~al.}{2024}]{sardone2024}
{Sardone} A.,  {Peter} A. H.~G.,  {Brooks} A.~M.,   {Kaczmarek} J.,  2024,
  \mn@doi [\apj] {10.3847/1538-4357/ad250f}, \href
  {https://ui.adsabs.harvard.edu/abs/2024ApJ...964..135S} {964, 135}

\bibitem[\protect\citeauthoryear{{Schlafly} \& {Finkbeiner}}{{Schlafly} \&
  {Finkbeiner}}{2011}]{Schlafly11}
{Schlafly} E.~F.,  {Finkbeiner} D.~P.,  2011, \mn@doi [\apj]
  {10.1088/0004-637X/737/2/103}, \href
  {http://adsabs.harvard.edu/abs/2011ApJ...737..103S} {737, 103}

\bibitem[\protect\citeauthoryear{Schlegel, Finkbeiner  \& Davis}{Schlegel
  et~al.}{1998}]{Schlegel1998}
Schlegel D.~J.,  Finkbeiner D.~P.,   Davis M.,  1998, \mn@doi [ApJ]
  {10.1086/305772}, 500, 525

\bibitem[\protect\citeauthoryear{{Schombert}, {Bothun}, {Schneider}  \&
  {McGaugh}}{{Schombert} et~al.}{1992}]{schombert1992}
{Schombert} J.~M.,  {Bothun} G.~D.,  {Schneider} S.~E.,   {McGaugh} S.~S.,
  1992, \mn@doi [\aj] {10.1086/116129}, \href
  {https://ui.adsabs.harvard.edu/abs/1992AJ....103.1107S} {103, 1107}

\bibitem[\protect\citeauthoryear{{Simon}}{{Simon}}{2019}]{simon2019}
{Simon} J.~D.,  2019, \mn@doi [\araa] {10.1146/annurev-astro-091918-104453},
  \href {https://ui.adsabs.harvard.edu/abs/2019ARA&A..57..375S} {57, 375}

\bibitem[\protect\citeauthoryear{{Slater} \& {Bell}}{{Slater} \&
  {Bell}}{2014}]{slater2014}
{Slater} C.~T.,  {Bell} E.~F.,  2014, \mn@doi [\apj]
  {10.1088/0004-637X/792/2/141}, \href
  {http://adsabs.harvard.edu/abs/2014ApJ...792..141S} {792, 141}

\bibitem[\protect\citeauthoryear{{Smercina}, {Bell}, {Price}, {D'Souza},
  {Slater}, {Bailin}, {Monachesi}  \& {Nidever}}{{Smercina}
  et~al.}{2018}]{smercina2018lonely}
{Smercina} A.,  {Bell} E.~F.,  {Price} P.~A.,  {D'Souza} R.,  {Slater} C.~T.,
  {Bailin} J.,  {Monachesi} A.,   {Nidever} D.,  2018, \mn@doi [\apj]
  {10.3847/1538-4357/aad2d6}, \href
  {https://ui.adsabs.harvard.edu/abs/2018ApJ...863..152S} {863, 152}

\bibitem[\protect\citeauthoryear{{Spekkens}, {Urbancic}, {Mason}, {Willman}  \&
  {Aguirre}}{{Spekkens} et~al.}{2014}]{Spekkens2014}
{Spekkens} K.,  {Urbancic} N.,  {Mason} B.~S.,  {Willman} B.,   {Aguirre}
  J.~E.,  2014, \mn@doi [\apjl] {10.1088/2041-8205/795/1/L5}, \href
  {https://ui.adsabs.harvard.edu/abs/2014ApJ...795L...5S} {795, L5}

\bibitem[\protect\citeauthoryear{{Spencer}, {Loebman}  \& {Yoachim}}{{Spencer}
  et~al.}{2014}]{Spencer2014}
{Spencer} M.,  {Loebman} S.,   {Yoachim} P.,  2014, \mn@doi [\apj]
  {10.1088/0004-637X/788/2/146}, \href
  {https://ui.adsabs.harvard.edu/abs/2014ApJ...788..146S} {788, 146}

\bibitem[\protect\citeauthoryear{{Stark} et~al.,}{{Stark}
  et~al.}{2016}]{Stark2016}
{Stark} D.~V.,  et~al., 2016, \mn@doi [\apj] {10.3847/0004-637X/832/2/126},
  \href {https://ui.adsabs.harvard.edu/abs/2016ApJ...832..126S} {832, 126}

\bibitem[\protect\citeauthoryear{{Steyrleithner}, {Hensler}  \&
  {Boselli}}{{Steyrleithner} et~al.}{2020}]{Steyrleithner2020}
{Steyrleithner} P.,  {Hensler} G.,   {Boselli} A.,  2020, \mn@doi [\mnras]
  {10.1093/mnras/staa775}, \href
  {https://ui.adsabs.harvard.edu/abs/2020MNRAS.494.1114S} {494, 1114}

\bibitem[\protect\citeauthoryear{{Stierwalt}, {Besla}, {Patton}, {Johnson},
  {Kallivayalil}, {Putman}, {Privon}  \& {Ross}}{{Stierwalt}
  et~al.}{2015}]{stierwalt2015}
{Stierwalt} S.,  {Besla} G.,  {Patton} D.,  {Johnson} K.,  {Kallivayalil} N.,
  {Putman} M.,  {Privon} G.,   {Ross} G.,  2015, \mn@doi [\apj]
  {10.1088/0004-637X/805/1/2}, \href
  {https://ui.adsabs.harvard.edu/abs/2015ApJ...805....2S} {805, 2}

\bibitem[\protect\citeauthoryear{{Strigari}, {Bullock}  \&
  {Kaplinghat}}{{Strigari} et~al.}{2007}]{Strigari_2007}
{Strigari} L.~E.,  {Bullock} J.~S.,   {Kaplinghat} M.,  2007, \mn@doi [\apjl]
  {10.1086/512976}, \href
  {https://ui.adsabs.harvard.edu/abs/2007ApJ...657L...1S} {657, L1}

\bibitem[\protect\citeauthoryear{{Tanaka}, {Chiba}, {Hayashi}, {Komiyama},
  {Okamoto}, {Cooper}, {Okamoto}  \& {Spitler}}{{Tanaka}
  et~al.}{2018}]{Tanaka2018}
{Tanaka} M.,  {Chiba} M.,  {Hayashi} K.,  {Komiyama} Y.,  {Okamoto} T.,
  {Cooper} A.~P.,  {Okamoto} S.,   {Spitler} L.,  2018, \mn@doi [\apj]
  {10.3847/1538-4357/aad9fe}, \href
  {https://ui.adsabs.harvard.edu/abs/2018ApJ...865..125T} {865, 125}

\bibitem[\protect\citeauthoryear{Tasitsiomi, Kravtsov, Wechsler  \&
  Primack}{Tasitsiomi et~al.}{2004}]{Tasitsiomi_2004}
Tasitsiomi A.,  Kravtsov A.~V.,  Wechsler R.~H.,   Primack J.~R.,  2004,
  \mn@doi [The Astrophysical Journal] {10.1086/423784}, 614, 533–546

\bibitem[\protect\citeauthoryear{{Theureau}, {Hanski}, {Coudreau}, {Hallet}  \&
  {Martin}}{{Theureau} et~al.}{2007}]{Theureau2007}
{Theureau} G.,  {Hanski} M.~O.,  {Coudreau} N.,  {Hallet} N.,   {Martin} J.~M.,
   2007, \mn@doi [\aap] {10.1051/0004-6361:20066187}, \href
  {https://ui.adsabs.harvard.edu/abs/2007A&A...465...71T} {465, 71}

\bibitem[\protect\citeauthoryear{{Tikhonov}, {Lebedev}  \&
  {Galazutdinova}}{{Tikhonov} et~al.}{2015}]{Tikhonov2015}
{Tikhonov} N.~A.,  {Lebedev} V.~S.,   {Galazutdinova} O.~A.,  2015, \mn@doi
  [Astronomy Letters] {10.1134/S1063773715060080}, \href
  {https://ui.adsabs.harvard.edu/abs/2015AstL...41..239T} {41, 239}

\bibitem[\protect\citeauthoryear{{Tollerud}, {Bullock}, {Strigari}  \&
  {Willman}}{{Tollerud} et~al.}{2008}]{Tollerud2008}
{Tollerud} E.~J.,  {Bullock} J.~S.,  {Strigari} L.~E.,   {Willman} B.,  2008,
  \mn@doi [\apj] {10.1086/592102}, \href
  {http://adsabs.harvard.edu/abs/2008ApJ...688..277T} {688, 277}

\bibitem[\protect\citeauthoryear{{Tully} et~al.,}{{Tully}
  et~al.}{2013}]{Tully2013}
{Tully} R.~B.,  et~al., 2013, \mn@doi [\aj] {10.1088/0004-6256/146/4/86}, \href
  {https://ui.adsabs.harvard.edu/abs/2013AJ....146...86T} {146, 86}

\bibitem[\protect\citeauthoryear{{Walsh}, {Willman}  \& {Jerjen}}{{Walsh}
  et~al.}{2009}]{Walsh2009}
{Walsh} S.~M.,  {Willman} B.,   {Jerjen} H.,  2009, \mn@doi [\aj]
  {10.1088/0004-6256/137/1/450}, \href
  {https://ui.adsabs.harvard.edu/abs/2009AJ....137..450W} {137, 450}

\bibitem[\protect\citeauthoryear{{Wang} et~al.,}{{Wang}
  et~al.}{2017}]{Wang2017}
{Wang} L.,  et~al., 2017, \mn@doi [\mnras] {10.1093/mnras/stx788}, \href
  {https://ui.adsabs.harvard.edu/abs/2017MNRAS.468.4579W} {468, 4579}

\bibitem[\protect\citeauthoryear{{Wang} et~al.,}{{Wang}
  et~al.}{2024}]{Wang2024}
{Wang} Y.,  et~al., 2024, \mn@doi [arXiv e-prints] {10.48550/arXiv.2404.14500},
  \href {https://ui.adsabs.harvard.edu/abs/2024arXiv240414500W} {p.
  arXiv:2404.14500}

\bibitem[\protect\citeauthoryear{{Wechsler} \& {Tinker}}{{Wechsler} \&
  {Tinker}}{2018}]{WechslerTinker2018}
{Wechsler} R.~H.,  {Tinker} J.~L.,  2018, \mn@doi [Annual Review of Astronomy
  and Astrophysics] {10.1146/annurev-astro-081817-051756}, \href
  {https://ui.adsabs.harvard.edu/\#abs/2018ARA&A..56..435W} {56, 435}

\bibitem[\protect\citeauthoryear{Weisz et~al.,}{Weisz et~al.}{2011}]{weisz2011}
Weisz D.~R.,  et~al., 2011, The Astrophysical Journal, 743, 8

\bibitem[\protect\citeauthoryear{{Weisz}, {Dolphin}, {Skillman}, {Holtzman},
  {Gilbert}, {Dalcanton}  \& {Williams}}{{Weisz} et~al.}{2014}]{weisz2014b}
{Weisz} D.~R.,  {Dolphin} A.~E.,  {Skillman} E.~D.,  {Holtzman} J.,  {Gilbert}
  K.~M.,  {Dalcanton} J.~J.,   {Williams} B.~F.,  2014, \mn@doi [\apj]
  {10.1088/0004-637X/789/2/147}, \href
  {http://adsabs.harvard.edu/abs/2014ApJ...789..147W} {789, 147}

\bibitem[\protect\citeauthoryear{{Wetzel}, {Tollerud}  \& {Weisz}}{{Wetzel}
  et~al.}{2015}]{wetzel2015}
{Wetzel} A.~R.,  {Tollerud} E.~J.,   {Weisz} D.~R.,  2015, \mn@doi [\apjl]
  {10.1088/2041-8205/808/1/L27}, \href
  {http://adsabs.harvard.edu/abs/2015ApJ...808L..27W} {808, L27}

\bibitem[\protect\citeauthoryear{Wetzel, Hopkins, Kim, Faucher-Gigu{\`e}re,
  Kere{\v{s}}  \& Quataert}{Wetzel et~al.}{2016}]{wetzel2016}
Wetzel A.~R.,  Hopkins P.~F.,  Kim J.-h.,  Faucher-Gigu{\`e}re C.-A.,
  Kere{\v{s}} D.,   Quataert E.,  2016, The Astrophysical Journal Letters, 827,
  L23

\bibitem[\protect\citeauthoryear{{Wheeler}, {Phillips}, {Cooper},
  {Boylan-Kolchin}  \& {Bullock}}{{Wheeler} et~al.}{2014}]{wheeler2014}
{Wheeler} C.,  {Phillips} J.~I.,  {Cooper} M.~C.,  {Boylan-Kolchin} M.,
  {Bullock} J.~S.,  2014, \mn@doi [\mnras] {10.1093/mnras/stu965}, \href
  {http://adsabs.harvard.edu/abs/2014MNRAS.442.1396W} {442, 1396}

\bibitem[\protect\citeauthoryear{{Willman} et~al.,}{{Willman}
  et~al.}{2005}]{Willman05a}
{Willman} B.,  et~al., 2005, \mn@doi [\aj] {10.1086/430214}, \href
  {http://adsabs.harvard.edu/abs/2005AJ....129.2692W} {129, 2692}

\bibitem[\protect\citeauthoryear{{Wright} et~al.,}{{Wright}
  et~al.}{2010}]{WISE2010A}
{Wright} E.~L.,  et~al., 2010, \mn@doi [\aj] {10.1088/0004-6256/140/6/1868},
  \href {https://ui.adsabs.harvard.edu/abs/2010AJ....140.1868W} {140, 1868}

\bibitem[\protect\citeauthoryear{{Wright}, {Brooks}, {Weisz}  \&
  {Christensen}}{{Wright} et~al.}{2019}]{wright2019}
{Wright} A.~C.,  {Brooks} A.~M.,  {Weisz} D.~R.,   {Christensen} C.~R.,  2019,
  \mn@doi [\mnras] {10.1093/mnras/sty2759}, \href
  {https://ui.adsabs.harvard.edu/abs/2019MNRAS.482.1176W} {482, 1176}

\bibitem[\protect\citeauthoryear{{Zhu} \& {Putman}}{{Zhu} \&
  {Putman}}{2023}]{zhu2023}
{Zhu} J.,  {Putman} M.~E.,  2023, \mn@doi [\mnras] {10.1093/mnras/stad695},
  \href {https://ui.adsabs.harvard.edu/abs/2023MNRAS.521.3765Z} {521, 3765}

\bibitem[\protect\citeauthoryear{{Zhu}, {Tonnesen}, {Bryan}  \& {Putman}}{{Zhu}
  et~al.}{2024}]{zhu2024}
{Zhu} J.,  {Tonnesen} S.,  {Bryan} G.~L.,   {Putman} M.~E.,  2024, \mn@doi
  [arXiv e-prints] {10.48550/arXiv.2404.00129}, \href
  {https://ui.adsabs.harvard.edu/abs/2024arXiv240400129Z} {p. arXiv:2404.00129}

\makeatother
\end{thebibliography}




\label{lastpage}
\end{document}